\documentclass[fleqn,usenatbib]{mnras}

\usepackage{newtxtext,newtxmath}

\usepackage[T1]{fontenc}
\usepackage{ae,aecompl}

\usepackage{multirow}

\usepackage{graphicx}	
\usepackage{amsmath,bm}	
\usepackage{amssymb}	
\usepackage{nicefrac}
\usepackage{hyperref} 
\usepackage[ruled]{algorithm2e}
\usepackage{algpseudocode}
\usepackage{tikz}
\usetikzlibrary{arrows,shapes,positioning,decorations,fit,calc,decorations.pathreplacing}
\usepackage{pgf}

\renewcommand{\AA}{\protect\hbox{$\overset{_{\circ}}{\text{A}}$}}
\newcommand{\Ha}{H$\alpha$}
\newcommand{\hii}{H{\sc II}}

\def\kpc{\mbox{kpc}}

\title[Hierarchical Bayesian approach for estimating physical properties]
{Hierarchical Bayesian approach for estimating physical properties in nearby galaxies: Age Maps (Paper II)}

\author[M.C. S\'anchez-Gil et al.]{
M. Carmen S\'anchez-Gil$^{1,2}$\thanks{E-mail: mcarmen.sanchez@uca.es},
Emilio J. Alfaro$^{2}$, Miguel Cervi\~{n}o$^{3,2,4}$, Enrique P\'erez$^{2}$,  
\newauthor  Joss Bland-Hawthorn$^{5}$, D.~Heath Jones$^{6}$
\\
$^{1}$Universidad de C\'adiz, Facultad de Ciencias, Puerto Real, Spain\\
$^{2}$Instituto de Astrof\'{i}sica de Andaluc\'ia (CSIC), 18008, Granada, Spain \\
$^3$Centro de Astrobiolog\'ia (CSIC/INTA), 28850 Torrej\'on de Ardoz, Madrid, Spain\\
$^{4}$Instituto de Astrof\'{i}sica de Canarias, c/v\'{i}a L\'actea s/n, 38205 La Laguna, Tenerife, Spain \\
$^{5}$Sydney Institute for Astronomy, School of Physics, University of Sydney, NSW 2006, Australia\\
$^{6}$English Language and Foundation Studies Centre, University of Newcastle, Callaghan NSW 2308, Australia
}

\date{Accepted 2018 November 12. Received  2018 October 26; in original form 2018 March 1}

\pubyear{2017}

\begin{document}
\label{firstpage}
\pagerange{\pageref{firstpage}--\pageref{lastpage}}
\maketitle

\begin{abstract}
One of the fundamental goals of modern astrophysics is to estimate the physical parameters of galaxies. 
We present a hierarchical Bayesian model to compute age maps from images in the \Ha\ line 
(taken with Taurus Tunable Filter, TTF), ultraviolet band (GALEX far UV, FUV), and infrared bands (Spitzer 24, 70, and 160 $\mu$m). 
We present the burst ages for young stellar populations  in a sample of nearby and nearly face-on galaxies. 
The \Ha\ to FUV flux ratio is a good relative indicator of the very recent star formation history (SFH). As a nascent star-forming region evolves, the \Ha\ line emission declines earlier than the UV continuum, leading to a decrease in the \Ha/FUV ratio. %
Using star-forming galaxy models, sampled with a probabilistic formalism, and
allowing for a variable fraction of ionizing photons in the clusters,
we obtain the corresponding theoretical ratio \Ha/FUV to compare with our observed flux ratios, and thus to estimate the ages of the observed regions. 
We take into account the mean uncertainties and the interrelationships between parameters when computing \Ha/FUV. 
We propose a Bayesian hierarchical model where a joint probability distribution is defined to determine the parameters (age, metallicity, IMF) from the observed data (the observed flux ratios \Ha/FUV).  The joint distribution of the parameters is described through  independent and identically distributed (i.i.d.) random variables generated through MCMC (Markov Chain Monte Carlo) techniques.
\end{abstract}

\begin{keywords}
methods: statistical, observational -- galaxies: spiral, starburst, star formation, structure.
\end{keywords}



\section{Introduction}
\label{sec:intro}

{
This work carries on the study of a sample of nearby galaxies, where star-forming regions are spatially resolved, in order to place the relationship between star formation, ultraviolet and \Ha\ emission on a stronger empirical foundation \citep[hereinafter Paper I]{2011MNRAS.415..753S}.  This paper focuses on the tools and  mathematical methodology applied, based on a Bayesian model that yields more accurate results. A first approach to  this Bayesian methodology can be found in \cite{2015JPhCS.633a2140S}. We refer to these two papers for more details about the motivation and interest in the study of the star formation history (SFH) and star formation rate (SFR) in galaxies, in particular by means of the comparison between \Ha\ and UV emission as a tracer of recent SFH.}

{
In this paper we continue with the study of the age maps started in Paper I, for three new galaxies, using the new age-dating methodology from Bayesian inference approach applied  pixel-wise. Appendix \ref{Sec:App3} contains the resulting age maps for the rest of the galaxies sample, corresponding to Paper I. 
}

{
Section \ref{sec:observations} describes the data and its reduction, the pixel-based mapping, as well as the extinction correction via the total Infrared (TIR) to FUV ratio. }

{
The stellar population model adopted and  its uncertainties when applied  pixel-wise is described in Section \ref{sec:SPM}. 
In our previous work \citep{2015JPhCS.633a2140S} we used the original {\sc Starburst99}  \citep{1999ApJS..123....3L, 2005ApJ...621..695V} code results. Here, {\sc Starburst99} has been modified to bring it closer with the theoretical model distribution assumed for the \Ha\ to FUV ratio (which also differs with the one  assumed in the previous work). Specifically, to obtain the number of ionizing photons, $Q(H)$, or the theoretical correlation $\rho$ between \Ha\ and FUV luminosities. 
}

{
Section 4 presents the new Bayesian inference methodologies proposed for age estimation generalized to estimate any parameter.
However, as we explain below, only the age variable is sensitive to the ratio \Ha/FUV. 
In Section \ref{sec:HBM} the age maps assume pixel independence, keeping the spatial resolution at pixel value, and providing a sample of the age posterior probability function given the observed flux ratios for each pixel of the image.
In Section \ref{sec:ImgSeg} we study how image segmentation may affect several issues, such as the possible dependence between adjacent pixels of sub-sampling effects of the IMF from stellar population models.
}

{
Finally, Section \ref{sec:results} presents an analysis of the age maps and the resulting age patterns in view of possible mechanisms of  galaxy structure and evolution, and  the generation of galactic spiral arms. 
Section \ref{sec:conclu} contains a brief discussion and the conclusions.
}

\section{Observations}
\label{sec:observations}

\begin{table}
	\centering
	\caption{Galaxy Parameters (Sourced from {\it NASA Extragalactic Database})}
	\begin{tabular}{lrrr}
		\hline
		\multirow{ 2}{*}{Galaxy} & NGC~1068 &NGC~ 5236 &NGC~5457  \\
		&  (M~77) &(M83) &(M101) \\
		\hline
		RA (J2000) &$02^h 42^m 40.7^s$ &$13h37m00.9s$ &$14^h 03^m 12.5^s$ \\[1.5ex]
		DEC (J2000) &$-00\degr 00' 48.0''$ & $-29^{\circ} 51' 56.0''$   & $+54^{\circ} 20' 56.0''$ \\[1.5ex]
		Type &(R)SA(rs)b&SAB(s)c & SAB(rs)cd \\[1.5ex]
		Redshift &0.003793 &0.001711 &0.000804 \\[1.5ex]
		Distance (Mpc) &14.4&$4.5\pm0.3$ & 6.7 \\[1.5ex]
		pc arcsec$^{-1}$& 69.8&21.82&32.5 \\[1.5ex]
		Inclination (deg) &$40\degr\pm3$ & $24^{\circ}$  &$18^{\circ}$ \\[1.5ex]
		Dim. (arcmin) & $7.1\times6.0$ &$2.9\times11.5$  & $28.8\times26.9$\\[1.5ex]
		M$_B$& $9.61$ &$8.20$ & $8.31$ \\
		\hline
		\label{tab:tab1}
	\end{tabular}
	\begin{flushleft}
		Distance reference:   \citet{1997Ap&SS.248..177B} for NGC~1068; M83 \citet{2003ApJ...590..256T};
		M101 \citet{2008ApJ...676..184T}
		-- Scale in pc per arcsec of the final images, and the age maps plots, which pixel scale is 1.5\arcsec/px. 
		-- Inclination angle reference:  \citet{1997Ap&SS.248..177B} for NGC~1068; M83 \citet{2012MNRAS.421.2917F};
		M101 \citet{1981A&A....93..106B}
	\end{flushleft}
\end{table}

\begin{figure*}
\begin{center}
\includegraphics[width=0.8\textwidth]{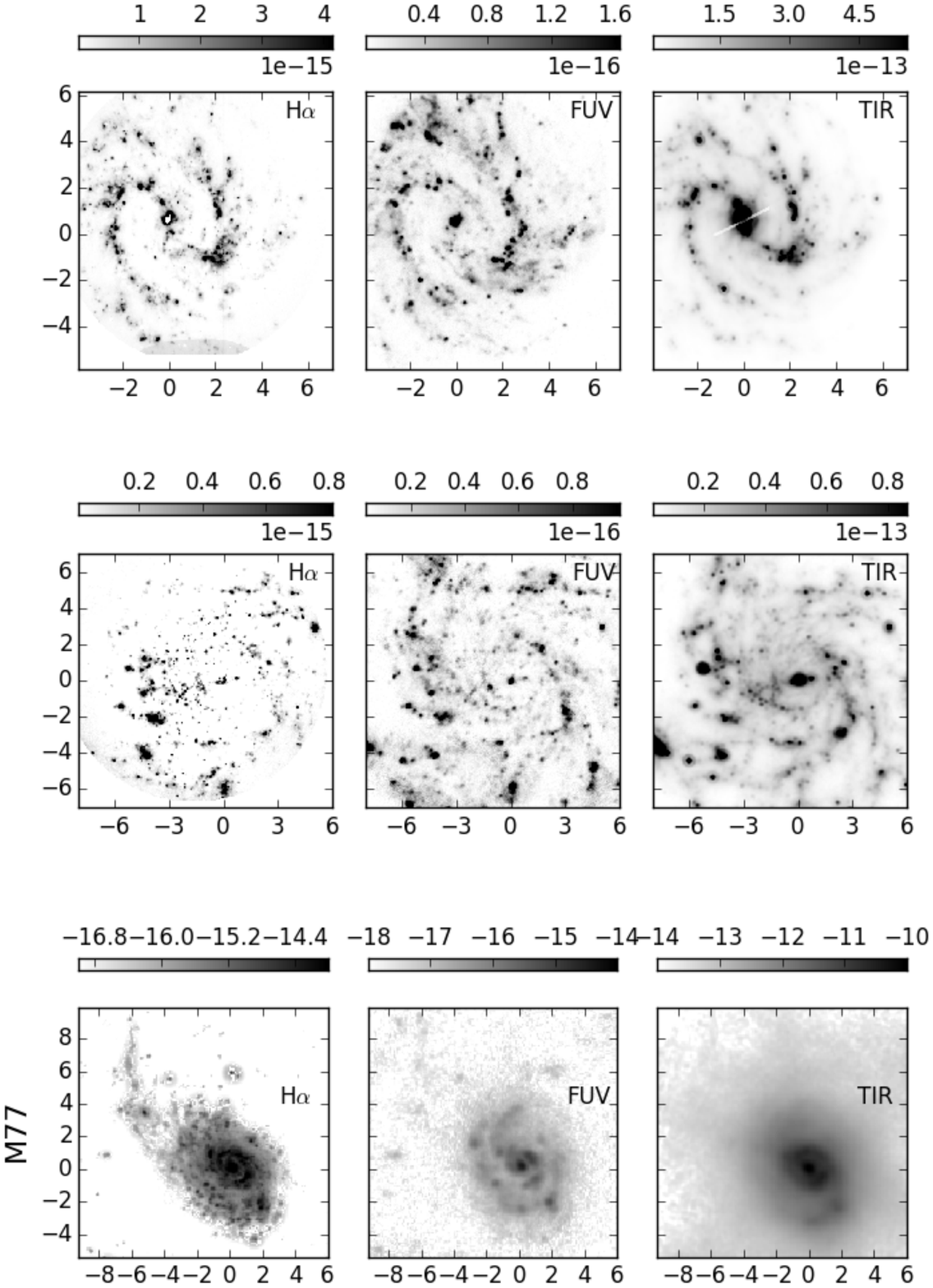}
\caption{The processed frames for the three galaxies. 
By row:  M83 (top), M101 (centre), and M77 (bottom).
By columns: 
\Ha\ images taken with the TTF at the WHT, in $erg\,s^{-1}cm^{-2}$ (left), 
FUV images from GALEX survey, in $\smash{erg\,s^{-1}cm^{-2}\AA^{-1}}$(centre), and MIPS/SPITZER infra red images, in units of $erg\,s^{-1}cm^{-2}$ (right). 
Images have been resampled to have identical size, orientation and pixel scale (1.5 arcsec pixel$^{-1}$).
The coordinates and axis scales are given in \kpc, with respect to the galaxy centre (North up, East left).
}
\label{fig:Fig1}
\end{center}
\end{figure*}

We study the case of three new galaxies in addition to the original sample in Paper I,  
which are also included in Appendix \ref{Sec:App1} for comparison between the present Bayesian methodology and our earlier empirical model.
This sample contains face-on nearby galaxies across a range of star-forming types. Low inclination angles mitigate the effects of extinction and scattering as well as wavelength shift in \Ha\ due to galactic rotation. 
Their proximity allows sufficient spatial resolution to resolve individual star forming structures within spiral arms.
A summary of  the main physical properties of the sample is given in Table~\ref{tab:tab1}. 

The \Ha\ images were taken with the Taurus Tunable Filter 
\citep[TTF,][]{1998PASA...15...44B,2002MNRAS.329..759J,2001ApJ...550..593J} on the William Herschel Telescope (WHT) during 1999 March 4$-$6.
Conditions were photometric with stable seeing of 1.0 arcsec.  TTF was tuned to a bandpass
of width $\Delta \lambda = 20 $\,\AA\  centred at $\lambda_{\rm c} = 6570$\,\AA. The intermediate-width 
R0 blocking filter ($\lambda_{\rm c}/\Delta\lambda = 6680/210$\,\AA) was used to remove
transmission from all but a single interference order. The pixel scale was 0.56 arcsec.

The Far UV images were obtained from the Nearby Galaxies Survey of the Galaxy Evolution Explorer mission
\citep[NGS survey, GALEX,][]{2005ApJ...619L...1M}. This survey contains well-resolved imaging (1.5 arcsec pix) of 296 and 
433 nearby galaxies for GR2/GR3 and GR4 releases, respectively, in two passbands: a narrower far-ultraviolet 
band (FUV; $\lambda_{\rm eff}/\Delta \lambda=1516 / 268$\,\AA), and a broader near-ultraviolet band 
(NUV; $\lambda_{\rm eff} = 2267 / 732$\,\AA). 

Ancillary 24, 70, and 160$\mu$ data from the Multiband Imaging Photometer for Spitzer (MIPS) were used to provide estimates of extinction, in the same way as in Paper I. 
For M83 and M101, we obtained IR data from the Spitzer Local Volume Legacy Survey\footnote{http://irsa.ipac.caltech.edu/data/SPITZER/LVL} (LVL). 
These images were resampled to a common 1.5 arcsec/px scale  (same as the 24$\mu$ MIPS and FUV images), 
and combined into an image of  total far infrared (TIR) flux, according to 
$F_{TIR} = \zeta_1 \nu F_{\nu}(24 \mu)+ \zeta_2 \nu F_{\nu}(70 \mu)+\zeta_3 \nu F_{\nu}(160 \mu)$, 
with [$\zeta_1$,$\zeta_2$,$\zeta_2$] = [1.559,0.7686,1.347] \citep{2002ApJ...576..159D}. 
For NGC~1068 only 70 and 160$\mu$ data were available from 
the Very Nearby Galaxy Survey\footnote{http://irsa.ipac.caltech.edu/data/Herschel/VNGS/overview.html} (VNGS). 
In this case the TIR flux was estimated from the latter as 
$S_{TIR} = c_{70}S_{70}+c_{160}S_{160}$, 
according to Eq. (4) of \citet{2013MNRAS.431.1956G} (where $c_{70}=0.999\pm0.023$ and $c_{160}=1.226\pm 0.017$, from their Table 3), 
which is a quite reliable fit, with a coefficient $R^2 = 0.97$.

Internal reddening is corrected using a straight relation between the extinction A$_{FUV}$  and  $F_{TIR}/F_{FUV}$ from    
Eq. (2) of \citep{2005ApJ...619L..51B}. 
The A(H$\alpha$) extinction factor was derived through the relation $A_{FUV} = 1.4 A(H \alpha)$ \citep{2005ApJ...619L..83B}.

With all images on a common scale of $1.5$ arcsec/px, our pixel-by-pixel technique becomes straightforward to implement. 
An example of the processed frames in \Ha, FUV and TIR  can be found in Fig.~\ref{fig:Fig1}. 
The top panels display the different images for M83. Artifacts of the data reduction in the centre of the \Ha\ and TIR images can be seen. Those pixels were masked in the final age map.  
NGC~1068, the bottom panels,
show the range of morphologies in a single galaxy at different wavelengths.
The strong effect of the central AGN is also quite evident. 

The estimated calibration uncertainties for the MIPS images are  2\%, 5\%, and 9\% for the 24, 70, and 160$\mu$ data respectively  
\citep{2007ApJ...655..863D}.
Average relative errors of the respective images are $E_{H\alpha}
\leq5\%$, $E_{FUV}\simeq15-25\%$, and $E_{TIR}\leq 10\%$, resulting in an 
overall uncertainty in F$_{H\alpha}/$F$_{FUV}$ (reddening corrected) flux ratio of $\sim$23-28\%.
More details on the data, data reduction, and extinction correction can be  found in  Paper I.

\section{Stellar Population Models}
\label{sec:SPM}

To obtain the number of ionizing photons $Q(H)$ and the stellar and nebular contribution to the FUV/GALEX band, we implement  the probabilistic formalism by \citet{2006A&A...451..475C} into {\sc starburst99} synthesis models \cite[version 7.0.1 August 2014]{1999ApJS..123....3L, 2005ApJ...621..695V}. 
The original {\sc starburst99} code has been modified to obtain  $Q(H)$ and the  FUV/GALEX band emission  for each of the stars along each isochrone (SB99 obtains such values for the ensemble after computing the total spectra of the cluster). We have included the FUV/GALEX filter response (as provided by the SVO filter service\footnote{\tt http://svo2.cab.inta-csic.es/theory/fps3/}) to customize SB99 to obtain this quantity for each star. Finally, we have obtained the nebular contribution to FUV/GALEX using the implementation of the nebular continuum used by SB99. $Q(H)$ values are converted to \Ha\ fluxes using the conversion factors provided by \cite{1995ApJS...96....9L}, but multiplied by a factor $f_{Q(H)} = 1-f_\mathrm{scp}$ to account for the escape of ionizing photons, $f_\mathrm{scp}$ \cite[e.g.][]{1991A&AS...88..399M}, 
hence our conversion formula is $L(\mathrm{H}\alpha) = 1.36 \times 10^{-12}\, Q(H)\, (1 - f_\mathrm{scp})$.
Simiarly, we multiply the nebular component in FUV/GALEX  by $(1 - f_\mathrm{scp})$ before adding it to the stellar component.

As a result, we obtain the isochrone table used by SB99 including,  for each star, the \Ha\ flux, the nebular contribution to the FUV/GALEX luminosity, the stellar FUV/GALEX luminosity, and the contribution to the total luminosity (i.e. its IMF-weighted value). Isochrones are computed in the age interval from 0.1 to 20 Myrs with a linear time step of 0.1 Myr for the Geneva evolutionary track set. We use standard mass loss rates for metalicity values of 0.040, 0.020, 0.008, 0.004, and 0.001, corresponding to the evolutionary tracks presented in
\citep{1992A&AS...96..269S,1993A&AS..101..415C,1993A&AS..102..339S,1993A&AS...98..523S}. 
Stellar libraries are from {\sc basel}  \citep{1997A&AS..125..229L} for intermediate and low mass stars, and from \cite{2002MNRAS.337.1309S} for massive and Wolf-Rayet (WR) stars when present in the cluster.   
The code was run using a \cite{1955ApJ...121..161S} IMF slope within the mass range of $1-120\ \mathrm{M}_\odot$ with a total mass of 3.14 M$_\odot$, which is the mean stellar mass in such IMF interval. In this way our results are naturally normalized to the total number of stars instead of the total mass, a requirement to compute properly the theoretical covariance matrix.

The resulting components in each isochrone table were added following prescriptions by \citet{2006A&A...451..475C}  to obtain the mean values, variances (expressed as effective number of stars, ${\cal{N}}_\mathrm{eff}$, where  
$1/{\cal{N}}_\mathrm{eff} = (\sigma/\mu)^2$, see \citealt{2002A&A...381...51C}), skewness ($\gamma_1$), and kurtosis ($\gamma_2$) of the stellar luminosity functions  of $L(\mathrm{H}\alpha)$  and $L(\mathrm{FUV})$, and the covariance between both luminosities\footnote{Note in papers previous to \cite{2006A&A...451..475C} the correlation coefficient was obtained under the incorrect assumption of Poissonian statistics in each isochrone component, instead of multinomial statistics. For the correct formula to compute variances and covariances between \Ha\ and FUV/GALEX luminosities for each possible case see \cite{2006A&A...451..475C,2013NewAR..57..123C}.}. We also obtained the Lowest Luminosity Limit (LLL) \citep{2003A&A...407..177C,2004A&A...413..145C} for \Ha\ and FUV, which gives the luminosity of the brightest individual star in the model for the given band; i.e. an observed cluster less luminous than the LLL could not be modeled by the mean value of the ensembles obtained by SSPs models, since there is a confusion between the emission of the ensemble and the emission of a single star, and the overall probability distribution function (PDF) of the theoretical integrated luminosity, and not just its mean value, must be taken into account \citep{2004A&A...413..145C}.

\begin{figure*}
\centering
\includegraphics[width=\textwidth]{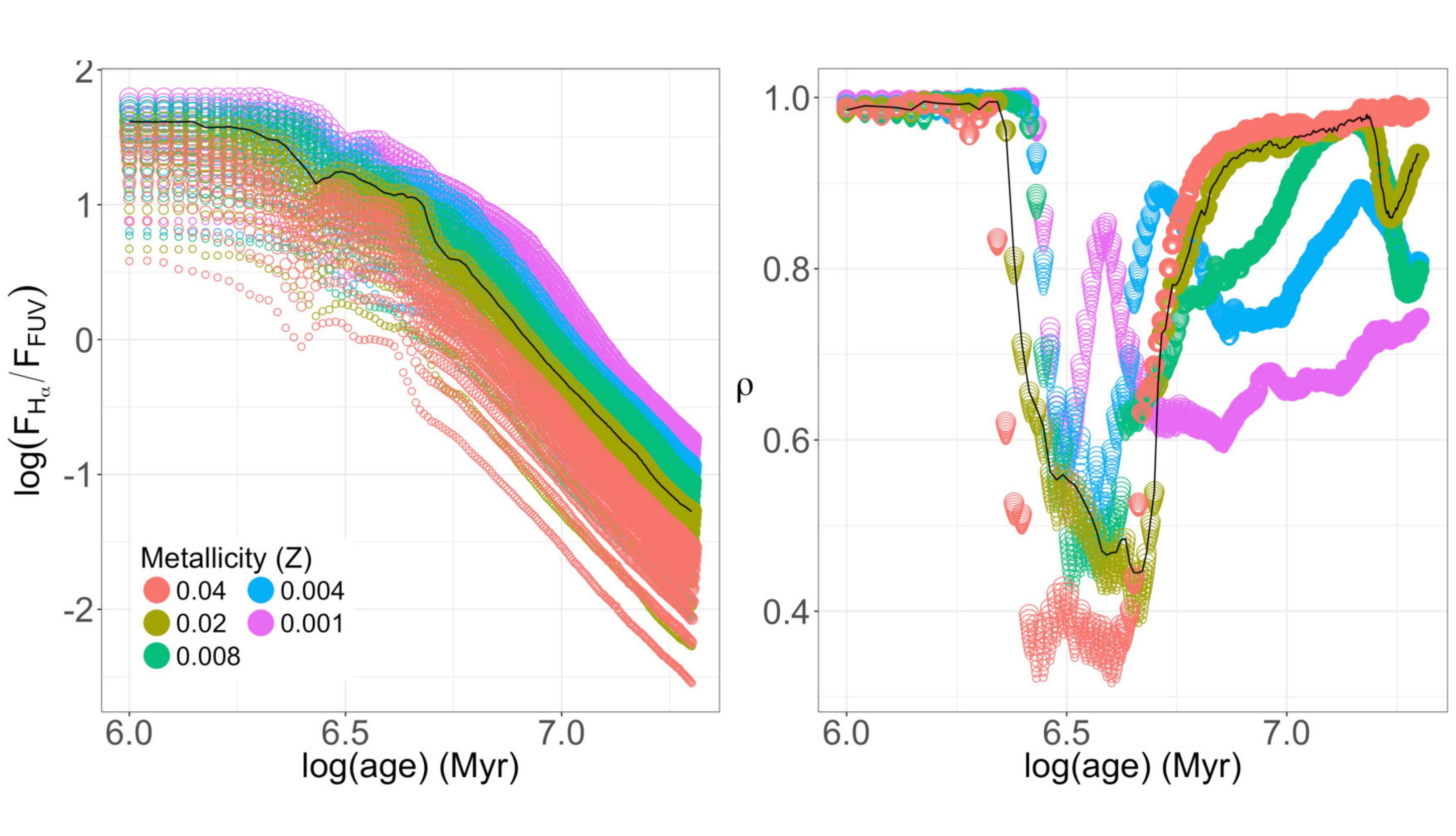}
\caption{Models from SB99. 
On the left, the model \Ha/FUV ratio vs.  age, ranging from 1 to 20 Myr, in log units. On the right, the theoretical correlation between \Ha\ and FUV luminosities vs.  age.
Both relationships are given for different combinations of the parameters: 
The metallicity $Z$ is represented with different colors, from $Z=0.001$ (purple points) to $Z=0.04$ (in red). The fraction of ionizing photons, $f_{Q(H)}$, is coded with different point sizes, ranging 10-100\% from the smallest to the largest.
The black line in both plots represents these model relationships for solar metallicity, $Z=0.02$, and $f_{Q(H)}=100\%$. 
}
\label{fig:Fig2}
\end{figure*}

The resulting evolution of the \Ha/FUV ratio mean values is shown in Fig.~\ref{fig:Fig2}, where we have tested that the mean fluxes and the ratio for the case of  $f_\mathrm{scp} = 0$ ($f_{Q(H)} = 1$) are coincident with SB99 results without modifications, which are also coincident with the parameters 
used in \citet{2011MNRAS.415..753S} with slight modifications due to the variation of stellar atmosphere models.

The right panel of Fig.~\ref{fig:Fig2} shows the resulting theoretical correlation coefficient between \Ha\ and FUV. The correlation coefficient is almost one up to 3 Myr, reflecting the fact that the same type of stars dominate both quantities. The correlation decreases abruptly during the Wolf-Rayet phase (WR) of the cluster, since \Ha\ is dominated by the WR stars, but the FUV is mainly produced by hot stars in the Main sequence. After the WR phase in the cluster, the correlation coefficient increases to values around $0.6$. It increases with  metallicity, which depends on the balance  between main sequence stars which still produce ionizing fluxes (O-B stars mainly) and those  that only produce FUV flux (B-A spectral types). The metallicity dependence is explained by the variation of the main sequence with metallicity: at low metallicity, the main sequence is hotter, hence FUV fluxes are produced by both ionizing and non-ionizing stars and the correlation coefficient is lower with respect to larger metallicities where FUV and ionizing stars are coincident. Finally, there is a small variation of the correlation coefficient with $f_{scp}$ since the escape of photons only affects the correlation coefficient by the nebular contribution to the FUV flux.

\begin{figure*}
\centering
\includegraphics[width=\textwidth]{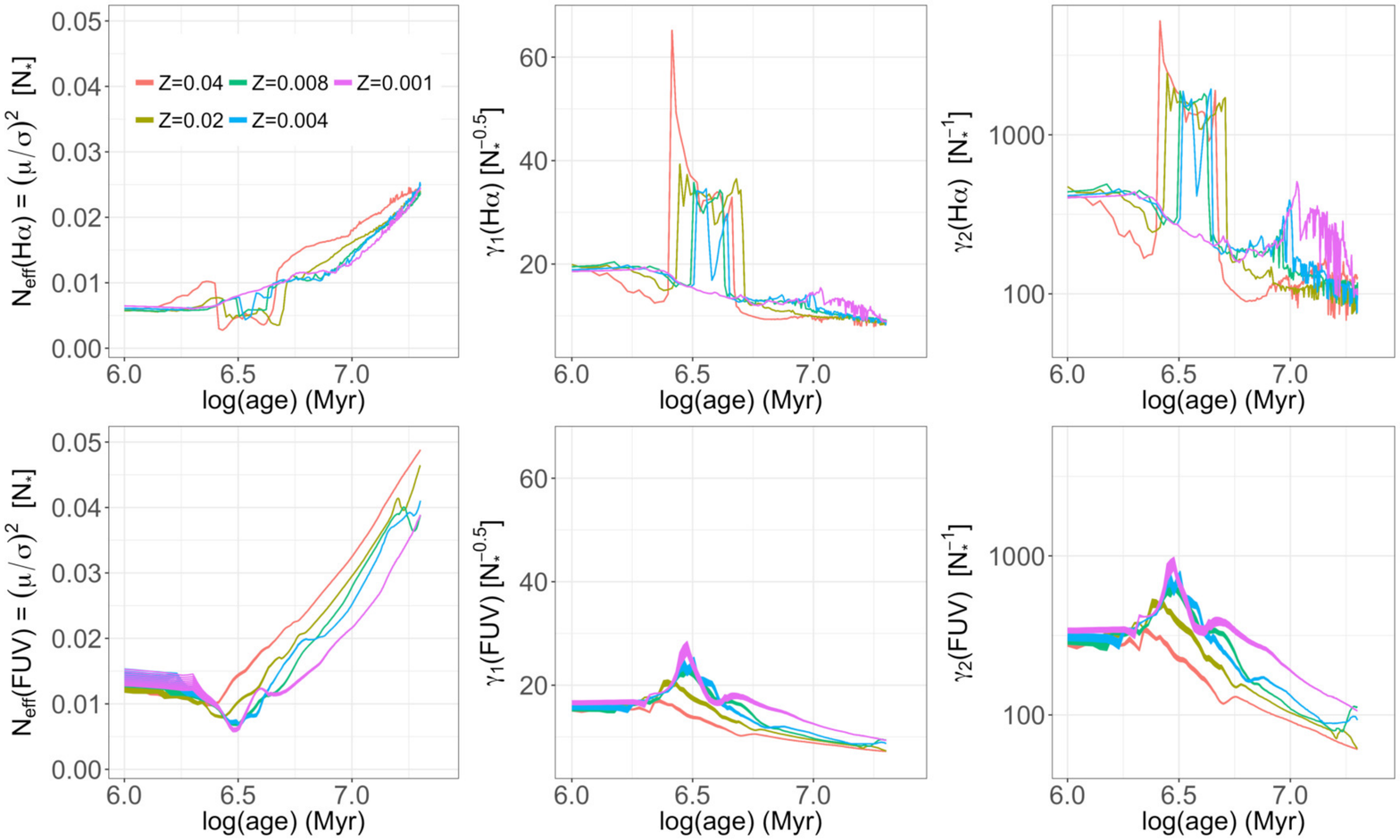}
\caption{Variance expressed in terms of effective number of stars (left), skewness (middle), and kurtosis (right) obtained from theoretical models. Top panels show the values for \Ha ~and bottom panels show the values for FUV as a function of the age. Metallicity follows the same color code as Fig.~\ref{fig:Fig2} 
}
\label{fig:Fig3}
\end{figure*}

Fig. \ref{fig:Fig3} shows the variance (expressed in terms of effective number ${\cal N}_\mathrm{eff}$ of stars), skewness ($\gamma_1$), and kurtosis ($\gamma_2$) obtained from theoretical models normalized to $N_* =1$. We recall that the effective number of stars scales with the number of stars; the skewness scales with the inverse of the square root of the number of stars, and the kurtosis scales with the inverse of the number of stars \citep{2006A&A...451..475C}. As reference, a relative dispersion of 4\% requires  that the used resolution element contains $6.25 \times 10^4$ stars; in that case $\gamma_1=0.08$ and $\gamma_2=0.008$ (assuming values of ${\cal N}_\mathrm{eff} = 0.01$, $\gamma_1= 20$, and  $\gamma_2 = 500$ when normalized to the number of stars). Such values imply that the underling distribution can be well represented by a Gaussian. In the case of a relative dispersion of 20\%,   only 2500 stars are required, but the values of $\gamma_1$ and  $\gamma_2$ increase to 0.4 and 0.2, respectively, and the underling distribution deviates from  Gaussian  (see Appendix  \ref{Sec:App0} for details).

The use of a pixel-wise age dating technique allows age mapping of the youngest stellar population  without prior assumptions about the spatial distribution of the star forming regions. 
It also provides a spatial characterization of the age distribution for HII regions in galaxies within the Local Volume through their spatially-resolved spiral arms or other galactic structures.
However, this pixel-based technique is subject to some systematic effects, including (i) the pixel-sized luminosities, and (ii) potential interaction between surrounding regions through adjacent pixels.  

\begin{figure*}
\centering
\includegraphics[width=\textwidth]{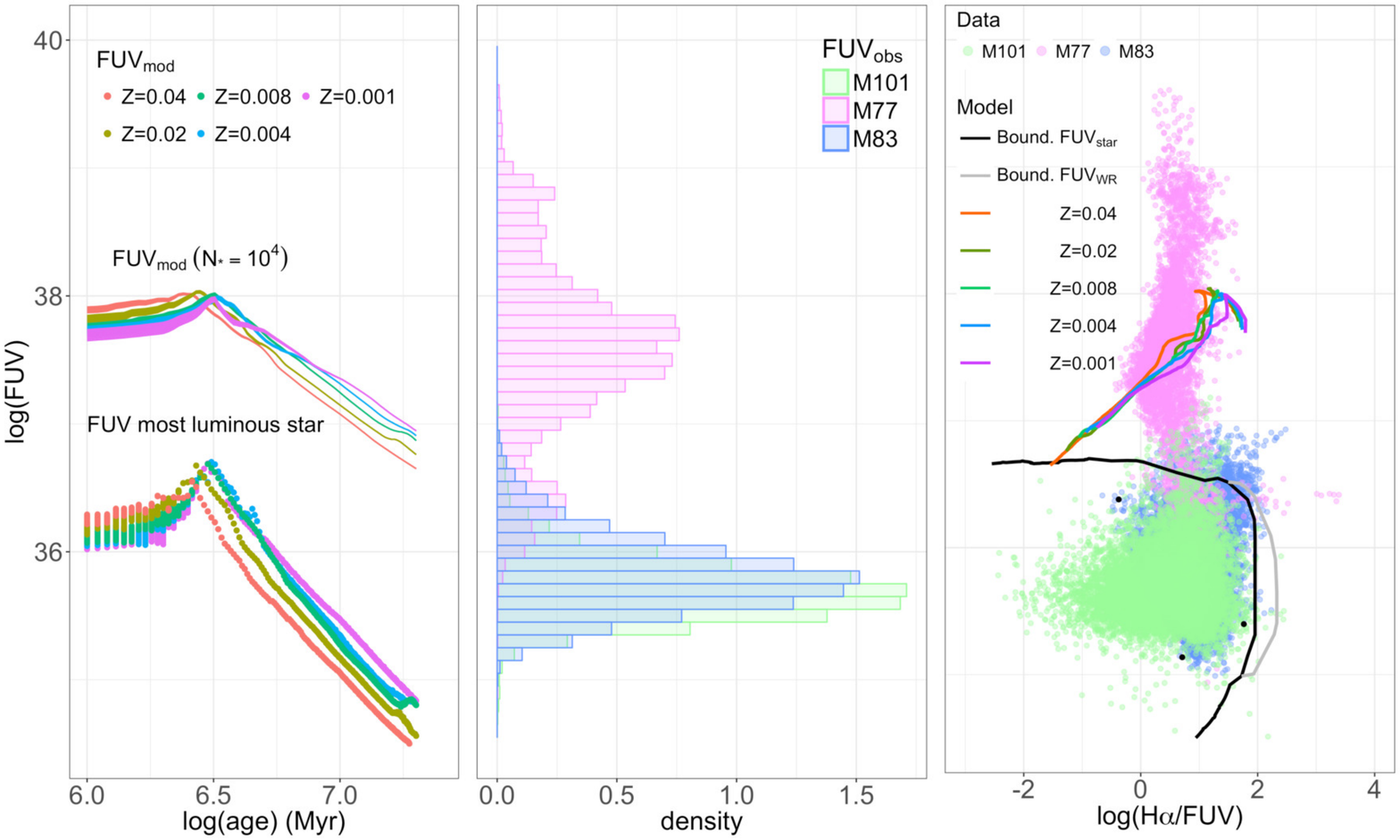}
\caption{Left: Time evolution of the FUV LLL for different metallicities; colors as in Fig. \ref{fig:Fig2}. The mean FUV integrated luminosity for $10^4$ stars is also shown for reference. 
Centre: Histograms of the FUV luminosities of the pixel-sized data of the three galaxies. 
Right: distribution of the pixel-sized data in the $\mathrm{H}\alpha/\mathrm{FUV}$ vs. $L(\mathrm{FUV})$ plane. Solid black and gray lines shows the boundaries of individual normal and WR stars respectively in this plane. Data points in between both lines corresponds to pixels-sized values that can be only explained by the presence of WR stars; data point with ratio larger than the boundary of the WR line requires some extra ionization source since are not compatible with the current models. The plot also shows the position of $\mathrm{H}\alpha/\mathrm{FUV}$ and the mean FUV integrated luminosity and for $10^4$ stars for the case of $f_\mathrm{scp}=0$ as reference. Finally, the three black point corresponds to the positions of the pixels shown in Fig. \ref{fig:Fig6}.}
\label{fig:Fig4}
\end{figure*}

The model validity was checked using a single stellar
population model in pixel-sized regions was checked by comparing our pixel-sized FUV luminosity with the possible LLL values in our age range.
We note that a similar test for  $L(\mathrm{H}\alpha)$ depends on  the factor $f_\mathrm{scp}$, so the LLL test is not decisive in this case.     
The result is shown in Fig. \ref{fig:Fig4}, where the left panel shows the value of the LLL for $L(\mathrm{FUV})$ at the five  metallicities; it also includes, as reference, the mean FUV integrated luminosity for $N_*=10^4$.
The middle panel shows the histogram of pixel $L(\mathrm{FUV})$ for the three galaxies, using the distances  quoted in Table \ref{tab:tab1}. Except for most pixels of M77, all the pixel-based values have luminosities below the LLL in FUV, for ages between 0 and 10 Myr (17 Myrs for the lower metallicity). 
The right panel shows the distribution of the pixels-based regions in the  $\mathrm{H}\alpha/\mathrm{FUV}$ vs. $L(\mathrm{FUV})$ plane. The color lines show the position of the mean values obtained with the synthesis models for the case of $10^4$ stars. The black line shows the region covered by individual normal stars and the gray line the location of individual WR stars. We note that any pixel-sized region with $2\leq\log (\mathrm{H}\alpha/\mathrm{FUV})\leq 2.4$  requires the presence of WR stars, and that pixel-sized regions with $\log (\mathrm{H}\alpha/\mathrm{FUV})> 2.4$ cannot be explained with synthesis models and require some extra ionization source. The extreme case that would be explained by the models are regions with $\log (\mathrm{H}\alpha/\mathrm{FUV})\sim 2.5$, which would correspond to regions completely dominated by nebular emission without stellar content, hence age determinations would not be performed. 
In M~83 those regions with an excess in this flux ratio actually correspond with those holding WR stars detected by \citet{2012ApJ...753...26K}.

Although we are clearly aware of the IMF sampling issues discussed in
\cite{2003A&A...407..177C,2004A&A...413..145C,2006A&A...451..475C,2013NewAR..57..123C}, we  apply a quasi ``standard'' SSP analysis to the age estimation.
We   calculate  model \Ha\ and FUV fluxes,  ages, masses, etc. assuming that the IMFs, 
even though well populated, contain an intrinsic scatter and that the integrated \Ha\ and FUV luminosities can be modeled by a 2D-Gaussian distribution, with the correlation coefficient described previously and a fixed standard deviation of 4\% for each theoretical luminosity. We have also computed the results for a standard deviation  of 20\% and this does not alter significantly the main results. Although not perfect, given the fail of the Gaussian approximation at larger standard deviations, it is a first step in the inclusion of sampling effects in the modelling of stellar clusters. A  detailed discussion of the relevance of sampling for this work is given in Appendix \ref{Sec:App0}.

We also examined the effect of adjacent pixels, to verify 
whether the amount of \Ha\  and FUV fluxes in a given pixel reflect the number of ionizing O and B stars in that pixel. 
In Paper I, we explored a range of different spatial binning scales. These results showed that age structures and gradients remain the same irrespective of the binning scale used (cf. Fig.6 of Paper I), confirming the robustness of  measurements to the effects of binning.
{In Section \ref{sec:ImgSeg}, we present a different approach to deal with the possible influence of adjacent pixels, 
as well as the low mass pixels below the LLL, 
under a Bayesian inference framework. 
We apply an image segmentation technique, to account for the possible effects of the spatial dependence in terms of adjacent regions (or pixels), grouping together neighbouring pixels that carry on average the same true value of the observed measured quantity (\Ha/FUV flux ratio). %
In this manner, we can address both systematic effects. Not only the dependence between adjacent regions but also the pixel-wise applicability of SSP models, since many of the grouped larger regions are now above the LLL threshold.}

{A similar approach can be found in \cite{2017MNRAS.466.3989C}. These authors implement a Bayesian Technique for Multi-image Analysis (BATMAN), focused on Integral Field Spectroscopy (IFS) data cubes. 
Unlike our fully Bayesian approach, BATMAN's  algorithm is rather a Bayesian approximation. The parameter estimation, carried out as usual, requires the selection of the most probable tesellation of the image performed by means of an iterative procedure to select the best model. 
Another significant difference is that the presence of gradients pose a  challenge to BATMAN's algorithm, since its segmentation model  does not consider the presence of gradients inside the regions. 
In our case the presence of age gradients, which are indeed expected, do not seem to affect the image segmentation. }

{A limitation of image segmentation techniques is the loss of spatial resolution. 
More complex and elaborate Bayesian models, such as latent Gaussian models for spatial modelling can be found \citep{Rue2017}. %
Not only the spatial correlation structure is estimated, with no loss of image resolution,  but also some fixed effects and  Gaussian random effects can be included in the linear predictor of the model fitting. 
However, these models are computationally challenging and time-consuming. An approximation for a faster inference is implemented in the \texttt{R-INLA}\footnote{http://www.r-inla.org/} package of \texttt{R}; however, it has a steep learning curve, requiring a deeper understanding of both the Bayesian and the latent Gaussian models for a proper implementation.
An example of its successful application to IFU spectra can be found in \cite{2018arXiv180206280G}. This type of data encodes some auto-correlated structures, and it is very important to evaluate the spatial information. 
Besides, the loss of spatial resolution in these cases is quite significant (see their Fig.~4 for a qualitative comparison with other image segmentation and spatial algorithms methods). 
This is not an issue with our images, that have high enough spatial resolution to resolve individual star forming structures within spiral arms, even after image segmentation. See bottom left panels of Figures~\ref{fig:Fig8}, \ref{fig:Fig10}, and \ref{fig:Fig12}.
In fact, the morphology of the segmented images traces  the same global and local spatial patterns as the pixel-based age maps.}

{Finally, we want to stress that there is an additional fundamental difference between our approach and those using IFU data (\citealt{2017MNRAS.466.3989C,2018arXiv180206280G} which also include a Bayesian analysis, or 
\citealt{2017MNRAS.471.3727D}, which  use  a Voronoi technique). In the case of IFU data a major issue is the requirement of increasing the low S/N of some spaxels and avoid data sparsity; in our case, we do not have such a problem, since photometric data have a larger S/N, and those pixels below a certain H$\alpha$ flux level are considered to belong to the background, so they are masked in the resulting maps.}

{In brief,  Section \ref{sec:ImgSeg} proposes a simple and fast fully Bayesian approach to take into account the spatial dependence by means of image segmentation. We will see that the loss of spatial resolution is not important, in part due to  the spatial characteristics of the data. We will show how this method provides good results within the aim of this work, the global determination of the age patterns along and across the spiral arms.  }

\section{A Bayesian framework for modeling the ratio images} 
\subsection{Hierarchical Bayesian Model}
\label{sec:HBM}

In this section, we address the problem of deriving the galaxy age map 
from observed flux ratio images, namely
$\hat{r} = \hat{F}_{H{\alpha}} / \hat{F}_{FUV}$, 
by establishing a probabilistic framework relating the random variables involved in the problem. These relationships 
will be formulated in terms of a joint probability distribution, given the 
observations and their uncertainties through a hierarchical Bayesian 
model~\citep[HBM;][]{Gelman03}. Specifically, we want to describe the
probability distribution of age given $\hat{r}$, which will be derived by
marginalization as we will see below.
{This procedure or algorithm is applied pixel by pixel throughout the image, keeping the spatial resolution of the flux ratio images, see top right panels of Figures~\ref{fig:Fig8}, \ref{fig:Fig10}, and \ref{fig:Fig12}.}
\begin{figure}
\tikzstyle{decision} = [rectangle, draw, fill = gray!20, rounded corners,
    text width=2.5em, text badly centered]
\tikzstyle{block} = [rectangle,rounded corners, draw, text width=2.5em, text centered, minimum height=2em]
\tikzstyle{line} = [draw, -latex']
\tikzstyle{cloud} = [draw, ellipse,fill=red!20, node distance=3cm,
    minimum height=2em]
     
\begin{tikzpicture}[
node distance = 1.5cm]

\node[decision](phi1) {$\phi_1$};
\node[decision, right = 1cm of phi1] (phi2){$\phi_2$};
\node[decision, right = 1cm of phi2](phi3)  {$\phi_3$};
\node[draw, rectangle,rounded corners, text width=4em, text centered, minimum height=2em, below of = phi1] (theta1) {$\theta_1 =$ age};
\node[draw, rectangle,rounded corners, text width=4em, text centered, minimum height=2em,below of = phi2]  (theta2) {$\theta_2=Z$};
\node[draw, rectangle,rounded corners, text width=5em, text centered, minimum height=2em,below of = phi3](theta3)  {$\theta_3=Q(H)$};

\node[draw,diamond,below = 1cm of theta2](ann)  {SB99$+$ANN}; 

\node[below = 1 cm of ann] (Rho0) {}; 
\node[block, right of= Rho0](LHa) {$L_{H\alpha}$}; 
\node[block, left  of = Rho0](LFUV)  {$L_{FUV}$}; 

\node[block, below = 1cm of Rho0] (Rho)  {$\rho$}; 
\node[block, right of = Rho](Ha) {$F_{H\alpha}$}; 
\node[block, right of = Ha](sigHa) {${\sigma}_{H\alpha}$}; 
\node[block, left of = Rho](FUV)  {$f_{FUV}$}; 
\node[block, left of = FUV](sigFuv) {${\sigma}_{FUV}$}; 

\node[decision,below of = sigHa] (hsigHa){$\hat{\sigma}_{H_{\alpha}}$};
\node[decision,below of = sigFuv](hsigFuv) {$\hat{\sigma}_{FUV}$};
\node[decision,minimum height=2em,below of = Rho](robs) {$r_{obs}$};

%
\path[line] (phi1) -> (theta1);
\path[] (phi1) edge  node[left] {hyperpriors()}  (theta1);
\path[line] (phi2) -> (theta2);
\path[line] (phi3) -> (theta3);

\draw ($(theta1.south)$) -- ++(0,-0.25)-- ++(3.8,0);
\draw[line] ($(theta2.south)$) -- ($(ann.north)$);
\draw ($(theta3.south)$) -- ++(0,-0.25);
\path[] (theta2) edge  node[left] {priors()}  (ann);

\draw[line] ($(ann.south)$)--($(Rho.north)$) ;
\draw[<-] ($(LFUV.north)$)-- ++(0,0.3)--++(3,0);
\draw[<-] ($(LHa.north)$)-- ++(0,0.3);

\path[line,dashed] (LHa) -> (Ha);
\path[line,dashed] (LFUV) -> (FUV);

\path[line,dashed] (FUV) -> (sigFuv);
\path[line,dashed] (Ha) -> (sigHa);
\path[line,dashed] (hsigHa) -> (sigHa);
\path[line,dashed] (hsigFuv) -> (sigFuv);

\draw[thick] ($(sigFuv.south)+(0.25,0)$)-- ++(0,-0.25)--++(5.5,0) ;
\draw[thick]  ($(FUV.south)$)-- ++(0,-0.25);
\draw[thick] ($(Ha.south)$)-- ++(0,-0.25);
\draw[thick]  ($(sigHa.south)+(-0.25,0)$)-- ++(0,-0.25);
\path[line,thick] (Rho) -> (robs);
\path[] (Rho) edge  node[left] {Likelihood}  (robs);

    
%
\node (rect) at (Rho) [yshift = -0.75cm,draw, rounded corners,black!25!blue!50,thick,
minimum width=7.5cm,minimum height=3.cm] (x) {};
\node[black!25!blue!50,anchor=south west] at (x.south west) {NSA}; 


\end{tikzpicture}
\caption{The hierarchical Bayesian model in plate 
notation. Blank shaded nodes represent random fixed (or observed) values, respectively. 
Arrows represent the kind of relationship between variables: probabilistic (solid arrows) and deterministic (dashed arrows).
In the central part of the graphical model we show
deterministic calculations done by SB99 scripts and the artificial neural
network (ANN).
The blue square at the bottom, encircles the variables involved in the Likelihood \eqref{Eq:Likelihood}, as well as the observed ratio which are the input for the NSA.}
\label{fig:Fig5} 
\end{figure}
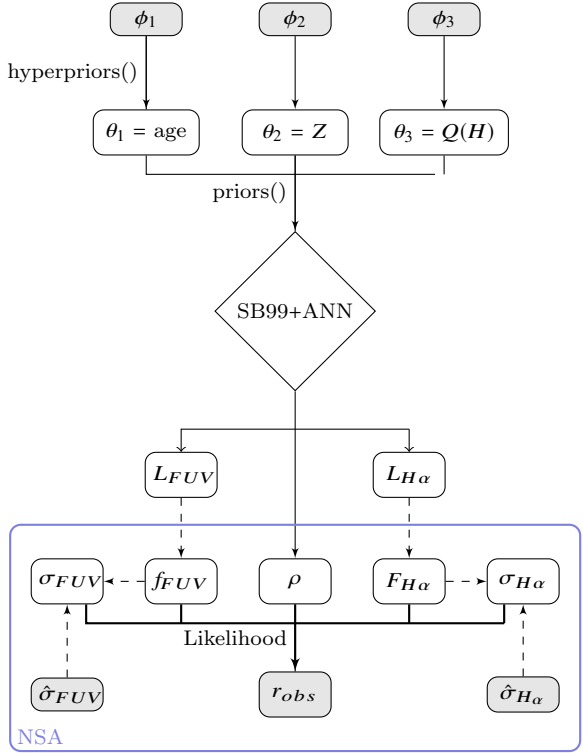

Figure~\ref{fig:Fig5} shows our HBM in plate notation, namely a 
graphical model representing the previous relationships. Nodes in
the graphical model are circles and squares representing random variables
and fixed values, respectively. Inside the circles and squares there are
numbers representing the dimension of values in each case,
and a shaded circle (square) means that the variable has been observed. 
Arrows represent the relationships (solid arrows probabilistic, dashed arrows
deterministic) between variables.

Let $\mathcal{H}$ represent the set of all population synthesis parameters held fixed,
such as the SFH (instantaneous), IMF~\citep[see][]{1955ApJ...121..161S}, evolutionary
tracks, as well as the extinction correction applied to the data. 
And let
$\bm\theta = (\theta_1, \theta_2, \theta_3) $ be the parameters to be estimated:
$\theta_1$ the age of the region under study (the image pixel), 
$\theta_2$ the metallicity and $\theta_3$ the
fraction of ionizing photons. 
Every parameter $\theta_i$ is connected with
the hyperparameter $\phi_i$ in the graphical model at Figure~\ref{fig:Fig5}. 
That is, there is a probability distribution once the $\phi_i$ is fixed. 
Specifically we set uniform prior distributions for all the parameters, 
$\theta_i \sim \mathcal{U}(\phi_i)$, where 
$ \phi_1= (1,20)$, corresponding to the age ($\theta_1$) ranging from 1 to 20 Myr;
$\phi_2 = (0.001, 0.02)$, for the metallicity ($\theta_2$). SB99 only covers five metallicity 
values, i.e.  $\nicefrac{1}{20}Z_{\odot} =  0.001$, 
$\nicefrac{2}{5}Z_{\odot} = 0.008$, $\nicefrac{1}{5}Z_{\odot} =  0.004$, 
$Z = Z_{\odot} = 0.02$ (solar) and $2Z_{\odot} =  0.04$;
and $\phi_3 = c(0.1,1)$, the fraction of ionizing  photons $Q(H)$ ($\theta_3$) ranging from 10\% to 100\%.
This is, priors for $\bm \theta$ were set as
\begin{eqnarray}\label{Eq:priors}
	\theta_1 & \sim & \mathcal{U}(0.1,20)  \nonumber\\ 
	\theta_2 & \sim & \mathcal{U}(0.001,0.04) \\ 
	\theta_3 &\sim  &  \mathcal{U}(0.1,1) \nonumber
\end{eqnarray}
Let us notice that we deal with a finite set of values for $\theta_i$, and therefore a finite number
of model flux ratios derived from them. See left panel in Figure~\ref{fig:Fig2}. 

As  described below, our main objective is to obtain a posterior distribution of the parameters $\theta_i$, using an iterative algorithm in our HBM.
The main consequence of using SB99 scripts and the iterative algorithm is the increase of  CPU time. 
In order to obtain a continuous range for the model flux ratios or parameters, rather than such discrete set of $\theta_i$ values, we set up an Artificial Neural Network (ANN) to interpolate the grid of parameters. 
A thorough introduction to ANN can be found in~\citet{Haykin1999}.

Specifically, our ANN is a \emph{multilayer perceptron} with four layers: two
hidden layers with $30$ and $50$ nodes, respectively. The input layer for the
parameters $\bm{\theta}$ and the output layer for the \Ha\ and FUV
luminosities. This topology was selected by doing cross-validation through
repeated random sub-sampling validation (90 percent of the dataset for
training and 10 percent for validating the ANN) and assessing the fit using
the mean squared error (MSE). 

Figure~\ref{fig:Fig5} shows the use of the ANN by dashed arrows linking the parameters $\theta_i$ with \Ha\ and FUV luminosities (i.e., a deterministic relationship characterized by the ANN). 
These dashed arrows also include the calculations to obtain the modeled \Ha\ and FUV fluxes, and therefore
the model flux ratio ${r} = {F}_{H{\alpha}} / {F}_{FUV}$ for a specific pixel, according to 
\begin{eqnarray}\label{eq:derivedfluxes}
	{F}_{H{\alpha}} &=& 4\pi D^2 {L}_{H{\alpha}}, \, 
	(erg\,s^{-1}cm^{-2}) \\
	{f}_{FUV} &=& 4\pi D^2 {L_{\lambda}}_{FUV}, \, 
	(erg\,s^{-1}cm^{-2}\AA^{-1})
\end{eqnarray}
with the same units as the observed data.

In order to explain completely the HBM, let us focus
on the right-hand side of  Figure~\ref{fig:Fig5}, showing the relationships
between unknown $r$ and observed $\hat{r}$ flux ratios. 
We apply this technique pixel by pixel (as described in Paper I), 
assuming a bivariate normal distribution for the observed fluxes as a result
of the convolution between the observational/model uncertainties and the unknown fluxes. 
Specifically, \\
\begin{equation}
	(\hat{F}_{H \alpha}, \hat{f}_{FUV})
	\sim
	\mathcal{N}((F_{H \alpha}, f_{FUV}), \Sigma),
\end{equation}
where $\Sigma$ is defined according to 
\begin{equation}
	\Sigma = 
	\begin{pmatrix}
		\sigma^2_{H \alpha} & \rho \sigma_{mod,H \alpha } \sigma_{mod,FUV} \\[1em]
		\rho \sigma_{mod,H \alpha } \sigma_{mod,FUV} & \sigma^2_{FUV}
	\end{pmatrix}
	\label{eq:sigma}
\end{equation}
$\rho$ is the theoretical correlation between the \Ha\ and FUV fluxes; 
and $\sigma^2 = \hat{\sigma}^2 + \sigma_{mod}^2$ accounts for both the observational and the model uncertainties.   

Therefore, 
$\hat{r} = \hat{F}_{H{\alpha}} / \hat{F}_{FUV}$ is the ratio of two
correlated normal random  variables, whose exact distribution is given by
\citet{Hinkley1969} as

\begin{align}
	\psi(r) \ \ & =  \ \ \frac{b(r)d(r)}{\sqrt{2\pi}{\sigma}_{H{\alpha}} {\sigma}_{FUV}a^3(r)}
	\left( 
	2\Phi\left( \frac{b(r)}{\sqrt{1-\rho^2}a(r)}\right) -1
	\right) +
	\nonumber\\[1em]
	 & \ \ \ \ + \frac{\sqrt{1-\rho^2}}{\pi{\sigma}_{H{\alpha}}\sigma_{FUV}a^2(r)}
	\exp\left\{-\frac{c}{2(1-\rho^2)}\right\},
	\label{eq:fddR}
\end{align}
where $r$ is the model flux ratio, and parameters $a(r)$, $b(r)$, $c$, and $d(r)$ are defined as 
\begin{eqnarray*}
	a(r) & = & \left( \frac{r^2}{{\sigma}^2_{H{\alpha}}}-\frac{2\rho r}{{\sigma}_{H{\alpha}} {\sigma}_{FUV}}+
	\frac{1}{{\sigma}^2_{FUV}}\right)^{1/2}
	\nonumber \\[7pt]
	b(r) & = & \frac{r{F}^2_{H{\alpha}}}{{\sigma}^2_{H\alpha}}-
	\frac{\rho ({F}_{H{\alpha}}+ r{f}_{FUV})}{{\sigma}_{H\alpha}{\sigma}_{FUV}}+
	\frac{{f}_{FUV}}{{\sigma}^2_{FUV}}
	\nonumber \\[7pt]
	c  & = & \frac{{F}^2_{H{\alpha}}}{{\sigma}^2_{H\alpha}}-
	\frac{2\rho {F}_{H{\alpha}} {f}_{FUV} }{{\sigma}_{H\alpha}{\sigma}_{FUV}}+
	\frac{{f}^2_{FUV}}{{\sigma}^2_{FUV}}
	\nonumber \\[7pt]
	d(r) & = & \exp\left\{\frac{b^2(r)-ca^2(r)}{2(1-\rho^2)a^2(r)}\right\}.
\end{eqnarray*}
Here $\rho$ is the correlation between ${F}_{H{\alpha}}$ and ${f}_{FUV}$  
and $\Phi$ is the cumulative density distribution of the standard normal. 
We recall that the values assumed for $\hat{\sigma}$ are 5\% for $F_{H \alpha}$ and 25\% for $\hat{f}_{FUV}$ (as shown in Section \ref{sec:observations}), and that $\sigma_{mod}$ has been taken as 4\% in both cases (see Section \ref{sec:SPM} and Appendix \ref{Sec:App0}).
Then, we can define the probabilistic relationship between unknown and observed 
flux ratios according to the likelihood
\begin{equation}\label{Eq:Likelihood}
	p(\hat{r} | \bm \theta, \mathcal{H}) = \psi (\hat{r} | F_{H \alpha}, f_{FUV}, \sigma_{H \alpha},
	\sigma_{FUV}, \rho) 
\end{equation}

Figure~\ref{fig:Fig5} explains how we obtain observed ratios from
parameters $\bm\theta$ through an HBM by using the graphical model. 
However our main objective is to move in the reverse order, 
i.e., to  obtain suitable parameters $\bm\theta$ from the observed ratios $\hat{r}$. 

The joint posterior probability distribution $p(\bm{\theta} | \hat{r})$ can  
be rewritten by using Bayes' theorem according to
\begin{equation}\label{eq:bayesdescomposition}
	p(\bm{\theta} | \hat{r}) = \frac{p(\hat{r} | \bm{\theta}) 
		p(\bm{\theta})}{p(\hat{r})} \propto  p(\hat{r} | \bm{\theta}) p(\bm{\theta}),
\end{equation}
where $p(\hat{r})$ is a normalization constant, which for our purposes can be
ignored. We will assume independence between parameters $\theta_i$, or
equivalently, the prior distribution $p(\bm{\theta}) = p(\theta_1)p
(\theta_2)p(\theta_3)$, where $p(\theta_i)$ is uniform distribution as defined
above. The likelihood, $p(\hat{r} | \bm{\theta})$, has been defined
previously in Equation~\eqref{Eq:Likelihood}.
%

%
\begin{figure*}
\includegraphics[width=0.9\textwidth]{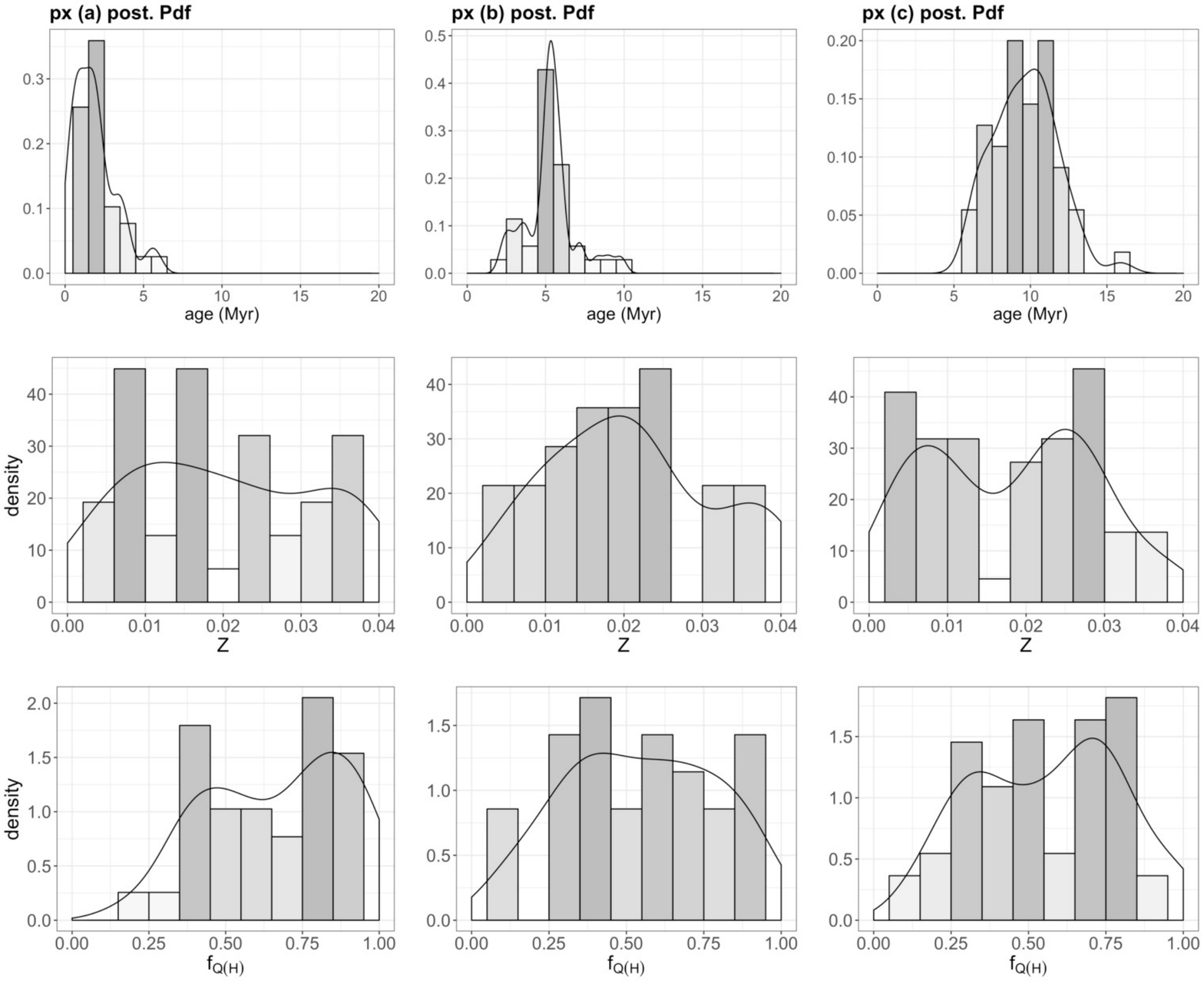}
\caption{Example marginal posterior probability distributions of  different parameters for three image pixels of different age. 
The black lines denote kernel density estimates. 
The scale fill gradient correspond to the density (from white lower to grey  higher). 
These plots highlight the sensitivity of the proposed methodology for the age estimation, unlike other parameters.}
\label{fig:Fig6}
\end{figure*}
\begin{figure*}
\includegraphics[width=0.9\textwidth]{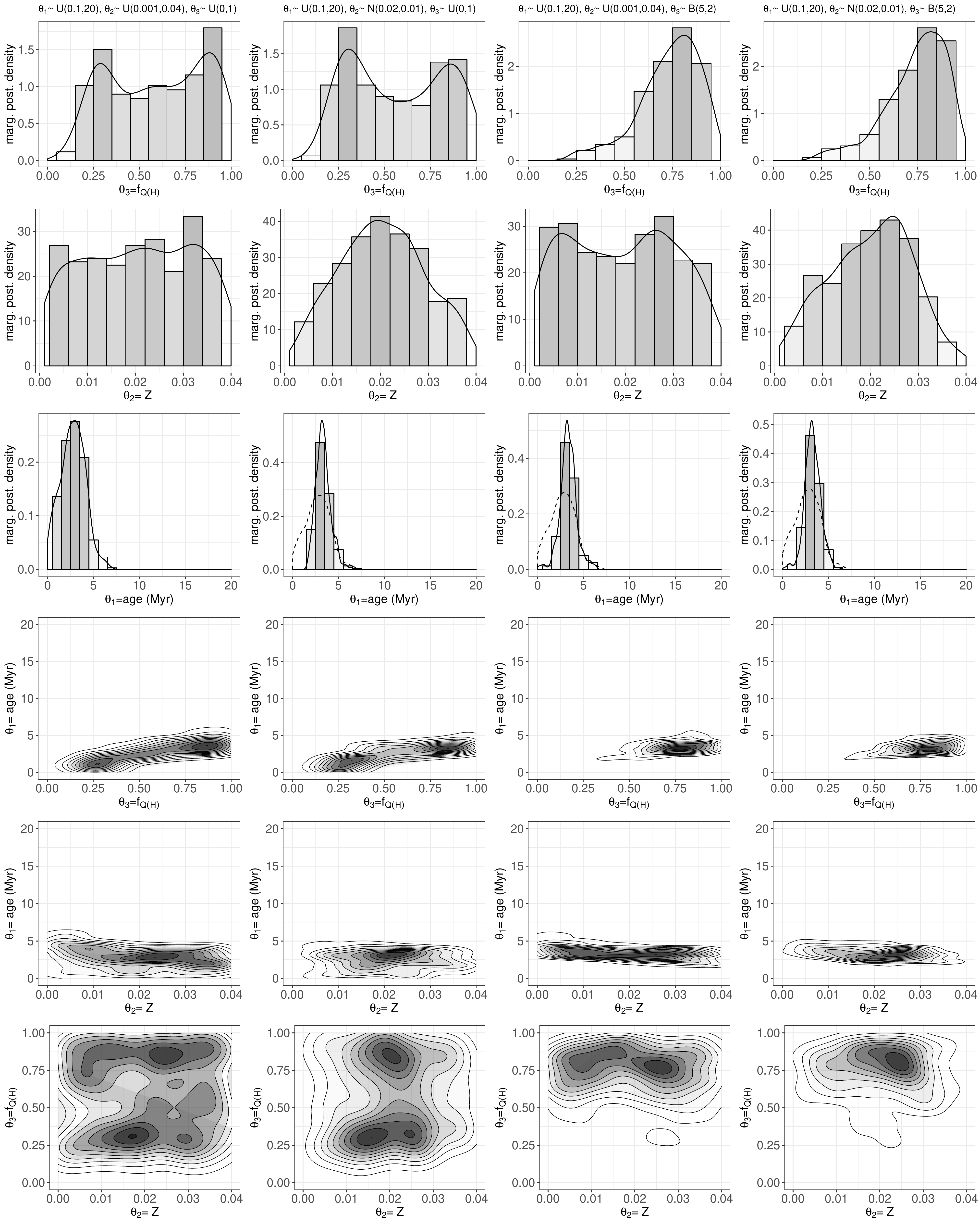}
\caption{
Posterior marginals and pairwise correlations for the three parameters, given different prior selections and a  flux ratio $\hat{r}=1.13$. 
The first column corresponds with uniform priors  for the three parameters, 
$\theta_1 \sim \mathcal{U}(0.1,20), \theta_2 \sim  \mathcal{U}(0.001,0.04)$, and  $\theta_3 \sim    \mathcal{U}(0.1,1)$
At the second column only the prior on metallicity is not uniformed distributed, 
$\theta_2 = Z \sim \mathcal{N}_{(0,\infty)}(0.02,0.01)$, i.e. a truncated Normal to non negative values.  
While at the third column $\theta_3 = f_{Q(H)} \sim B(5,2)$, a Beta distribution assuming highest values of $f_{Q(H)}$  to be more likely. 
Last column shows the results when both priors are not uniformly distributed. 
The first three rows are the same as in Figure~\ref{fig:Fig6}: 
the black solid lines denote the corresponding kernel density estimates. The dashed lines denote the density  when all priors are uniformly distributed, for comparison purposes. 
The three last panels rows  display how correlated are the posterior marginals for the different parameters.  
}
\label{fig:Fig6b}
\end{figure*}

In the Bayesian framework, inference proceeds by
estimating the posterior distribution  $p (\bm{\theta} | \hat{r})$ and then
marginalizing to obtain the age distribution of the region under study
according to
\begin{equation}\label{eq:marginalizationPost}
	p(\mbox{age} | \hat{r}) = p(\theta_1 | \hat{r}) = 
	\iint p(\bm{\theta} | \hat{r}) \, d \theta_2 d \theta_3
\end{equation}
where  the posterior distribution $p(\bm{\theta}|\hat{r})$ was characterized
by an independent and identical distributed sample obtained by using the
Nested  Sampling Algorithm~\citep[NSA;][]{skilling2006}. 
Although we refer to these authors, a brief summary is explained in Appendix~\ref{Sec:App2}.

Finally, we obtain the posterior distribution of the parameter of interest 
(age) by marginalizing nuisance parameters in the posterior distribution as 
set in Equation~\eqref{eq:marginalizationPost}. 
Marginalized distributions for the other two parameters can be obtained in the same way. 

Figure~\ref{fig:Fig6} shows an example of the resulting posterior distributions.  
For the case of M83, we have chosen three pixels of different estimated ages
($1.2$, $5.5$ and $10$ Myr) and plotted (left to right respectively) the histograms of the samples from the marginalized distributions.

On the top, we have the marginalized posterior probability distributions of the age parameter. 
From an initial, non-informative, uniform prior  $\mathcal{U}(0.1,20)$ Myr, we can observe how well defined is the estimated posterior distribution. 
In general, the median or the (uni)modal values are close from each other, close to symmetrical distributions, and well differentiated. 
As mentioned in Section \ref{sec:intro}, the ratio \Ha/FUV is very sensitive to age variations for young SF regions. 

The middle and bottom panels of Figure \ref{fig:Fig6} show the marginalized posterior distributions of the metallicity and the fraction of ionizing photons, respectively.
The assumed priors for these parameters were also uniform.  However, it is remarkable that their posterior distributions are too spread and flat, remaining nearly uniform in many cases. 
This result should indicate that this method is not able to determine these parameters. 
In fact, the metallicity is a parameter quite difficult to estimate. 

{
In order to check how much the selection of priors in these other properties may influence the age estimate, we compute the posterior probability density function $p(\bm{\theta} | \hat{r})$ %
for different priors. 
The results, for a given flux ratio of $\hat{r}=1.13$ (a young region), are shown in Figure \ref{fig:Fig6b}. 
The first column corresponds to the uniform priors assumed  along this work. 
In the second column only the prior on metallicity is not uniform, $\theta_2 = Z \sim \mathcal{N}_{(0,\infty)}(0.02,0.01)$ truncated Normal, negative values are not allowed.
Similarly, in the third column for the fraction of ionizing photons, $\theta_3 = f_{Q(H)} \sim B(5,2)$ Beta distribution, assuming higher values of $f_{Q(H)}$  to be more likely. 
The fourth column shows the results when both priors are not uniform distributed. 
For other flux ratio values  or different prior selections, 
e.g. $\theta_2 \sim B^*(2,2)$, or $B^*(5,2)$ (Beta distribution transformed from the unit interval $[0,1]$ to $[0.001,0.04]$), 
$\theta_3 \sim B(2,2)$, or $U(0.5,1)$, etc, the results were similar. 
}

{
Figure \ref{fig:Fig6b} shows that, except for the age, the posterior PDF of the other parameters is dominated by the prior. 
Besides, the influence of the selection of the prior distribution  in these other parameters is almost negligible on the age estimate. 
The mode (and median) of the different age posterior PDFs are pretty close. In fact, it seems that adding some prior information results in narrower and taller posterior density functions around the mode.
}

{
The three last rows display how correlated are the posterior marginals for the different parameters.  Notice how the age correlates with the fraction of ionizing photons and the metallicity, whereas these two parameters are not only nearly uncorrelated but nearly independent, as well as their priors. 
These parameters are not determined by the \Ha/FUV ratio. Therefore, their posteriors remain dominated by the priors distributions assumed, as it is clearly seen in the figure. Any posterior sample occupies nearly the whole space of parameters, except for the age.  
We confirm that this methodology, based on the \Ha/FUV flux ratio, is robust and efficient for dating  \hii\ regions but not for determining other parameters such as the metallicity.  
}

\subsection{Image Segmentation}
\label{sec:ImgSeg}

{ 
In the following, we describe an image segmentation technique to assess the possible effects of spatial dependence in terms of adjacent regions (or pixels), grouping together as a single region those neighbouring pixels that carry the same average true value of the measured quantity.}

{ 
In Paper I  this was achieved by calculating the age maps after re-scaling the \Ha/FUV flux ratio images at $3 \times 3$ or $6 \times 6$ pixel binning. We found that not only the main structures and global age patterns remain unchanged, but also some local age gradients. 
This level of pixel bining is in agreement with the results from several variograms computed in M83, where we found that the level of spatial dependence is roughly under bins of $5 \times 5$ pixels. }

{ 
Both segmentation techniques imply a loss of resolution, but it is worst in the rebinning case.  The proposed image segmentation technique, based also on a hierarchical Bayesian approach, is an improvement compared to re-binning the images as in Paper I. }

We segment the \Ha/FUV ratio image, in terms of homogeneous values, to model the effects of the spatial dependence (in terms of adjacent pixels). 
We work with regions resulting from clustering several pixels.
This approach is driven by the assumption that pixels with similar H$\alpha$/FUV ratios will share similar inferred properties. 
Segmentation maps serve to identify structures sharing common  properties relevant to the interpretation of the age map. 

The purpose of image segmentation is to cluster pixels into homogeneous classes, without prior definition of those classes, based only on spatial coherence.
We present a fully Bayesian approach, based on the Potts model for the image reconstruction \citep[][Chapter 8]{10.1007/978-1-4614-8687-9}. 

Consider the ``true'' image as a random bidimensional array  $\bm x = \{ x_i, i\in \mathcal{I}\}$
whose elements are indexed by the lattice $\mathcal{I}$, the location of the pixels, and related through a neighborhood relation. 
The four nearest neighbors of the $i-$th pixel $(m,n)$ are $(m,n-1),(m,n+1),(m-1,n)$, and $(m+1,n)$ respectively, denoted as $j\sim i$.

This neighborhood relation is translated into a probabilistic dependence by means of Markov Random Field (MRF), where each $x_i$ takes a finite set of values. 
The conditional distribution of any pixel $i \in \mathcal{I}$, given the rest of the pixels of the image, depends only on the values of their neighbors, denoted by $n(i)$, 
$p(x_i |x_{-i}) = p(x_i | x_{n(i)})$.

Denote the observed flux by $\bm y$, considered  ``noisy'' in the sense that the measured flux of a pixel is not observed exactly but with some perturbation (instrument noise, reduction process, etc). 
Both objects $\bm x$ and $\bm y$ are arrays, with each entry of $\bm x$ taking a finite number of values,
for numerical convenience.Each entry of $\bm y$ takes real values. 

The aim is to draw inference on the ``true'' image $\bm x$, given an observed noisy image $\bm y$. 
We are thus interested in the posterior distribution of $\bm x$ given $\bm y$, provided by  Bayes' theorem 
\begin{equation}
	p(\bm x|\bm y) \propto f(\bm y|\bm x)p(\bm x).
\end{equation}

We assume a Potts model for the prior on $\bm x$, 
a specific family of distributions inspired from particle physics
in order to structure images and other spatial structures in terms of local homogeneity 
\citep{RevModPhys.54.235}. 
\begin{equation}
	p(\bm x|\beta) = \frac{1}{Z(\beta)} exp\left\{ \beta \sum_{j \sim i} \mathbb{I}_{x_j =x_i} \right\}
	\label{Eq:Potts}
\end{equation}
where 
${Z(\beta) = \sum_{i \in \mathcal{I}} exp\left\{ \beta \sum_{j \sim i} \mathbb{I}_{x_j =x_i} \right\}}$
is the normalizing constant of the Potts model with $G$ categories.

The likelihood $f(\bm y|\bm x)$ describes the link between the  observed image and the underlying classification 
of homogeneous flux ratios. That is, it gives the distribution of  the noise. 
We will make the assumption that the observations in $\bm y$ are conditionally independent of $\bm x$ and  Gaussian, 
\begin{equation}
	f(\bm y|\bm x,\sigma,\mu_1,\ldots,\mu_G) = \prod_{i \in \mathcal{I}} \frac{1}{\sqrt{2\pi}\sigma} 
	exp\left\{ -\frac{1}{2\sigma^2}(y_i- \mu_{x_i})^2 \right\}.
	\label{Eq:noisyimage}
\end{equation}

For a fully Bayesian approach, we have to give the distribution of the hyperparameters $\beta$, $\sigma$, and $\mu_1,\ldots,\mu_G$, respectively. Since there is no additional information about any of these nuisance parameters we assume uniform and independent hyper priors.

The Potts model parameter $\beta$, which represents the strength between neighbouring pixels,  
is uniform distributed over $\beta \sim \mathcal{U}(0,2)$, 
\citep[e.g.][and references within]{10.1007/978-1-4614-8687-9, Stoehr17}. 
Above this critical value $\beta_c = 2.269$  (for the case of four neighbour relation) 
the Markov chain is no longer irreducible, converging to one of two different stationary distributions, depending on the starting point. The distribution becomes multimodal, known as phase transition in particle physics.

The mean value of each homogeneous class $\bm \mu = (\mu_1,\ldots,\mu_G) \sim \mathcal{U}(\bm \mu; y_{min}\leq \mu_1 \leq \ldots \leq \mu_G \leq y_{max})$. 
This is a generalization of the treatment for an image where $y$ represents its noisy version of the true colour or grey level (and not necessarily an integer). 

The recorded values of $\bm y$ represent the ratio of the \Ha\ and FUV fluxes
( e.g. in the range  $(-2,3)$ for M83 example). 
We classify this ratio image into $G=5$ homogeneous regions or classes, with mean $\mu_g$. The number of classes is inspired in the previous work in Paper I. 

For the noise variance it is assumed an uniform prior on $log(\sigma)$, or equivalently $p(\sigma^2)\propto \sigma^{-2}$.  

Finally, the posterior distribution for $\bm x$ is
\begin{equation}
	P(x_{i}=g|\bm y,\beta,\sigma^{2},\bm \mu) \propto 
	exp \left\{
	\beta \sum_{j\sim i} \mathbb{I}_{x_j = x_i}
	- \frac{1}{2\sigma^2}(y_i - {\mu_g})^2
	\right\}
	\label{Eq:postpott}
\end{equation}

Appendix \ref{Sec:App1} describes in more detail the implementation of a hybrid Gibbs algorithm for sampling. 
The underlying algorithm addresses the reconstruction of an image distributed from
a Potts model based on a noisy version of this image. 
The purpose of image segmentation 
is to cluster pixels into homogeneous classes without preference and based only on the spatial coherence of the structure. 

Once the image has been classified into $G$ different homogeneous classes, 
with $\mu_{g}$ the mean value of the ratio \Ha/FUV for each class,  we can apply both methodologies: 
the one  used in Paper I and the one presented here. 
In Section \ref{sec:HBM}, we compare these ratio values with the SB99 model to assign an age range to that homogeneous region. 
In particular, for M83 these mean values are: 
$\mu_1=0.3045$, corresponding to ages $>6$ Myr; 
$\mu_2=0.6139$ and $\mu_3=0.8807$ for the age range  $5-6$ Myr; 
$\mu_4=1.151$ for $3-4$ Myr; 
and $\mu_5=1.5195$ for ages $1-2$ Myr.

The resulting age map is also discrete (bottom-left plot in Fig. \ref{fig:Fig8}),
but the structures of the age patterns are  more consistent with the \Ha/FUV image, as seen comparing the flux ratio image (top-left) with the age map obtained in Paper I (bottom-right).
When  compared, the age patterns and structures of these two discrete age maps are quite similar, the main difference being just the age range assignment. 
It attests the robustness of the results in Paper I, and their consistency with the present analysis. 
{The loss of resolution in the resulting segmented age map is rather negligible under the point of view of the  study of age patterns.}
The main aim of these studies is to get the age structures and patterns along and across the spiral arms, rather than giving an absolute age. 

{We can conclude that the effect of spatial dependence with adjacent regions does not affect the study of the age patterns when this dating technique is applied pixel by pixel. 
The \Ha/FUV ratio proves to be a robust estimator of the age, local gradients, and global age patterns remain unchanged despite some loss of spatial resolution. 
}

\section{Results: Bayesian Age Maps} 
\label{sec:results}

\begin{figure*}
\includegraphics[width=0.95\textwidth]{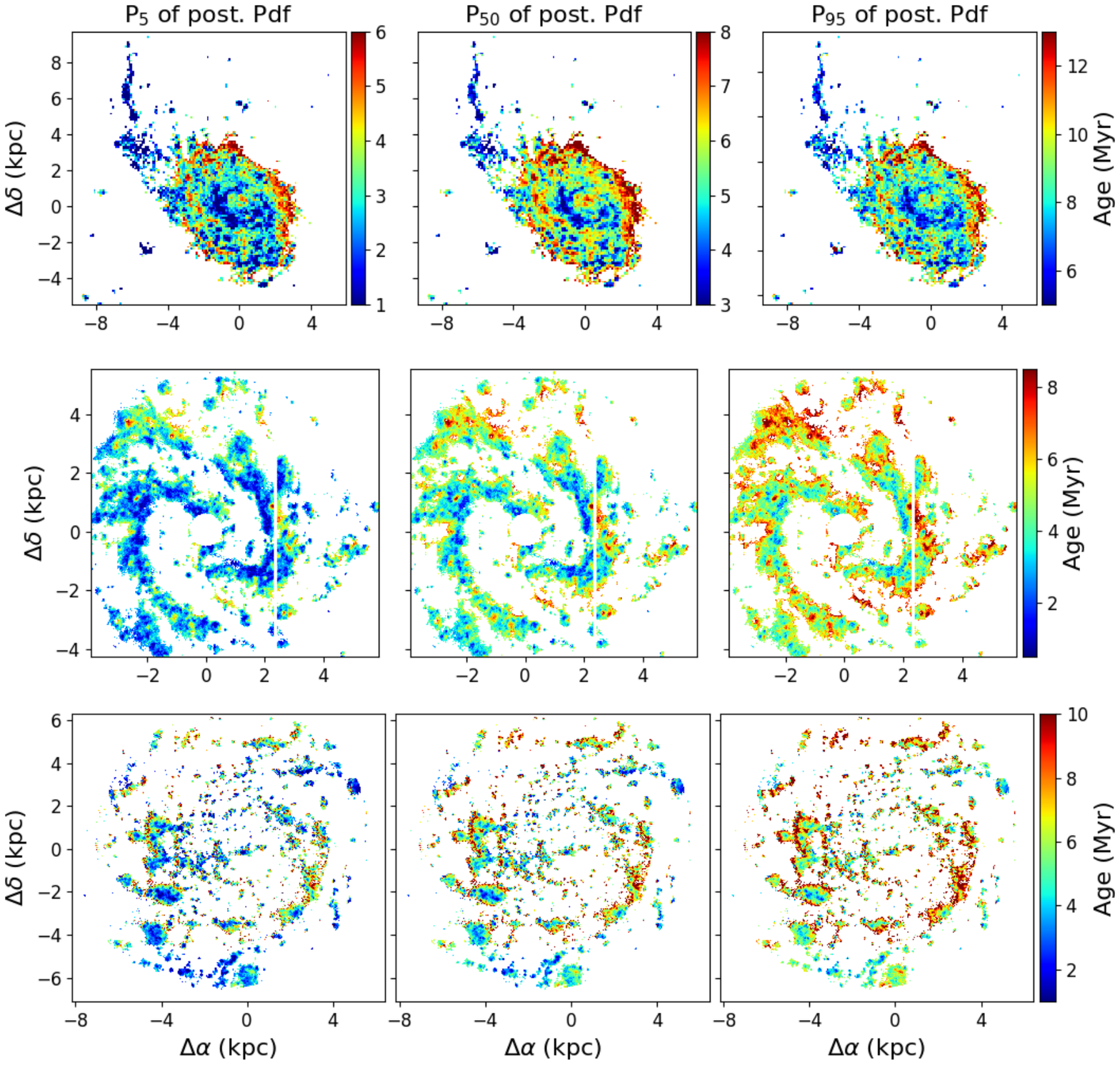}	
\caption{Comparison between age maps obtained in Section \ref{sec:HBM}, showing the central 90\%  credible interval of the age posterior distribution. 
I.e. the true value of the age will be in between these values with a probability of 90\%. 
They provide an idea of the age estimation uncertainty.
Age maps for NGC~1068 are on the top, each one with their own age scale, highligthing the age pattern structure remains nearly the same. 
M83 is in the middle panels,  and the bottom panels correspond with  M101. 
The latters share the same age scale to exhibit as the general age pattern structure remains, except an average `zero' point.}
\label{fig:Fig7}
\end{figure*}

Figures \ref{fig:Fig8},  \ref{fig:Fig10}, and  \ref{fig:Fig12}, 
show the resulting age maps for the galaxies of this study, applying the three  proposed age dating methodologies. In them we compare the two methods presented at the current work (discussed in Sections \ref{sec:HBM}) and \ref{sec:ImgSeg}, as well as the previous method (Paper I).

The top left panels show the \Ha/FUV ratio images for each galaxy, after a $3\sigma$ noise filter masking the background or fainter  \Ha\ emission pixels. 
The remaining pixels with strong \Ha\ emission, understood as whole or partial  \hii\ regions,  define the spiral arms and star forming regions clearly. 
Relative age patterns are already within this ratio image, where the higher \Ha\ to FUV values  correspond with the younger regions, whereas the darker pixels of lower ratio values represent older regions. 
This distinction between `younger' and `older'  is under a very young age scale framework, since \Ha\ emission is mainly available up to $\sim$15-20 Myr. That is, we are studying the spatial distribution of the very recent star forming regions.   

The bottom left panels, \textit{(ii)}, display the age maps following the Bayesian age dating  methodology of Section \ref{sec:ImgSeg}, assuming dependence between adjacent pixels. The \Ha to FUV flux ratio  image has been segmented  into  five or four  heterogeneous average 
$<F_{H\alpha} / F_{FUV}>$   regions.   
In this manner, we deal with issues affecting  age dating by means of a SSP model, such as IMF subsampling, luminosity or mass thresholds, or spatial influence from adjacent regions. In so doing, we lose spatial resolution into discretized age maps. 

Top right panels,\textit{(i)}, are the age maps obtained with the  Bayesian approach presented in Section  \ref{sec:HBM}, assuming spatial independence of the fluxes from each pixel. 
The higher advantage of this last technique is not only the great resolution of the age maps, but also obtaining a posterior age distribution function in each pixel of the image. 
We are able to give probabilities or select any statistical moment since a sample of the posterior probability function for the age is given. 
To match this with the same criterion of Paper I,  and for plotting purposes, each pixel of the image shows the Mode of the age posterior probability distribution.

Bottom right panels, \textit{(iii)}, are the age maps applying the  previous methodology of Paper I. 
In this case we also get a 'discrete' or categorized age map, into four age ranges:
$<4$, $4-6$, $6-9$, and $>9$ Myr. 
The color scale has been matched to the other Bayesian age maps for comparison.
We take the median age for each age range but the $>9$ Myr range by $9$ Myr. 

Figs. \ref{fig:Fig8},  \ref{fig:Fig10}, and  \ref{fig:Fig12} show the results from the previous Paper I methodology \textit{(iii)} to be robust. The look broadly the same, although the new age maps fits better with the observed structures in the \Ha/FUV flux ratio image. 
Moreover,  the present methodologies let us to get the posterior PDF  of the age, and so a continuous age map, as well as statistical estimator of probabilities.  
That is. it also improves not only qualitatively the resolution of the age patterns, 
resulting  in richer age patterns structures, but the quality of the information and the results. 
Once the data are given, we get the conditional probability of the model parameters of the observed data, and not only a single value estimator. 

Figure \ref{fig:Fig7} gives the uncertainties for the age estimation in the \textit{(i)} age maps, when the posterior probability distribution is available.
It shows the central 90\%  credible interval of the age posterior distribution; i.e., the true value of the age will be in between these values with a probability of 90\%.  
On the left it shows the 5-th percentile of the age distribution. 
The central age map corresponds to the median age, P50, and the 95-th percentile on the left. 
After such comparison we check out the robustness of the methodology for age dating. 

The corresponding age maps for NGC~1068 are on the top, each with their own age range scale. Despite the different age scales, the general age patterns are much the same. The only difference is a  zero point offset in age.  
The middle and bottom panels show the corresponding percentiles for M83 and M101, respectively. For these maps we leave the same age scale to highlight the offset point between them. Once again the general age patterns remain the same. 
On average, this zero point value is less than $\pm 2 $ Myr for the three galaxies, even $ \lesssim \pm 1 $ Myr  in the case of M83.

Finally, in Appendix \ref{app:App3} we show the age maps \text{(i)} and \text{(iii)} for the galaxy sample studied in Paper I, for the different methodologies to be compared. Several percentiles of the distribution are shown to give the corresponding age uncertainties.  
We can infer the same physical analysis and conclusions from the new age maps as was done in Paper I. 
The general age patterns results are the same, with the exception of the resulting continuous age maps what improves the resolution of the age patterns and structures, as well as the potential  of the Bayesian approach.


\subsection{M83}

\begin{figure*}
\includegraphics[width=0.95\textwidth]{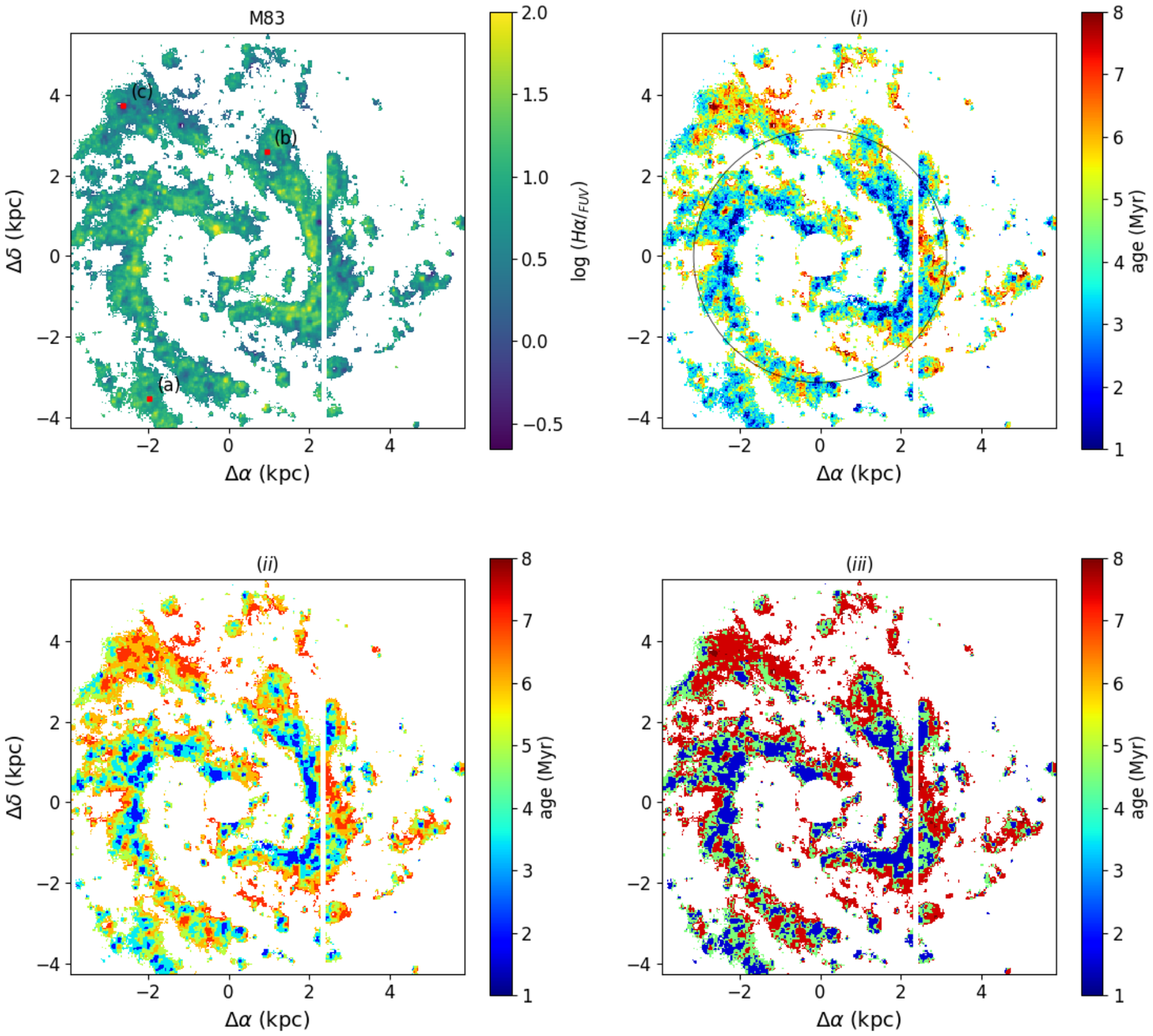}
\caption{ \textit{Top left:} Image of the \Ha\ to FUV ratio for M83 (in log units). 
The higher values of $F_{H\alpha} / F_{FUV}$ flux ratio (in yellow)  denote the youngest SF regions. 
While the lower ratios indicate 'older' ages; 
\textit{Top right (i):} Age map for M83 obtained with the presented Bayesian approach in Section \ref{sec:HBM}, assuming independence between pixels, and taking the mode of the posterior 
probability distribution of the age parameter at each pixel of the image. 
The black circle indicates the corotation radius, at $\sim 3$ kpc (see text); 
\textit{Bottom left (ii):} Age map obtained with the Bayesian approach in Section \ref{sec:ImgSeg}, assuming dependence between adjacent pixels and after an image segmentation of the $F_{H\alpha} / F_{FUV}$ flux ratio, into five heterogeneous regions. 
The color scale has been matched to the previous age map, at the top left;
\textit{Bottom right (iii):} Age map for M83, applying the  previous methodology of Paper I. 
In this case we get a 'discrete' or categorized age map into four classes or age ranges: $<4$, $4-6$, $6-9$ and $>9$ Myr. 
The color scale has been matched to the Bayesian age maps for comparison purposes 
( taking the median age for  each age range, except for the last one it is taken the minimum).
{The vertical white stripes of the images correspond with some artifacts from the \Ha\ image, so they were masked out.}
}
\label{fig:Fig8}
\end{figure*}
\begin{figure*}
\includegraphics[width=0.95\textwidth]{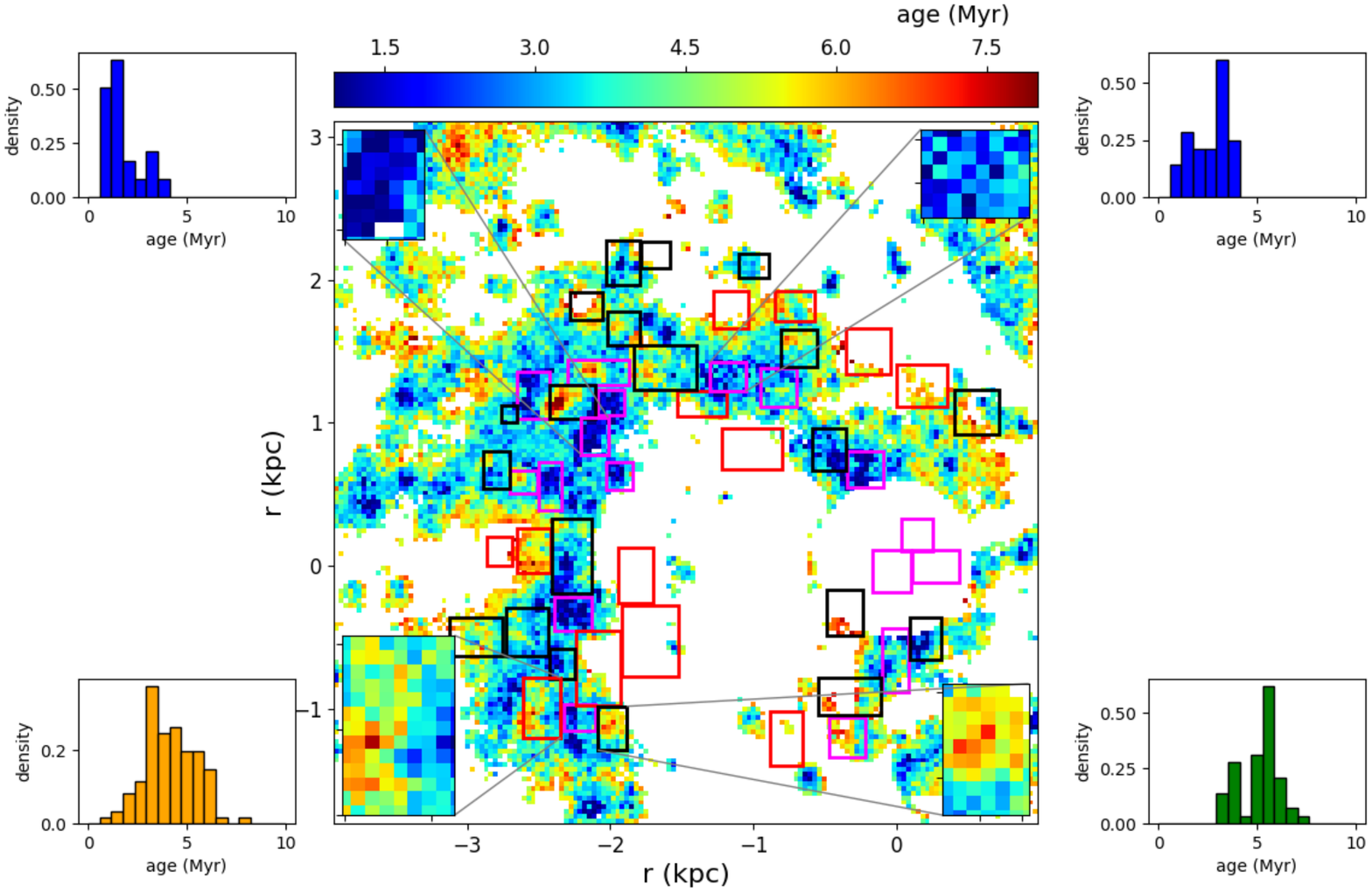}	
\caption{Zoom of the nuclear region and part of the eastern spiral arm and inter-arm region of M83, showing 50 regions from  \citep{2012ApJ...753...26K} for comparison. These regions were age-dated with data  from HST/WFC3, with the aim of studying the relationship of the spatial variations of stellar ages with the evolution of the galaxy and the  star formation triggering mechanisms.
The regions with  ages $1-10$ Myr are plotted with magenta boxes, 
ages of $10-20$ Myr in black,  and for ages greater than 20 Myr in red.}
\label{fig:Fig9}
\end{figure*}

M83 (NGC 5236), or the `Southern Pinwheel' galaxy, is a nearby barred-spiral galaxy. 
The nearest galaxy in the sample (4.5 Mpc; \citet{2002A&A...385...21K}),
its proximity and low inclination angle gives spatial resolution, not only to resolve its spiral arms but also to observe a wealth of detail in individual star-forming structures within the arms. 

M83 is a metal-rich spiral galaxy,  with a radial metallicity gradient which flattens at large radii \citep{2016ApJ...830...64B}. 
The observable used in our study is not sensitive to metallicity, we could not see nay gradient or pattern in the metallicity maps. 
The middle panels of Fig. \ref{fig:Fig6} show the marginal posterior probability distribution of the metallicity for three pixels \textit{(a)},  \textit{(b)}, and  \textit{(c)} with different flux ratio values and location within the disk (see top left plot at Fig. \ref{fig:Fig8}).
The \Ha/FUV flux ratio is quite sensitive to age variations, but metallicity  behaves  as a free parameter in a our model, free to fit the data.

The resulting age maps  \textit{(i)},\textit{(ii)}, and \textit{(iii)}, see Fig. \ref{fig:Fig8},  are  similar to that obtained for M~51 in Paper I, dominated by a young population of stars with less than 6-7 Myr. 
The age structure exhibits gradients across the spiral arms,  
with the younger stars toward the inner edge while the older stars are located approaching the outer edges within the corotation radius, 
as expected from the density wave theory \citep{1969ApJ...158..123R,2009ApJ...694..512M}.
The corotation radius of  $\sim 3$ kpc ($\sim 2.3'-2.4'$  \citep{2014PASJ...66...46H} at a distance of 4.5 Mpc) is plotted as a black circle in the top right panel of Fig. \ref{fig:Fig8}. 
We can also observe the inverse age pattern outside corotation; i.e., the younger population are preferentially located radially in the outer side of the arms. 
The eastern arm exhibits pretty well how the youngest population, bluest pixels, change from the inner to outer arm side when crossing the corotation radius. 

Fig. \ref{fig:Fig9} displays 50 regions of an average size $260$ pc $\times \, 280$ pc, which cover the nuclear region and part of the eastern spiral arm and inter-arm region \citep[][see their fig.1]{2012ApJ...753...26K}.
These authors selected these regions with the aim of studying the spatial variations of stellar ages in M83, looking for evidence of the evolution of the galaxy and star formation triggers.
Their age scale is similar to ours, with regions classified into 3 age ranges: 
encompassing $1-10$ Myr, which covers the majority of our age range. These are plotted with magenta boxes at Fig. \ref{fig:Fig9}. 
It can be seen that  many of these regions also correspond with our younger ages. 
Two examples are magnified at the top left and top right sides. The blue histograms represent the age distribution of two of these regions, with ages less than 5 Myr.   

Regions with ages of $10-20$ Myr are shown with black boxes. They correspond to our oldest ages,  since this is the limit for \Ha\ emission. 
We observe in Fig. \ref{fig:Fig9} how many of these regions are indeed dominated by our older ages (orange and red pixels). 
But there are also some of these regions more likely to be intermediate ages, such as the case in the bottom left of the plot.

Outlined red boxes show the third group, with ages greater than 20 Myr. Most of these do not show \Ha\ in our maps, as they correspond to the post-nebular phase.
In general we find quite a good agreement between the two results.

For the youngest star forming regions, our age maps find a similar scenario as that described by \citet{2012ApJ...753...26K}, and in agreement with density wave theory. 
These authors also found that younger (10 Myr) stars are found mainly  in concentrated aggregates along the active star forming regions in the spiral arm. Intermediate age stars are located downstream, on the opposite side from the dust lane, as expected based on density wave models, and the older stars more dispersed (see their fig. 12). 
\citet{2012ApJ...753...26K} argue stars form primarily in star clusters and then disperse on short timescales to form the field population. 
Moreover, Wolf-Rayet stars, which are taken into account in our SSP models, correlate with the position of many of the youngest regions. 


\citet{2010MNRAS.409..396D} study  the mechanisms triggering star formation in galaxies and their evolution.
They discuss the locations of age-dated stellar clusters as a  possible discriminant of the origin/source of the excitation mechanism for the spiral structure. 
Under the assumption that stellar clusters form predominantly within the spiral arms (higher gas density regions), the distribution of the age-dated clusters through out a spiral could give some clues to the mechanism for spiral arm formation. 
These authors found  diverse spatial distributions for clusters  of different ages, depending on the underlying dynamics of the galaxy and the spiral excitation mechanism. 

Despite the different age scales (the age in \citet{2010MNRAS.409..396D} varies from $\sim 2 - 130$ Myr, see their fig.2), we  note that their models for a  global age pattern still agree with our age maps for the youngest stellar populations (up to around 15 or 20 Myr).

For both M83 and M101 (Figs. \ref{fig:Fig8} and \ref{fig:Fig10}),  we can identify the age distribution for a galaxy model of a constant pattern speed with a bar, and a floculent spiral, respectively (cf. their figure 2). 
\citet{2010MNRAS.409..396D} describe the younger stars in the spiral arms or bar with older stars downstream in the interarm regions for the former model. 
The distribution of stellar clusters is more complicated in floculent spirals, because each segment of a spiral arm tends to contain clusters of a similar age. This can be observed in Fig. at \ref{fig:Fig10} for  M101, mostly  at the  extreme southern segments.

\subsection{M101}

\begin{figure*}
\includegraphics[width=0.95\textwidth]{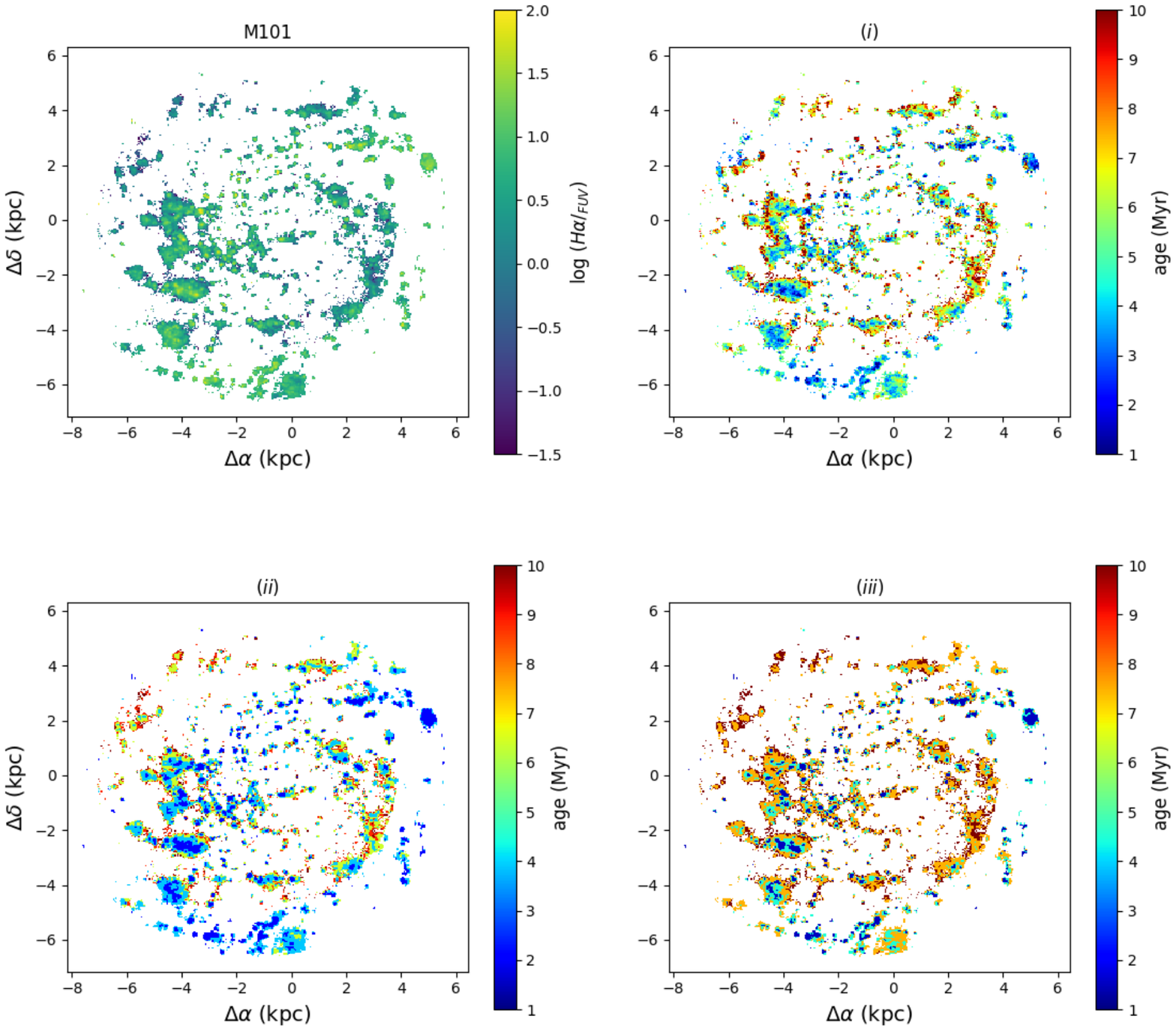}	
\caption{As Fig.~\ref{fig:Fig8}, but for M101.}
\label{fig:Fig10}
\end{figure*}
\begin{figure*}
\includegraphics[width=0.95\textwidth]{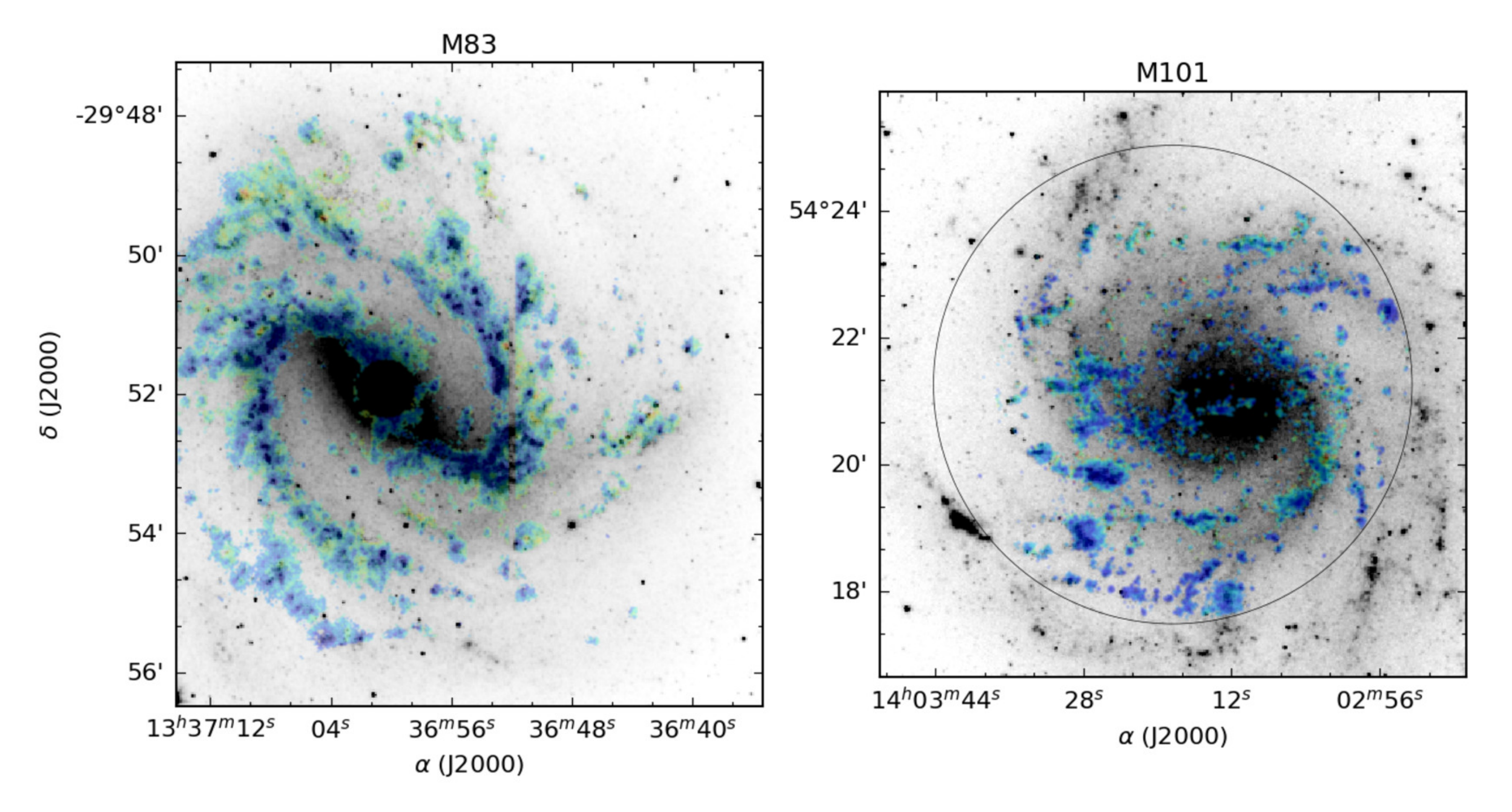}	
\caption{Comparison  between the resulting morphology of star forming regions  for M83 and M101. Despite both being spiral galaxies, the former has a well defined spiral estructure, whereas the latter has flocculent spiral arms segments. 
The background image is a  3.6 $\mu$m IRAC image from the \textit{Spitzer} survey, where it is effectively observed the characteristic spiral morphology in both cases.
Black circle defines the TTF field of view for the \Ha\ image, leaving outside part of the disk and spiral arms of M101.}
\label{fig:Fig11}
\end{figure*}

M101 is a nearby, face-on, giant spiral galaxy. At a distance of $6.7$ kpc  \citep{1981A&A....93..106B}, it provides an enough spatial resolution to study the underlying stellar populations. 
It is also an excellent laboratory for studying stars, as OB type stars, blue giants, yellow supergiants, and red supergiants, as well as stellar clusters, \hii\ regions, and supernova remnants  \citep[][and references within]{2013AJ....146..114G,2014AJ....148...58G}.

\citet{2014AJ....148...58G} explore the M101 SFH as a function of the radial distance and its effect on the emission properties and stellar populations. 
They find that the mass fraction for stars younger than 16 Myr (roughly comparable to our age range) is $15\%-35\%$ in the inner regions, compared to less than $5\%$ in the outer regions.
This percentage is greater than $50\%$ in the inner regions for stars younger than 35 Myr. 
That is, the inner regions are  dominated by a young stellar populations.  
Comparing the results in their figure 8 within the reach of our data, 8 kpc, we see that 85\% and 95\% of stellar mass fraction have ages younger that 15 or 20 Myr under 10 kpc in radius, in good agreement with our results.   

As in Fig. \ref{fig:Fig8} for M83, Fig. \ref{fig:Fig10} shows the age maps for M101; \textit{(i)} and \textit{(ii)}, derived with the methods introduced  in this paper, and \textit{(iii)} as in Paper I.
A common characteristic feature of these age maps is their morphologies, with a structure more flocculent than spiral.
Furthermore, we detect some an age gradient tending toward younger stars in the `outer' disk. It is not actually the outer disc but the middle region, since we only observe around 8 kpc, due to the circular Taurus aperture. 

According to \citet{2013ApJ...769..127L}, the radial age profile of M101 presents a younger bulge, comprising an older inner region of the disk with steeper age gradient, and an outer disk region with a flatter gradient. These gradients and ages are not comparable with our age maps, since they are in Gyr units. 
However, at youngest age scales there are similar trends. 
The interaction with another galaxy of the M101 group and consequent  gas accretion could trigger this star formation in the outer regions. 
The low gas metallicity gradient of M101 with respect to other spiral galaxies also  supports an interacting or recent merger scenario 
\citep[][and references within]{2013ApJ...769..127L}. 

\citet{2013ApJ...769..127L} also describe spurious arms full of \hii\ regions much younger than the interarm in the inner disk (albeit within a different framework, of wider age scales in Gyr and providing the whole SFH of M101).

Fig. \ref{fig:Fig11} shows a comparison  between the  morphologies of the star forming regions in M83 and M101. 
The background IRAC 3.6 $\mu$m  images show clearly their spiral structure. 
However, while the age map of M83 presents a well defined structure, whereas the corresponding  M101 image shows flocculent spiral arms segments. 
The black circle in the M101 image defines the TTF field of view for the \Ha\ image. 
Despite the regions it leaves outside the field of view,  this does not affect the resulting global morphology for this galaxy. The inner disk has a clear flocculent structure. 

The difference between the morphologies of the star forming regions of both M83 and M101 is interesting. It is likely related to the processes triggering  the star formation, as well as the subsequent evolution.


\subsection{NGC~1068}

\begin{figure*}
\includegraphics[width=0.95\textwidth]{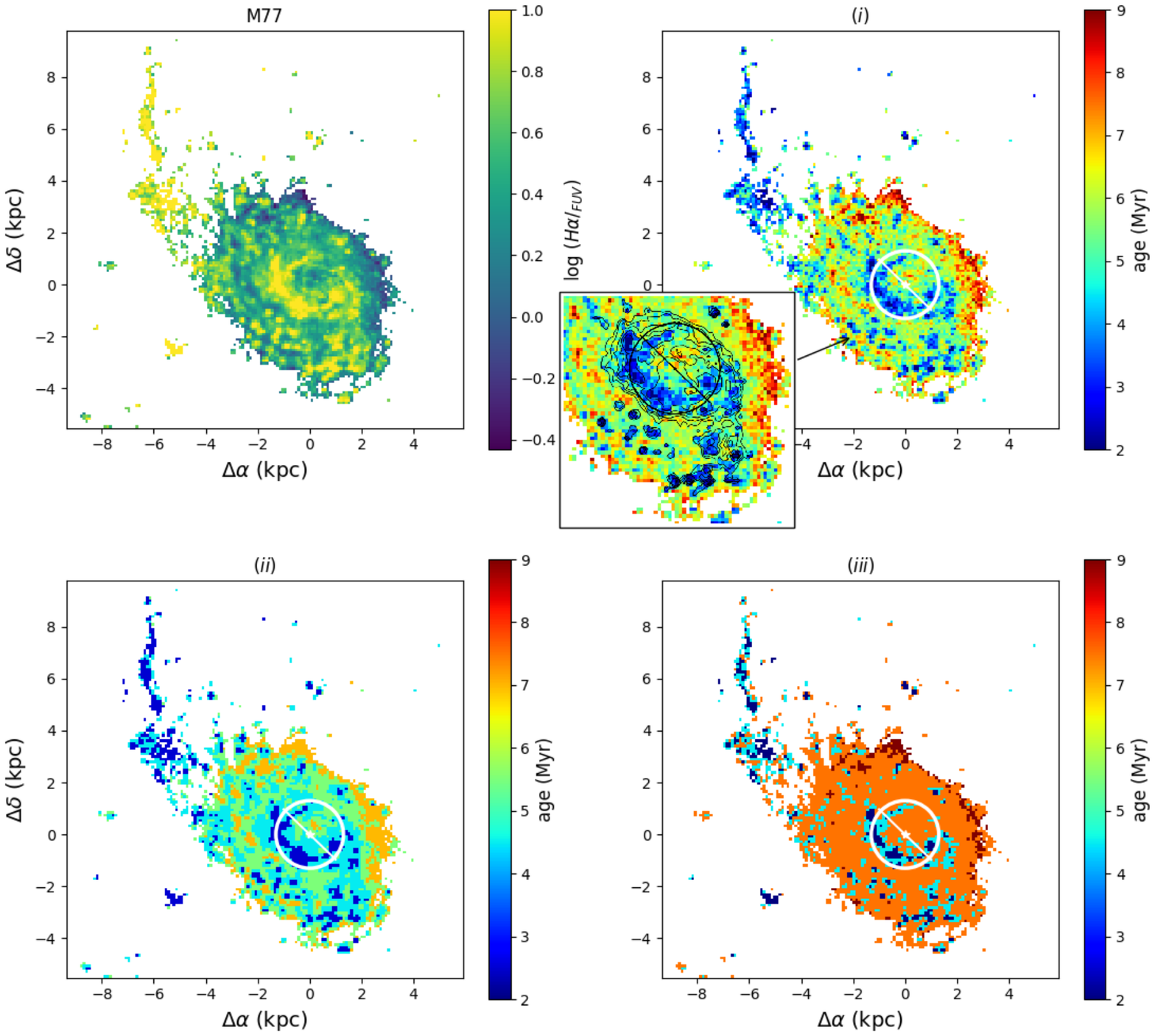}	
\caption{As Fig.~\ref{fig:Fig8}, but for NGC~1068.
The AGN location is marked with a star in all the age maps, as well as the 3 kpc bar \citep{1997Ap&SS.248....9B}, outlined by a white line, and the $\approx 1.3$ kpc corotation radius \citep{2014A&A...567A.125G} with a white (black in the zoomed central region) circle. 
The central part of the disk is zoomed in the \textit{(i)} age map, plotting the highest levels, greater than  percentiles $P_{75}$,$P_{80}$,$P_{85}$,$P_{90}$, and $P_{95}$, of the age contours.
These contours define the starburst ring, and some of the south-west spiral arm and concentrated knots of SF. }
\label{fig:Fig12}
\end{figure*}

NGC~1068 is an early-type Sb, barred spiral galaxy, 
and the closest  \citep[14.4 Mpc, $1'' = 72$ pc;][]{1997Ap&SS.248....9B} luminous Seyfert 2 galaxy. 
It is considered the prototype Seyfert 2 galaxy \citep{1974ApJ...192..581K}. 
Its brightness, nuclear activity, proximity, and orientation 
make it  an excellent laboratory (in a single physical framework) to study the Seyfert nucleus, inner disk structure, as well as the unifying, dusty torus model for Seyfert galaxies \citep[e.g.][]{1982MNRAS.200.1067O,1998MNRAS.300..388D, 2001ApJ...546..866B, 2014A&A...565A..71L}. 

Likely the best studied active galaxy in the local universe, it has been subject of numerous studies, at many different wavelengths. 
Signs of the AGN are evident at nearly all wavelengths, such as the bright cones of photoionized gas detected both in optical and X-rays wavelenghts  \citep[][and references within]{2003AJ....126.2185V}.  
It has a large-scale oval and a nuclear bar with a pseudo-bulge which is very massive with respect to its central black hole  \citep[e.g.][]{2013ARA&A..51..511K}.
The ionization source of the gas in the circumnuclear region is due mainly to the AGN,   
so  age determination with the proposed methodologies basedon  \Ha/FUV  is not reliable for this innermost region. 
The location of the AGN  is marked with a star in the age maps of Fig. \ref{fig:Fig11}. 
Diagnostic line ratio plots have shown that photoionization is the preferred mechanism for high excitation gas in NGC~1068 (Nishimura et al. 1984; Evan \& Dopita 1987; Bergeron et al. 1989). 

Once the background has been subtracted, the age maps (see Fig. \ref{fig:Fig12}) do not show  well-defined spiral arms as M83, indicating a lower intensity of SF activity across the disk.
The circumnuclear ring or pseudo-ring (the star-forming or starburst ring)  has been clearly resolved into tightly-wound spiral arms at the ends of the 3 kpc bar \citep{1997Ap&SS.248....9B}, outlined by the white line segment at Fig. \ref{fig:Fig12}.
This starburst ring is clearly defined in our age maps  by the bluest, and therefore youngest, structures inside and around the corotation radius, outlined with a white/black circumference at Fig. \ref{fig:Fig12}, at $r \sim 1.3$ kpc \citep{2014A&A...567A.125G}.
The \hii\ regions in the circumnuclear ring are as bright as a string of M82-type galaxies, and compete in terms of bolometric luminosity with the nucleus \citep{1998MNRAS.300..388D}. 
\citet{1984ApJ...282..427T} 
found that approximately the half of the bolometric luminosity of NGC~1068, 
$\sim3\times 10^{11} L_{\odot}$, corresponds to the Seyfert nucleus whereas the other half to the starburst region, where most of the CO(3-2) flux in the disk is detected 
\citep{2014A&A...567A.125G}.
The existence of the circumnuclear ring is attributed to gas settling between the inner Lindbland resonances (ILRs) as a result of the action of a barred gravitational potential \citep{1988ApJ...334..573T}. Therefore vigorous SF is expected to occur as a direct result of the increased cloud density in the ILR.

The three age maps in Fig. \ref{fig:Fig12}, coincide with the location of the youngest regions, the bluest ones, mainly within the brightest regions of the ring, around the corotation radius, 
with some knots of high SF regions throughout the disk, 
and at  north-east plume out to a radius of $\sim 7-8$ kpc from the nucleus. 
However, multiline imaging and long-slit spectroscopy of the gas in this north-east complex or filament, roughly aligned with the ionization cone on smaller scale, revealed contributions from the AGN hard radiation to the high ionization of the gas in this region \citep{2003AJ....126.2185V}. If photionization is not the main source of the gas ionization, the ages determined by  \Ha/FUV  are not reliable, as it also happens within the circumnuclear region, due to the AGN effects.

In the rest of the disk, the younger population is concentrated in the inner ring, as mentioned above. It seems to extend along the south-west spiral arm at larger scale, as well as it concentrates in some knots of SF regions across the disk. 
The intermediate ages are within the disk, and the older ones are found at the edges. 

The relevance  or  strength of the  ring luminosity, being  host of the youngest population in the disk, is checked by plotting  contour levels from the \Ha\ image. The highest levels appear only around this ring. 
To see this feature in more detail, the central part of the disk is zoomed in the \textit{(i)} age map of Fig. \ref{fig:Fig12}, combining the ages and highest levels of age contours.
Only  those levels greater than the $P_{75}$ percentile are shown, which coincide with the ring and some of the south-west spiral arm, but not in the plume.

\section{Discussion}
\label{sec:conclu}

In this paper we present two Bayesian inference algorithms that allow us to evaluate the age of the stellar population of a galaxy at pixel resolution. 
The methodology of section \ref{sec:HBM} provides the posterior probability density function for different parameters of the stellar population synthesis models, including age, metallicity and the fraction of ionizing photons. Only the former results in sensitive to the \Ha/FUV flux ratio.
We focused on the age variable, with a unimodal distribution and well defined central value in most cases, Fig. \ref{fig:Fig6}. 
The accuracy of age estimation in the cases studied  is better than $\pm 2$ Myr. 
However total uncertainty could be strongly influenced when applying synthetic stellar models to single pixels, more than by the own model uncertainties. 
This imprecision is  affected by the mass of the underlying stellar population, whose parameters are estimated. 
The measurement of this uncertainty is not trivial when the total brightness of the pixel is lower than the Lower Luminosity Limit (LLL) of the model \citep{2006A&A...451..475C}, and is beyond the scope of this work. 

The other Bayesian approach developed in Section \ref{sec:ImgSeg} try to mitigate the effect of these luminosity/mass thresholds or IMF sub-sampling issues, when SSP models are directly applied to pixel-wise regions. 
On the other hand, we get a discretization of the age map, and so higher resolution is lost. Nevertheless the main structures remain recognizable, or otherwise unchanged. 

In general, the age patterns we show appear to be quite robust to sampling and only sensitive to the \Ha/FUV flux ratio. 
So the importance of these two new techniques is, on the one hand, the huge potential provided by the Bayesian inference and holding the age posterior probability function (a sample of the posterior). 
And in the other case, to  deal  with the uncertainties derived when applying SSP models to pixel-wise size regions. 
The present work focuses mainly on the formulation of a new methodology to estimate the age from UV, optical, and infrared images by Bayesian inference.
 
The number of arms shown, the sharpness and width of these arms, as well as the separation and location of large star-formation complexes along the arms are some of the features that can give us some first hints about the nature of the physical mechanisms within a spiral galaxy. 
However, the morphology by itself can sometimes be misleading.  Thus we also need  physical observables, drawing from their galactic spatial distribution to reconcile model predictions from different spiral pattern generators, \citet{2010MNRAS.409..396D}. 
Age map analysis is one of the best ways to test these aims.

These proposed methodologies only refer to the youngest stellar populations (i.e. the distribution or location of the youngest stellar regions, less than 20 Myr), from the very recent to less recent episodes of SF along the galactic disk. These are mainly in the arms and interarm regions of the grand-design galaxies. 
The use of age maps to elucidate the origin of the spiral arms is outside the scope of this paper. 
However we are able to highlight how age maps show very distinctive morphologies characteristics for three spiral galaxies. 


\section*{Acknowledgements}
We acknowledge financial support from the Spanish Ministry of Economy and Competitiveness through grants AYA2017-88007-C3-1-P, AYA2016-75931-C2-1-P, AYA2015-68012-C2-1, AYA2014-57490-P, AYA2014-58861-C3-1-P, AYA2016-77846-P, 
AYA2013-40611-P,  and  from the Consejer\'{\i}a de Educaci\'on y Ciencia (Junta de Andaluc\'{\i}a) through TIC-101, TIC-4075 and TIC-114.





\begin{thebibliography}{99}
%
\bibitem[de Amorim et al.(2017)]{2017MNRAS.471.3727D} de Amorim, A.~L., Garc{\'{\i}}a-Benito, R., Cid Fernandes, R., et al.\ 2017, \mnras, 471, 3727 
%
\bibitem[Barker et al.(2008)]{2008A&A...484..711B} Barker, S., de Grijs, R., \& Cervi{\~n}o, M.\ 2008, \aap, 484, 711 
%
%
\bibitem[Bland-Hawthorn \& Jones(1998)]{1998PASA...15...44B} Bland-Hawthorn, J., \& Jones, D.~H.\ 1998, \pasa, 15, 44 
%
%
\bibitem[Bland-Hawthorn et al.(1997)]{1997Ap&SS.248..177B} Bland-Hawthorn, J., Lumsden, S.~L., Voit, G.~M., Cecil, G.~N., \& Weisheit, J.~C.\ 1997, \apss, 248, 177 
%
\bibitem[Bland-Hawthorn et al.(1997)]{1997Ap&SS.248....9B} Bland-Hawthorn, J., Gallimore, J.~F., Tacconi, L.~J., et al.\ 1997, \apss, 248, 9 
%
\bibitem[Bosma et al.(1981)]{1981A&A....93..106B} Bosma, A., Goss, W.~M., \& Allen, R.~J.\ 1981, \aap, 93, 106 
%
\bibitem[Boissier et al.(2005)]{2005ApJ...619L..83B} Boissier, S., Gil de Paz, A., Madore, B.~F., et al.\ 2005, \apjl, 619, L83 
%
\bibitem[Bresolin et al.(2016)]{2016ApJ...830...64B} Bresolin, F., Kudritzki, R.-P., Urbaneja, M.~A., et al.\ 2016, \apj, 830, 64 
%
\bibitem[Bruhweiler et al.(2001)]{2001ApJ...546..866B} Bruhweiler, F.~C., Miskey, C.~L., Smith, A.~M., Landsman, W., \& Malumuth, E.\ 2001, \apj, 546, 866 
%
\bibitem[\protect\citeauthoryear{Buat \& Xu}{1996}]{BuatXu1996} Buat, V., \& Xu, C.\ 1996, \aap, 306, 61 
%
\bibitem[\protect\citeauthoryear{Buat et al.}{1999}]{Buat1999} Buat, V., Donas, J., Milliard, B., \& Xu, C.\ 1999, \aap, 352, 371 
%
\bibitem[Buat et al.(2005)]{2005ApJ...619L..51B} Buat, V., Iglesias-P{\'a}ramo, J., Seibert, M., et al.\ 2005, \apjl, 619, L51 
%
\bibitem[Casado et al.(2017)]{2017MNRAS.466.3989C} Casado, J., Ascasibar, Y., Garc{\'{\i}}a-Benito, R., et al.\ 2017, \mnras, 466, 3989 
%
\bibitem[Cervi{\~n}o et al.(2002)]{2002A&A...381...51C} Cervi{\~n}o, M., Valls-Gabaud, D., Luridiana, V., \& Mas-Hesse, J.~M.\ 2002, \aap, 381, 51 
%
\bibitem[Cervi{\~n}o et al.(2003)]{2003A&A...407..177C} Cervi{\~n}o, M., Luridiana, V., P{\'e}rez, E., V{\'{\i}}lchez, J.~M., \& Valls-Gabaud, D.\ 2003, \aap, 407, 177 
%
\bibitem[Cervi{\~n}o \& Luridiana(2004)]{2004A&A...413..145C} Cervi{\~n}o, M., \& Luridiana, V.\ 2004, \aap, 413, 145 
%
\bibitem[\protect\citeauthoryear{Cervi{\~n}o \& Luridiana}{2006}]{2006A&A...451..475C} Cervi{\~n}o, M. \& Luridiana, V.\ 2006, \aap, 451, 475 
%
\bibitem[Cervi{\~n}o(2013)]{2013NewAR..57..123C} Cervi{\~n}o, M.\ 2013, \nar, 57, 123 
%
\bibitem[Charbonnel et al.(1993)]{1993A&AS..101..415C} Charbonnel, C., Meynet, G., Maeder, A., Schaller, G., \& Schaerer, D.\ 1993, \aaps, 101, 415 
%
%
\bibitem[Dale et al.(2007)]{2007ApJ...655..863D} Dale, D.~A., Gil de Paz, A., Gordon, K.~D., et al.\ 2007, \apj, 655, 863 
%
\bibitem[Dale \& Helou(2002)]{2002ApJ...576..159D} Dale, D.~A., \& Helou, G.\ 2002, \apj, 576, 159 
%
\bibitem[Davies et al.(1998)]{1998MNRAS.300..388D} Davies, R.~I., Sugai, H., \& Ward, M.~J.\ 1998, \mnras, 300, 388 
%
\bibitem[Dobbs \& Pringle(2010)]{2010MNRAS.409..396D} Dobbs, C.~L., \& Pringle, J.~E.\ 2010, \mnras, 409, 396 
%
%
\bibitem[Foyle et al.(2012)]{2012MNRAS.421.2917F} Foyle, K., Wilson, C.~D., Mentuch, E., et al.\ 2012, \mnras, 421, 2917 
%
\bibitem[Galametz et al.(2013)]{2013MNRAS.431.1956G} Galametz, M., Kennicutt, R.~C., Calzetti, D., et al.\ 2013, \mnras, 431, 1956 
%
\bibitem[Garc{\'{\i}}a-Burillo et al.(2014)]{2014A&A...567A.125G} Garc{\'{\i}}a-Burillo, S., Combes, F., Usero, A., et al.\ 2014, \aap, 567, A125 
%
\bibitem[\protect\citeauthoryear{Gelman et al.}{2003}]{Gelman03} Andrew Gelman, John B. Carlin, Hal S. Stern \&  Donald B. Rubin 2003, Bayesian Data Analysis  Chapman and Hall/CRC, 158488388X 
%
\bibitem[Gonz{\'a}lez-Gait{\'a}n et al.(2018)]{2018arXiv180206280G} Gonz{\'a}lez-Gait{\'a}n, S., de Souza, R.~S., Krone-Martins, A., et al.\ 2018, arXiv:1802.06280 
%
\bibitem[\protect\citeauthoryear{Glazebrook et al.}{1999}]{Glazebrook1999} Glazebrook, K., Blake, C., Economou, F., Lilly, S., \& Colless, M.\ 1999, \mnras, 306, 843 
%
\bibitem[\protect\citeauthoryear{Gordon et al.}{2000}]{Gordon2000} Gordon, K.~D., Clayton, G.~C., Witt, A.~N., \& Misselt, K.~A.\ 2000, \apj, 533, 236 
%
\bibitem[Grammer \& Humphreys(2013)]{2013AJ....146..114G} Grammer, S., \& Humphreys, R.~M.\ 2013, \aj, 146, 114 
%
\bibitem[Grammer \& Humphreys(2014)]{2014AJ....148...58G} Grammer, S., \& Humphreys, R.~M.\ 2014, \aj, 148, 58 
%
\bibitem[\protect\citeauthoryear{Grebel}{2000}]{Grebel2000} Grebel, E.~K.\ 2000, Star Formation from the Small to the Large Scale, 445, 87 
%
\bibitem[\protect\citeauthoryear{Haykin, S.}{1999}]{Haykin1999} Haykin, Simon \ 1999, Neural Networks, A Comprehensive Foundatio, Prentice-Hall, 0-13-273350-1
%
\bibitem[\protect\citeauthoryear{Hinkley, D. V.}{1969}]{Hinkley1969} Hinkley, D. V. \ 1969, On the Ratio of Two Correlated Normal Random Variables, Biometrika Vol. 56, No. 3 (Dec., 1969), pp. 635-639
%
\bibitem[Hirota et al.(2014)]{2014PASJ...66...46H} Hirota, A., Kuno, N., Baba, J., et al.\ 2014, \pasj, 66, 46 
%
%
\bibitem[Jones et al.(2002)]{2002MNRAS.329..759J} Jones, D.~H., Shopbell, P.~L., \& Bland-Hawthorn, J.\ 2002, \mnras, 329, 759 
%
\bibitem[Jones \& Bland-Hawthorn(2001)]{2001ApJ...550..593J} Jones, D.~H., \& Bland-Hawthorn, J.\ 2001, \apj, 550, 593 
%
\bibitem[Karachentsev et al.(2002)]{2002A&A...385...21K} Karachentsev, I.~D., Sharina, M.~E., Dolphin, A.~E., et al.\ 2002, \aap, 385, 21 
%
\bibitem[\protect\citeauthoryear{Kennicutt}{1998}]{Kennicutt1998} Kennicutt, R.~C., Jr.\ 1998, \apj, 498, 541 
%
\bibitem[Khachikian \& Weedman(1974)]{1974ApJ...192..581K} Khachikian, E.~Y., \& Weedman, D.~W.\ 1974, \apj, 192, 581 
%
\bibitem[Kim et al.(2012)]{2012ApJ...753...26K} Kim, H., Whitmore, B.~C., Chandar, R., et al.\ 2012, \apj, 753, 26 
%
\bibitem[Kormendy \& Ho(2013)]{2013ARA&A..51..511K} Kormendy, J., \& Ho, L.~C.\ 2013, \araa, 51, 511 
%
\bibitem[Leitherer et al.(1999)]{1999ApJS..123....3L} Leitherer, C., Schaerer, D., Goldader, J.~D., et al.\ 1999, \apjs, 123, 3 
%
\bibitem[Leitherer \& Heckman(1995)]{1995ApJS...96....9L} Leitherer, C., \& Heckman, T.~M.\ 1995, \apjs, 96, 9 
%
\bibitem[Lejeune et al.(1997)]{1997A&AS..125..229L} Lejeune, T., Cuisinier, F., \& Buser, R.\ 1997, \aaps, 125, 229 
%
%
\bibitem[Lin et al.(2013)]{2013ApJ...769..127L} Lin, L., Zou, H., Kong, X., et al.\ 2013, \apj, 769, 127 
%
\bibitem[L{\'o}pez-Gonzaga et al.(2014)]{2014A&A...565A..71L} L{\'o}pez-Gonzaga, N., Jaffe, W., Burtscher, L., Tristram, K.~R.~W., \& Meisenheimer, K.\ 2014, \aap, 565, A71 
%
%
\bibitem[Mas-Hesse \& Kunth(1991)]{1991A&AS...88..399M} Mas-Hesse, J.~M., \& Kunth, D.\ 1991, \aaps, 88, 399 
%
\bibitem[Martin et al.(2005)]{2005ApJ...619L...1M} Martin, D.~C., Fanson, J., Schiminovich, D., et al.\ 2005, \apjl, 619, L1 
%
\bibitem[Mart{\'{\i}}nez-Garc{\'{\i}}a et al.(2009)]{2009ApJ...694..512M} Mart{\'{\i}}nez-Garc{\'{\i}}a, E.~E., Gonz{\'a}lez-L{\'o}pezlira, R.~A., \& Bruzual-A, G.\ 2009, \apj, 694, 512 
%
\bibitem[Marin \& Robert (2014)]{10.1007/978-1-4614-8687-9} Marin, Jean-Michel and Christian P. Robert. 2014, Bayesian Essentials with R. New York: Springer 10.1007/978-1-4614-8687-9
%
%
\bibitem[Orr \& Browne(1982)]{1982MNRAS.200.1067O} Orr, M.~J.~L., \& Browne, I.~W.~A.\ 1982, \mnras, 200, 1067 
%
\bibitem[Roberts(1969)]{1969ApJ...158..123R} Roberts, W.~W.\ 1969, \apj, 158, 123 
%
%
\bibitem[\protect\citeauthoryear{Rue et al.}{2017}]{Rue2017}
H\r{a}vard Rue, Andrea Riebler, Sigrunn H. S\o rbye, Janine B. Illian, Daniel P. Simpson, Finn K. Lindgren. 
Annual Review of Statistics and Its Application 2017 4:1, 395-421 
%
\bibitem[Salpeter(1955)]{1955ApJ...121..161S} Salpeter, E.~E.\ 1955, \apj, 121, 161 
%
%
\bibitem[S{\'a}nchez-Gil et al.(2011)]{2011MNRAS.415..753S} S{\'a}nchez-Gil, M.~C., Jones, D.~H., P{\'e}rez, E., et al.\ 2011, \mnras, 415, 753 
%
\bibitem[S{\'a}nchez Gil et al.(2015)]{2015JPhCS.633a2140S} S{\'a}nchez Gil, M.~C., Berihuete, A., Alfaro, E.~J., P{\'e}rez, E., \& Sarro, L.~M.\ 2015, Journal of Physics Conference Series, 633, 012140 
%
\bibitem[Schaller et al.(1992)]{1992A&AS...96..269S} Schaller, G., Schaerer, D., Meynet, G., \& Maeder, A.\ 1992, \aaps, 96, 269 
%
%
\bibitem[Schaerer et al.(1993a)]{1993A&AS..102..339S} Schaerer, D., Charbonnel, C., Meynet, G., Maeder, A., \& Schaller, G.\ 1993(a), \aaps, 102, 339 
%
\bibitem[Schaerer et al.(1993b)]{1993A&AS...98..523S} Schaerer, D., Meynet, G., Maeder, A., \& Schaller, G.\ 1993(b), \aaps, 98, 523 
%
%
\bibitem[\protect\citeauthoryear{Skilling, J.}{2006}]{skilling2006} Skilling, John. Nested sampling for general Bayesian computation. Bayesian Anal. 1 (2006), no. 4, 833--859. doi:10.1214/06-BA127. http://projecteuclid.org/euclid.ba/1340370944.
%
\bibitem[Smith et al.(2002)]{2002MNRAS.337.1309S} Smith, L.~J., Norris, R.~P.~F., \& Crowther, P.~A.\ 2002, \mnras, 337, 1309 
%
Julien Stoehr. A review on statistical inference methods for discrete Markov random fields. 2017.
\bibitem[Stoehr (2017)]{Stoehr17} Julien Stoehr \ 2017. A review on statistical inference methods for discrete Markov random fields.  Pre-print 
\url{https://hal.archives-ouvertes.fr/hal-01462078v2},
\url{arXiv:1704.03331}
%
\bibitem[Telesco \& Decher(1988)]{1988ApJ...334..573T} Telesco, C.~M., \& Decher, R.\ 1988, \apj, 334, 573 
%
\bibitem[\protect\citeauthoryear{Telesco et al.}{1984}]{1984ApJ...282..427T} Telesco, C.~M., Becklin, E.~E., Wynn-Williams, C.~G., \& Harper, D.~A.\ 1984, \apj, 282, 427
%
\bibitem[Thim et al.(2003)]{2003ApJ...590..256T} Thim, F., Tammann, G.~A., Saha, A., et al.\ 2003, \apj, 590, 256 
%
\bibitem[Tully et al.(2008)]{2008ApJ...676..184T} Tully, R.~B., Shaya, E.~J., Karachentsev, I.~D., et al.\ 2008, \apj, 676, 184-205 
%
\bibitem[V{\'a}zquez \& Leitherer(2005)]{2005ApJ...621..695V} V{\'a}zquez, G.~A., \& Leitherer, C.\ 2005, \apj, 621, 695 
%
\bibitem[Veilleux et al.(2003)]{2003AJ....126.2185V} Veilleux, S., Shopbell, P.~L., Rupke, D.~S., Bland-Hawthorn, J., \& Cecil, G.\ 2003, \aj, 126, 2185 
%
\bibitem[Wu (1982)]{RevModPhys.54.235} Wu, F. Y. \ 1982, The Potts model, Rev. Mod. Phys., 54, 1, 235--268 
%




\end{thebibliography}




\appendix

\newpage 
\onecolumn
\section{Sampling effects and the Ratio of two correlated normal random variables}
\label{Sec:App0}
Along this work we assume that the underling distributions of \Ha ~and FUV luminosities can be described as a Gaussian distribution, therefore their ratio distribution is the ratio of two correlated Gaussian distributions, this is we  take into account the correlation between the \Ha\ and the FUV flux from the same source. 

Let us first show the mathematical steps to obtain the ratio distribution analytically, to compare it with the results of synthesis models. More details can be found in \citet{Hinkley1969}. 
Given two normal random variables  
$X \sim \mathcal{N}(\mu_x,\sigma_{x})$ and  $Y \sim \mathcal{N}(\mu_y,\sigma_{y})$, which are correlated $\rho \ne 0$, and  
assuming the joint density of $(x,y)$ is $g(x,y)$ a bivariate normal distribution.

The  probability distribution function of the ratio $Z=X/Y$ is obtained by marginalizing the joint distribution function $g(x,y)$, and replacing $x$ by $zy$ into the Eq.\eqref{Eq:Pz1}.

\begin{eqnarray}
		p_z(z) &=&  \int_{-\infty}^{\infty} |y| g(zy,y) \mathrm{d}y = 
		\label{Eq:Pz1}\\[1em]
		&=& \int_{-\infty}^{\infty} |y| \frac{1}{2\pi\sigma_x\sigma_y\sqrt{1-\rho^2}} 
		exp\left\{ \frac{-1}{2(1-\rho^2)} 
		\left(\frac{(zy-\mu_x)^2}{\sigma_x^2}-\frac{2\rho(zy-\mu_x)(y-\mu_y)}{\sigma_x\sigma_y}
		+\frac{(y-\mu_y)^2}{\sigma_y^2} \right)\right\} \mathrm{d}y  =
		\nonumber\\[1em]
		&=& \int_{-\infty}^{\infty} 
		\frac{|y|}{2\pi\sigma_x\sigma_y\sqrt{1-\rho^2}}  
		exp\left\{  \frac{-1}{2(1-\rho^2)} \left[
		y^2 \left(\frac{z^2}{\sigma_x^2} - \frac{2\rho z}{\sigma_x\sigma_y}+ \frac{1}{\sigma_y^2} \right)  
		- 2y \left(\frac{z\mu_x}{\sigma_x^2}-\frac{\rho(z\mu_y+\mu_x)}{\sigma_x\sigma_y}+\frac{\mu_y}{\sigma_y^2} \right) + 
		\left( \frac{\mu_x^2}{\sigma_x^2} - \frac{2\rho\mu_x\mu_y}{\sigma_x\sigma_y} + \frac{\mu_y^2}{\sigma_y^2} \right)
		\right]\right\} \mathrm{d}y
		\nonumber \\\nonumber 
\end{eqnarray}
Defining the following parameters \eqref{Eq:Params}, as a function of the ratio $z$ and the parameters of the bivariate normal distributions, 

\begin{align}
a(z) &= \,\,\, \sqrt{\frac{z^2}{\sigma_x^2} - \frac{2\rho z}{\sigma_x\sigma_y}+ \frac{1}{\sigma_y^2}}
& b(z) &= \,\,\, \frac{z\mu_x}{\sigma_x^2}-\frac{\rho(z\mu_y+\mu_x)}{\sigma_x\sigma_y}+\frac{\mu_y}{\sigma_y^2}
\nonumber\\[1em]
c \,\,\, &= \,\,\, \frac{\mu_x^2}{\sigma_x^2} - \frac{2\rho\mu_x\mu_y}{\sigma_x\sigma_y} + \frac{\mu_y^2}{\sigma_y^2}
& d(z) &= \,\,\, exp\left\{\frac{b^2(z)-ca^2(z)}{2(1-\rho^2)a^2(z)}\right\} 
\label{Eq:Params}
\\\nonumber
\end{align}
Eq.~\eqref{Eq:Pz1} is simplified into Eq.~\eqref{Eq:Pz2}

\begin{eqnarray}
		p_z(z) &=& 
        \frac{1}{2\pi\sigma_x\sigma_y\sqrt{1-\rho^2}} \int_{-\infty}^{\infty} |y| 
		exp\left\{ -\frac{1}{2} \left( \frac{y^2 a^2(z)}{(1-\rho^2)}  -2 \frac{y b(z)}{(1-\rho^2)} + \frac{c}{(1-\rho^2)}\right)\right\} \mathrm{d}y =  \nonumber\\[1em]
		&=&  \frac{1}{2\pi\sigma_x\sigma_y\sqrt{1-\rho^2}} \int_{-\infty}^{\infty} |y|  
		exp\left\{ -\frac{1}{2} \left(  
		\left( \frac{y a(z)}{\sqrt{1-\rho^2}} - \frac{b(z)}{a(z)\sqrt{1-\rho^2}} \right)^2+ 
		\left(\frac{c}{(1-\rho^2)} - \frac{b^2(z)}{(1-\rho^2)a^2(z)}  \right) 
		\right)\right\} \mathrm{d}y = 
		\nonumber\\[2em]
		&=&  \frac{d(z)}{2\pi\sigma_x\sigma_y\sqrt{1-\rho^2}}  
		\int_{-\infty}^{\infty} |y| exp\left\{ -\frac{1}{2} 
		\left( \frac{y a(z)}{\sqrt{1-\rho^2}} - \frac{b(z)}{a(z)\sqrt{1-\rho^2}} \right)^2
		\right\} \mathrm{d}y 
		\label{Eq:Pz2}
\end{eqnarray}

We separate the integral into the positive and negative ranges in Eq.~\eqref{Eq:Pz2}, and apply the change of variable  
$ u = {(y a^2(z)-b(z))}/{a(z)\sqrt{1-\rho^2}}$
        
\begin{eqnarray}
	p_z(z) &=&  
    	\frac{d(z)}{2\pi\sigma_x\sigma_y\sqrt{1-\rho^2}}  \left(
		\int_{-\infty}^{0} 
        -\mathbf{y} \, exp\left\{ -\frac{1}{2} 
		\left( \overbrace{\frac{y a^2(z)-b(z)}{a(z)\sqrt{1-\rho^2}}}^{u} \right)^2
		\right\} 
        \mathrm{d}y +
		\int_{0}^{\infty} \mathbf{y} \, exp\left\{ -\frac{1}{2} 
		\left( 
			\overbrace{\frac{y a^2(z)-b(z)}{a(z)\sqrt{1-\rho^2}}}^{u} 
		\right)^2 \right\} 
        \mathrm{d}y 
		\right)  
        \nonumber\\[1em]
        &=&  \frac{d(z)}{2\pi\sigma_x\sigma_y\sqrt{1-\rho^2}}  
        \left(  \ \ 
		\int_{-\infty}^{\frac{-b(z)}{a(z)\sqrt{1-\rho^2}}} 
		- \mathbf{\left(u + \frac{b(z)}{a(z)\sqrt{1-\rho^2}} \right) 
			\frac{\sqrt{1-\rho^2}}{a(z)}} \, exp\left\{-\frac{1}{2}u^2\right\} 
		\frac{\sqrt{1-\rho^2}}{a(z)} \mathrm{d}u \ \ \ \ + 
        \right. 
		\nonumber\\[1em]
		&& \qquad\qquad\qquad\qquad	+ \ \   \left. \int_{\frac{-b(z)}{a(z)\sqrt{1-\rho^2}}}^{\infty}  
		\mathbf{\left(u + \frac{b(z)}{a(z)\sqrt{1-\rho^2}} \right) 
			\frac{\sqrt{1-\rho^2}}{a(z)}} \, exp\left\{-\frac{1}{2}u^2\right\} 
		\frac{\sqrt{1-\rho^2}}{a(z)} \mathrm{d}u  \ \ 
		\right)  \ \ = 
		\nonumber\\[2em]
		&=&  \frac{d(z)\sqrt{1-\rho^2}}{2\pi\sigma_x\sigma_y a^2(z)}  \left(
		\int_{-\infty}^{\frac{-b(z)}{a(z)\sqrt{1-\rho^2}}} -u \, e^{-\frac{1}{2}u^2}  - 
		\frac{b(z)}{a(z)\sqrt{1-\rho^2}}\, e^{-\frac{1}{2}u^2} \mathrm{d}u \ \ +
		\ \ \int_{\frac{-b(z)}{a(z)\sqrt{1-\rho^2}}}^{\infty} u \, e^{-\frac{1}{2}u^2}  +
		\frac{b(z)}{a(z)\sqrt{1-\rho^2}}\, e^{-\frac{1}{2}u^2} \mathrm{d}u \right) = 
		\nonumber\\[2em]
		&=& \frac{d(z)\sqrt{1-\rho^2}}{2\pi\sigma_x\sigma_y a^2(z)}  
		\frac{b(z)}{a(z)\sqrt{1-\rho^2}}
		\left( \ \ 
		-  \int_{-\infty}^{\frac{-b(z)}{a(z)\sqrt{1-\rho^2}}} e^{-\frac{1}{2}u^2} 			\mathrm{d}u  \ \ \ \  + \ \ \ \ 
		 \int_{\frac{-b(z)}{a(z)\sqrt{1-\rho^2}}}^{\infty} e^{-\frac{1}{2}u^2} 			\mathrm{d}u \ \ 
		\right) \ \ \ \ + 
		\label{Eq:Pz3}\\[2em]
		&&  \qquad
		+ \ \  \frac{d(z)\sqrt{1-\rho^2}}{2\pi\sigma_x\sigma_y a^2(z)} 
		\left( \ \ 
		 \int_{-\infty}^{\frac{-b(z)}{a(z)\sqrt{1-\rho^2}}} -u \, e^{-\frac{1}{2}u^2} 		\mathrm{d}u
		\ \ + \ \ 
	    \int_{\frac{-b(z)}{a(z)\sqrt{1-\rho^2}}}^{\infty} u \, e^{-\frac{1}{2}u^2}   		\mathrm{d}u  \ \ 
		\right) =  
		\label{Eq:Pz4}
        \\ \nonumber
\end{eqnarray}

The integrals between the parenthesis in  Eq. \eqref{Eq:Pz3} are the  cumulative distribution function (CDF) of the standard normal distribution, $\Phi(x)$, by a factor of $\sqrt{2\pi}$.
And the two integrals in  Eq. \eqref{Eq:Pz4} are finite and immediate integrals.
They are simplified into Eq. \eqref{Eq:Pz5}, by means of the relation 
$-\Phi(-a) + \Phi(a)= \Phi(a)-1 + \Phi(a) = 2\Phi(a)-1$.

\begin{eqnarray}
	p_z(z) &=& 
    	\frac{b(z)d(z)}{\sqrt{2\pi}\sigma_x\sigma_y a^3(z)}  
		\left(  - \Phi\left(\frac{-b(z)}{a(z)\sqrt{1-\rho^2}}\right) +
		 \Phi\left(\frac{b(z)}{a(z)\sqrt{1-\rho^2}}\right)\right) \, +
		\frac{d(z)\sqrt{1-\rho^2}}{2\pi\sigma_x\sigma_y a^2(z)} 2  exp\left\{-			\frac{1}{2}\frac{b^2(z)}{(1-\rho^2)a^2(z)}\right\} \ \ = 
		\nonumber\\[2em]
		&=& \frac{b(z)d(z)}{\sqrt{2\pi}\sigma_x\sigma_y a^3(z)}  
		\left( 2\Phi\left(\frac{b(z)}{a(z)\sqrt{1-\rho^2}}\right) - 1 \right) \, +
		\frac{\sqrt{1-\rho^2}}{\pi\sigma_x\sigma_y a^2(z)} exp\left\{\frac{-c}{2(1-\rho^2)}\right\} 		\label{Eq:Pz5}
        \\ \nonumber
\end{eqnarray}

\begin{figure}
\centering
\includegraphics[width=0.85\textwidth]{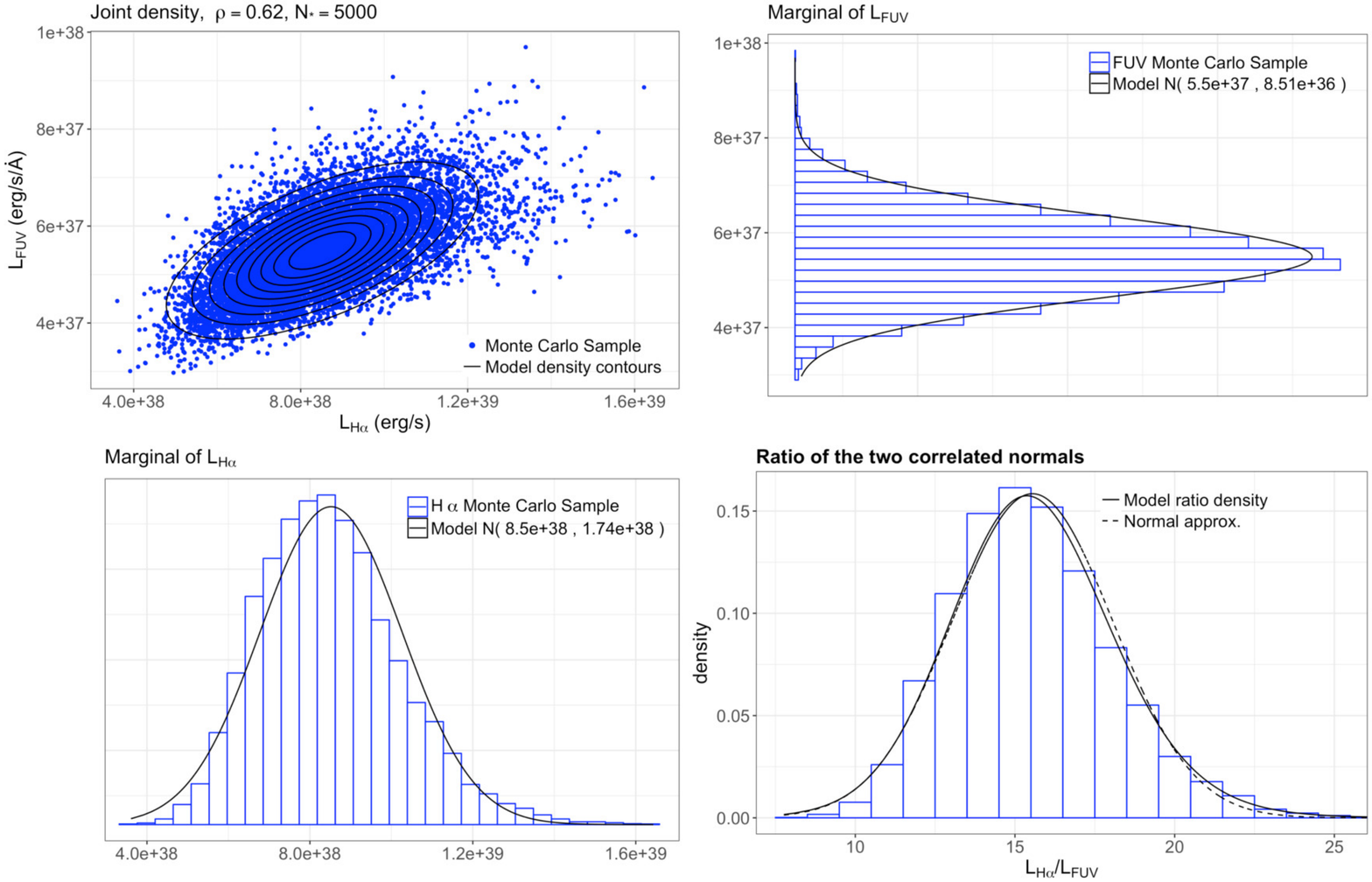}
\caption{Joint distribution of \Ha and FUV (top-left), histograms of the distribution of FUV (top-right) and \Ha (bottom-left), and the ratio distribution obtained from the simulations for an age of 2.8 Myr obtained from 10$^4$ Monte Carlo simulations with 5000 stars each. Over-plotted is a Gaussian distribution with the mean and variance obtained in Sect.~\ref{sec:SPM}. We also plot the analytic ratio distribution.}
\label{fig:FigA1}
\end{figure} 
\begin{figure}
\centering
\includegraphics[width=0.8\textwidth]{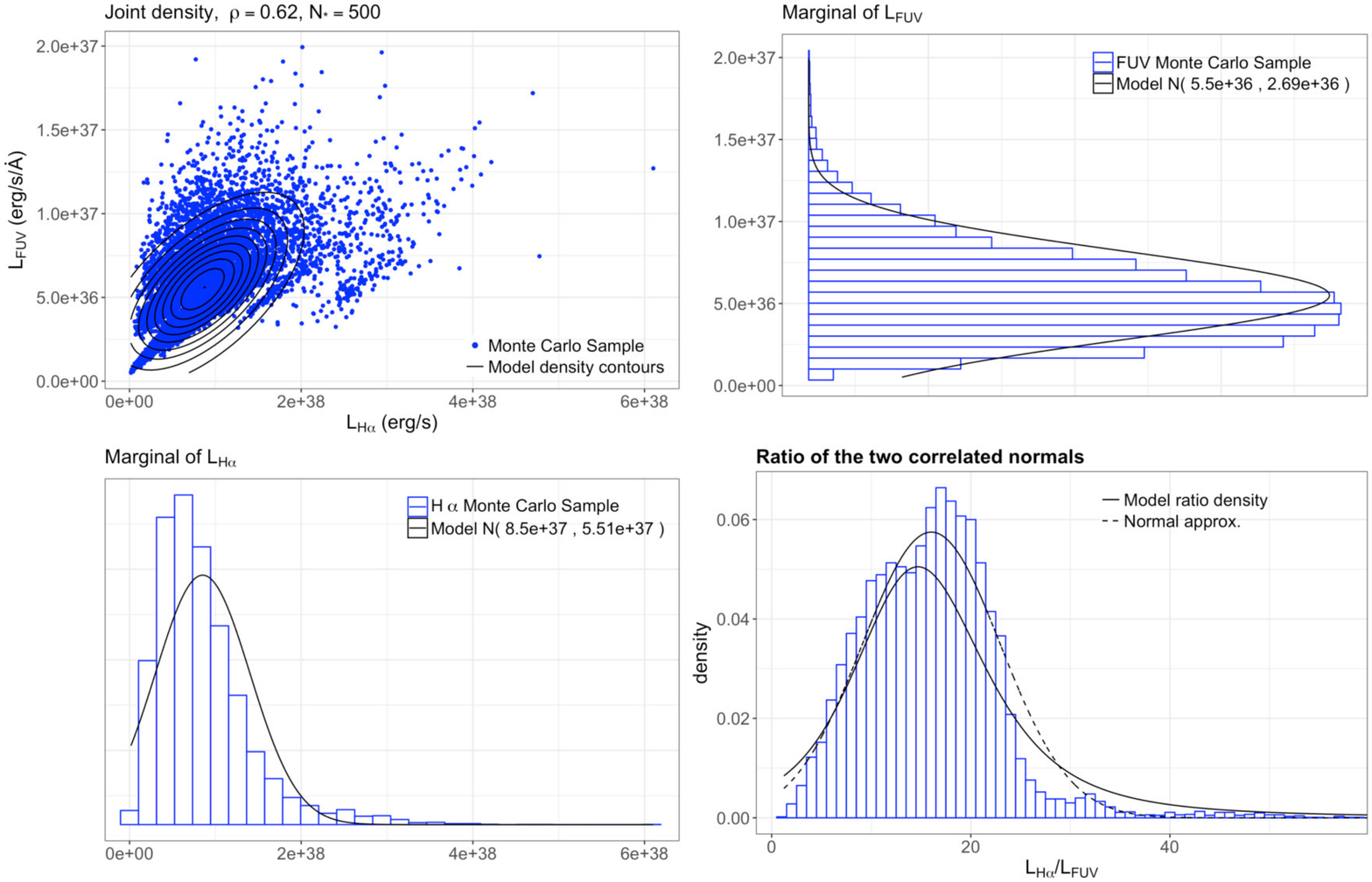}
\caption{As Fig.~\ref{fig:FigA1}, but for pixels with 500 stars.}
\label{fig:FigA2}
\end{figure} 

Once the analytical Eq.\eqref{Eq:Pz5} has been found, it is compared with the results of our synthesis models. To do so, we have computed simulations which assume the same IMF as used in Sect. \ref{sec:SPM} \citep{1955ApJ...121..161S} with mass limits of $1-120$ M$_\odot$, $Z=0.020$, and $f_\mathrm{esc} = 0$. 
We have computed two sets of $10^4$ Monte Carlo simulations for the IMF. The first set contains 5000 stars per simulation ( see Fig. \ref{fig:FigA1}) and the second one 500 stars (see Fig. \ref{fig:FigA2}). Once the IMFs are obtained we follow the evolution of each cluster. 
Figs. \ref{fig:FigA1} and \ref{fig:FigA2} show the joint distribution of \Ha\ and FUV luminosity (top-left), the histograms of the marginal distributions of FUV (top-right) and \Ha (bottom-left), and the \Ha/FUV ratio distribution obtained from the simulations for an age of 2.8 Myr (bottom-right). 
We note that this age corresponds to the worse values of skewness $\gamma_1$ and kurtosis $\gamma_2$, where more deviation of the Gaussian case is expected (c.f. Fig.~\ref{fig:Fig3}). 
The associated Gaussian distributions, with mean and variance obtained in Sect.~\ref{sec:SPM} for the corresponding number of stars in each set, are overplotted in the histograms as well. We also overplot a Gaussian approximation for the \Ha/FUV ratio with mean equals to the ratio of the mean values of \Ha\ and FUV, and its dispersion is calculated by standard error propagation analysis (see, e.g. appendix in \citealt{2002A&A...381...51C}).

Figure \ref{fig:FigA1}, with $5\times 10^3$ stars per simulation (similar to $1.5\times 10^4$ M$_\odot$ in the $1-120$ M$_\odot$ range), shows that the Gaussian approximation for the integrated luminosities and the \Ha/FUV ratio analytical  distribution are in good agreement for Monte Carlo simulations. 
This result is consistent with the associated values of $\gamma_1(H\alpha) = 0.55$  and $\gamma_2(H\alpha) =0.48$ for \Ha, and $\gamma_1(\mathrm{FUV})=0.26$ and $\gamma_2(\mathrm{FUV})=0.09$ for FUV, associated to this number of stars. 
FUV is approximated by a Gaussian due to its lower $\gamma_1$ and $\gamma_2$ values. 
We also note that for such number of stars, metallicity and age, the relative standard deviation is around a 20\%, which is about the maximum test value we have assumed in our analysis. 
We check that the ratio distribution is also similar to a Gaussian distribution with a relative standard deviation of a 16\%.

Figure~\ref{fig:FigA2} shows the latter analysis but for $5\times 10^2$ stars per simulation (similar to $1.5\times 10^3$ M$_\odot$ in the $1-120$ M$_\odot$ range). In this case, the mean value of $\log L(FUV) = 36.7$ is close to the LLL for this age. 
The luminosity distributions show an asymmetry with a high luminosity tail,  and skewness and kurtosis values, $\gamma_1(H\alpha) = 1.74$ and $\gamma_2(H\alpha) =4.80$, that make more appreciable the detach from gaussianity. 
The ratio analytical distribution also departs from the Monte Carlo simulations, and differences with the Gaussian case are also more relevant. 
However the global behavior of Monte Carlo distribution is still captured by the \Ha/FUV ratio distribution. 
A peculiar feature of the ratio distribution obtained from Monte Carlo simulations is the presence of three local maximums. 
This is an artifact due to the assignation of close atmosphere model in the computation of the isochrones and it is implicit in SB99 models, what produces an artificial discretization of colors (see a discussion in \citealt{2006A&A...451..475C}). 
Such effect is also visible in the joint distribution of \Ha\ and FUV which shows a discrete and well-aligned structure. 
In the case of lower number of stars (equivalent to lower masses or lower luminosities for a given age) the detach from a Gaussian distribution is larger, and a proper analysis would require to interpolate atmosphere models to avoid numerical artifacts in the Monte Carlo distributions.

Given that in our pixel-by-pixel analysis we find some pixel values below the LLL, what is the error in our age determinations due to sampling effects?
Notwithstanding, we can still evaluate qualitatively this effect by comparing the synthesis models results with the used isochrones,  which is a general technique for cases of extreme sampling (see, e.g. \citealt{2008A&A...484..711B}).

\begin{figure}
\centering
\includegraphics[width=0.9\textwidth]{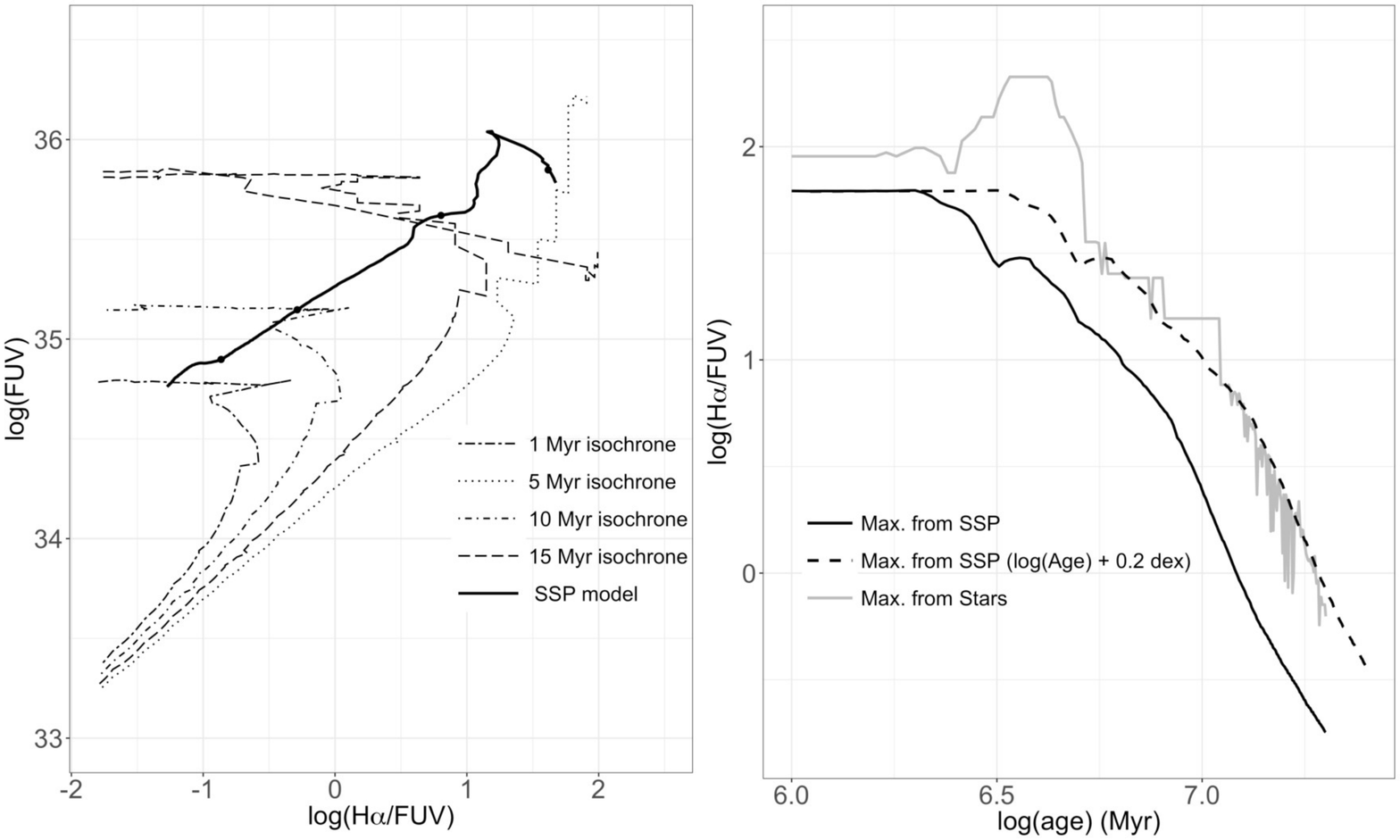}
\caption{Left: Comparison of the position of SB99 isochrones at 1, 5, 10 and 15 Myr with Z=0.020 and synthesis models results in the \Ha/FUV vs. FUV plane. Black symbols shows the synthesis models results at the quoted ages for the isochrones. The stairs-like behavior is due to the use of close atmosphere model approach in SB99 computations. Right: Maximum values of the  \Ha/FUV ratio from the full isochrone and SSP models set.}
\label{fig:FigA3}
\end{figure}

The general situation is shown in the left panel of Fig.{\ref{fig:FigA3}}, which shows isochrones for $Z=0.020$ and ages of 1, 5, 10, and 15 Myrs. 
Synthesis models computations are overplotted, the same ages are shown as black points, and the FUV luminosity has been multiplied by 100 for clarity. 
The isochrones exhibit a step-like behavior due to the close-atmosphere assignation approach, as described above. 
It can be observed that \Ha/FUV ratio is close to the main sequence turn-off defined by the isocrones at the corresponding ages. 
SSP results are at the left of the main sequence turn-off except during the WR phase (5 Myrs), where they are at the right of the turn-off. 
Such situation produces an asymmetry in ages inferences in the most extreme case of sampling effects (each pixel containing a single star).

In this situation, the ages inferred from the use of mean values of synthesis models and the \Ha/FUV ratio would be severely overestimated. This is because the \Ha/FUV ratio of stars at a given age would span towards low values. 
On the other hand, the same ages would only be moderately underestimated since the possible maximum \Ha/FUV ratio for a given age is similar to the maximum stellar value. I.e., the actual age would be much younger (lower) than the inferred one, but for sure it cannot be much older (larger) than our inference. 
Right panel in \ref{fig:FigA3} shows the comparison of the maximum \Ha/FUV ratio that a single star would reach as a function of the age. The maximum mean value of \Ha/FUV is obtained from synthesis models results considering all possible cases for metallicity and $f_\mathrm{scp}$ considered in this work. 
Except for ages lower than the WR phase age, the use of the mean value would obtained by standard SSP codes produce, in the worse case, an underestimate of $\log (\mathrm{Age})$ of 0.2 dex.

For a final evaluation of sampling effects with our methodology, we have computed sets of $10^4$ Monte Carlo simulations for each metallicity and the considered age range. Assuming the number of stars is in the range from $50$ to $5 \times 10^4$, following a power law with exponent -1 in order to increases the relevance of sampling effects and assuming too a flat distribution of $f_\mathrm{scp}$. 
The implicit distributions are similar to the those presented in Sect.~\ref{sec:HBM}, except by the inclusion of the distribution in the number of stars and the use of discrete values in the Z distribution. Despite there is still the problem of using close atmosphere model approach, this set of simulations allows us to obtain a first order comparison of the method  performance of Sect.~\ref{sec:HBM}. 

The method followed here is a na\"ive direct inversion of the problems from Monte Carlo simulations. This is, the resulting distributions are obtained from Monte Carlo simulations in the rectangular region defined by $H\alpha_\mathrm{obs} \pm \sigma_\mathrm{obs}(H\alpha)$  and $FUV_\mathrm{obs}\pm \sigma_\mathrm{obs}(FUV)$. 
We note that it is only a first order approximation (a correct analysis would require to weight the region by a 2D uncorrelated Gaussian distribution, and to interpolate atmosphere models, which is outside the scope of this paper).

\begin{figure}
\centering
\includegraphics[width=0.9\textwidth]{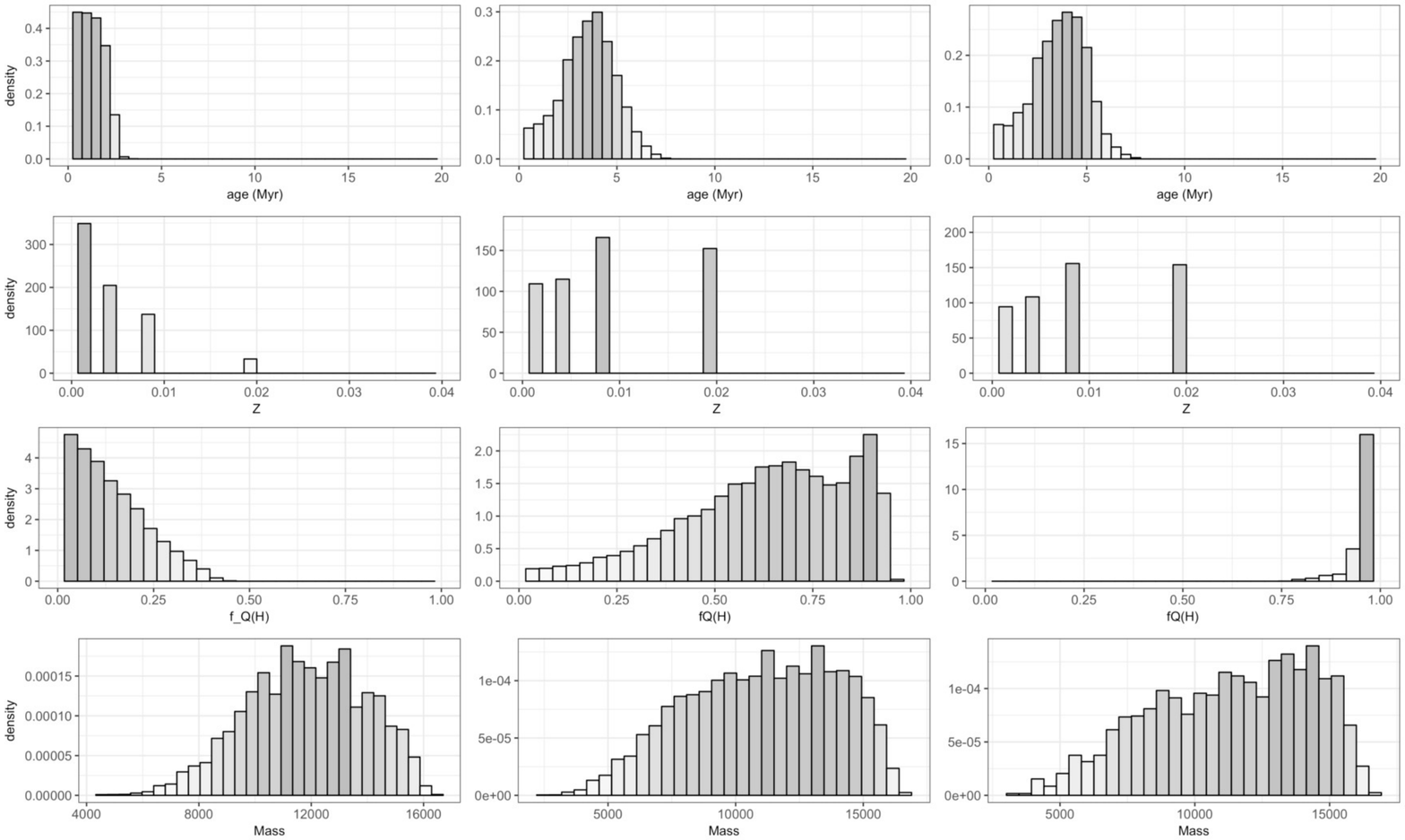}
\caption{Samples of the probability distributions of the  different
parameters obtained forma a direct inversion of Monte Carlo simulations, for some pixels of the image with different modal ages. 
These plots can be compared with the ones in Fig.~\ref{fig:Fig6}.}
\label{fig:FigA4}
\end{figure}

Fig.~\ref{fig:FigA4} displays the distribution of ages for the same points shown in Fig.~\ref{fig:Fig6}. We note that the age distribution is similar for large (young) and small (old) \Ha/FUV values, although Monte Carlo simulations show a fat tail that extends towards lower ages, and a narrow tail for larger ages. 
As explained above, this feature dues to the asymmetry between isochrones and synthesis models computations, as well as our assumption for the distribution of the number of stars which favors extreme sampling effects.
The situation is more extreme for the intermediate  \Ha/FUV values. For well sampled clusters there is an equilibrium between normal and WR stars, whereas for sub-sampled pixels normal stars (hence lower ages) are preferred.

%

\section{Nested sampling algorithm for sampling the marginal posterior distribution for the age parameter}\label{Sec:App2}

\begin{figure}
\centering
\begin{tikzpicture}[
    auto,
    node distance=2cm,
    semithick,
    bend angle=10,
    graybox/.style = {draw=gray!20, fill=gray!20, rounded corners},
    line/.style = {draw=black, thick},
    box/.style = {circle, draw=blue!50, fill=blue!20, minimum size=4mm}
    ]
 
    	\coordinate (Z1) at (-5cm, 1cm);
    	\coordinate (Z2) at (-5cm, 0cm);
    	\coordinate (Z3) at (-5cm, -1cm);
 	
	\coordinate (S1) at (-3cm, 1.5cm);
	\coordinate (S2) at (-3cm,  .5cm);
    	\coordinate (S3) at (-3cm,-1.5cm);

	\coordinate (M1) at (0cm, 2cm);
    	\coordinate (M2) at (0cm, 1cm);
    	\coordinate (M3) at (0cm, 0cm);
    	\coordinate (M4) at (0cm, -2cm);
 
    	\coordinate (T1) at (2cm, 0.75cm);
    	\coordinate (T2) at (2cm, -0.75cm);
 
    	\node (BBox) [graybox, minimum width=1cm, minimum height=4cm] at (-3cm, 0.cm) {};
    	\node [above] at (-3cm, 2.cm) {Layer 2};
     	\node [above] at (-3cm, 2.5cm) {(Hidden)};

	\node (BBox2) [graybox, minimum width=1cm, minimum height=5cm] at (0cm, 0.cm) {};
    	\node [above] at (0cm, 2.5cm) {Layer 3};
     	\node [above] at (0cm, 3.cm) {(Hidden)};

    	\node [above] at (-5cm, 2.cm) {Layer 1};
    	\node [above] at (-5cm, 2.5cm) {(Input)};
    	\node [above] at (2cm, 2.cm) {Layer 4};
    	\node [above] at (2cm, 2.5cm) {(Output)};
    \node (Zbox1)  [box] at (Z1)  {$\theta_1$};
    \node (Zbox2)  [box] at (Z2)  {$\theta_2$};
    \node (Zbox3)  [box] at (Z3)  {$\theta_3$};

    \node (Sbox1) [circle,draw=black!50] at (S1) {$x_{11}$};
    \node (Sbox2) [circle,draw=black!50] at (S2) {$x_{12}$};
    \node (Sbox3) [circle,draw=black!50] at (S3) {$x_{1n}$};

    \node (Mbox1) [circle,draw=black!50] at (M1) {$x_{21}$};
    \node (Mbox2) [circle,draw=black!50] at (M2) {$x_{22}$};
    \node (Mbox3) [circle,draw=black!50] at (M3) {$x_{23}$};
    \node (Mbox4) [circle,draw=black!50] at (M4) {$x_{2m}$};
     
    \node (Tbox1) [box] at (T1) {F$_{H\alpha}$};
    \node (Tbox2) [box] at (T2) {f$_{FUV}$};
     
    \fill [black] (0cm,  -.8cm) circle (1.2pt);
    \fill [black] (0cm,   -1cm) circle (1.2pt);
    \fill [black] (0cm, -1.2cm) circle (1.2pt);
 
    \fill [black] (-3cm, -.3cm) circle (1.2pt);
    \fill [black] (-3cm, -.5cm) circle (1.2pt);
    \fill [black] (-3cm, -.7cm) circle (1.2pt);

    	\draw[->,thin] (Zbox1) -- node [above] {} (Sbox1);
    	\draw[->,thin] (Zbox1) -- node [above] {} (Sbox2);
    	\draw[->,thin] (Zbox1) -- node [below] {} (Sbox3);

	\draw[->,thin] (Zbox2) -- node [above] {} (Sbox1);
    	\draw[->,thin] (Zbox2) -- node [above] {} (Sbox2);
    	\draw[->,thin] (Zbox2) -- node [below] {} (Sbox3);

	\draw[->,thin] (Zbox3) -- node [above] {} (Sbox1);
    	\draw[->,thin] (Zbox3) -- node [above] {} (Sbox2);
    	\draw[->,thin] (Zbox3) -- node [below] {} (Sbox3);

    \draw[->,very thin] (Sbox1) -- node [above] {} (Mbox1);
    \draw[->,very thin] (Sbox1) -- node [above] {} (Mbox2);
    \draw[->,very thin] (Sbox1) -- node [above, pos=.7] {} (Mbox3);
    \draw[->,very thin] (Sbox1) -- node [below, pos=.3] {} (Mbox4);

    \draw[->,very thin] (Sbox2) -- node [above] {} (Mbox1);
    \draw[->,very thin] (Sbox2) -- node [above] {} (Mbox2);
    \draw[->,very thin] (Sbox2) -- node [above, pos=.7] {} (Mbox3);
    \draw[->,very thin] (Sbox2) -- node [below, pos=.3] {} (Mbox4);

    \draw[->,very thin] (Sbox3) -- node [above] {} (Mbox1);
    \draw[->,very thin] (Sbox3) -- node [above] {} (Mbox2);
    \draw[->,very thin] (Sbox3) -- node [above, pos=.7] {} (Mbox3);
    \draw[->,very thin] (Sbox3) -- node [below, pos=.3] {} (Mbox4);

    	\draw[->,thin] (Mbox1) -- node [above] {} (Tbox1);
	\draw[->,thin] (Mbox1) -- node [above] {} (Tbox2);

    	\draw[->,thin] (Mbox2) -- node [below] {} (Tbox1);
    	\draw[->,thin] (Mbox2) -- node [below] {} (Tbox2);
    
	\draw[->,thin] (Mbox3) -- node [below] {} (Tbox1);
	\draw[->,thin] (Mbox3) -- node [below] {} (Tbox2);

	\draw[->,thin] (Mbox4) -- node [below] {} (Tbox1);
	\draw[->,thin] (Mbox4) -- node [below] {} (Tbox2);

   \end{tikzpicture}
\includegraphics[width=0.45\textwidth]{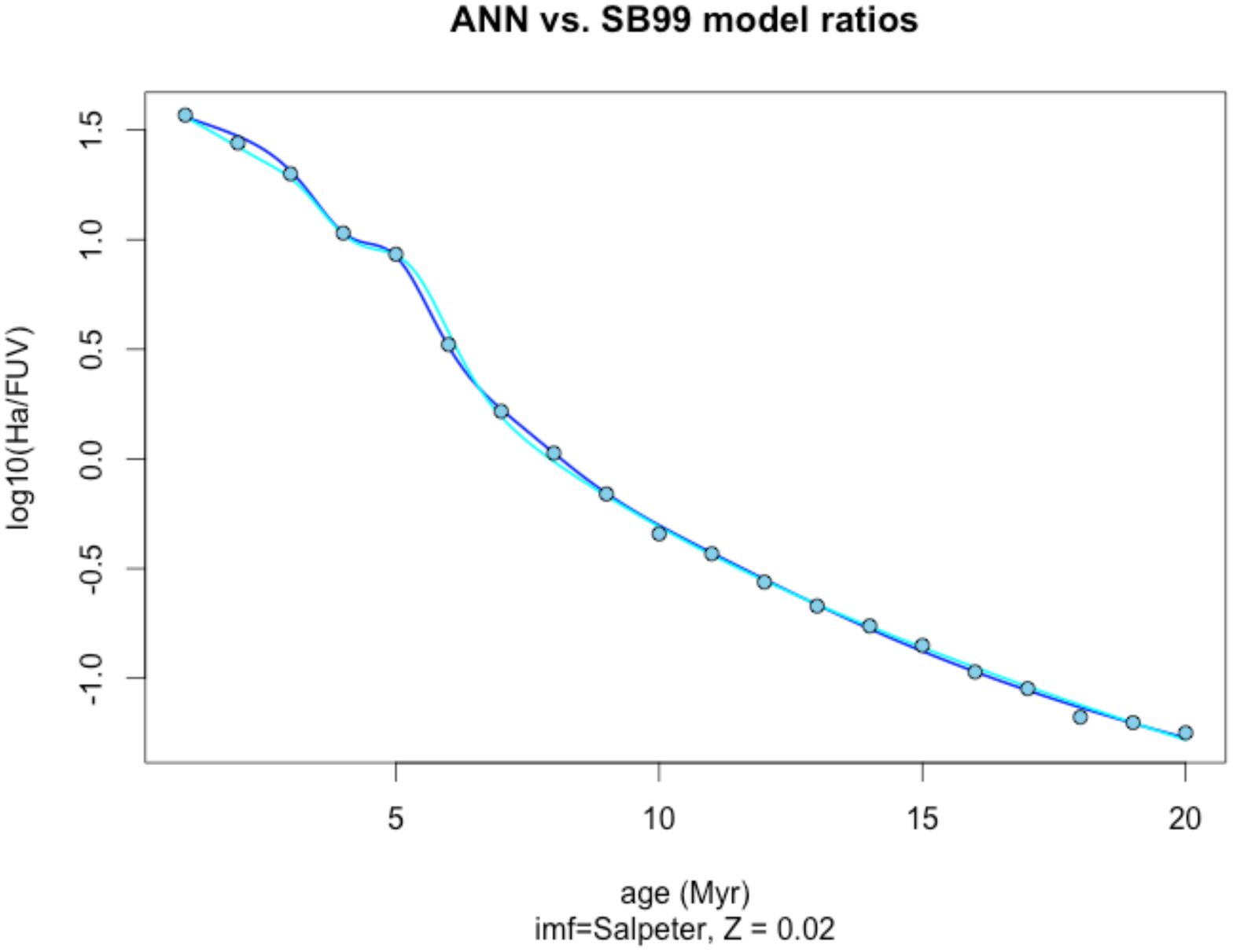}
\caption{
\textit{On the left:} Multilayer Perceptron for the estimation of the \Ha\ and FUV luminosities from the SED given by SB99 (output layer), depending on the input parameters $\bm{\theta} = (\theta_1,\theta_2,\theta_3) = (age,IMF,Z)$. The ANN was developed with two hidden layers with n=20 and m=50 nodes respectively, and it was trained with 70~\% and validate with 30~\% of the grid points. The uncertainties due this interpolation were consider negligible.
\textit{On the right:} Comparison of the ANN (interpolation for the \Ha\ to FUV flux ratio given different parameters than in SB99. Blue line corresponds to the ratios from the ANN given straightforward  the model ratios, whereas the cyan line when the  \Ha\ and FUV fluxes are calculated separately. They are in good agreement.} 
\label{fig:FigA5}
\end{figure}

\begin{figure}
\centering
\includegraphics[width=\textwidth]{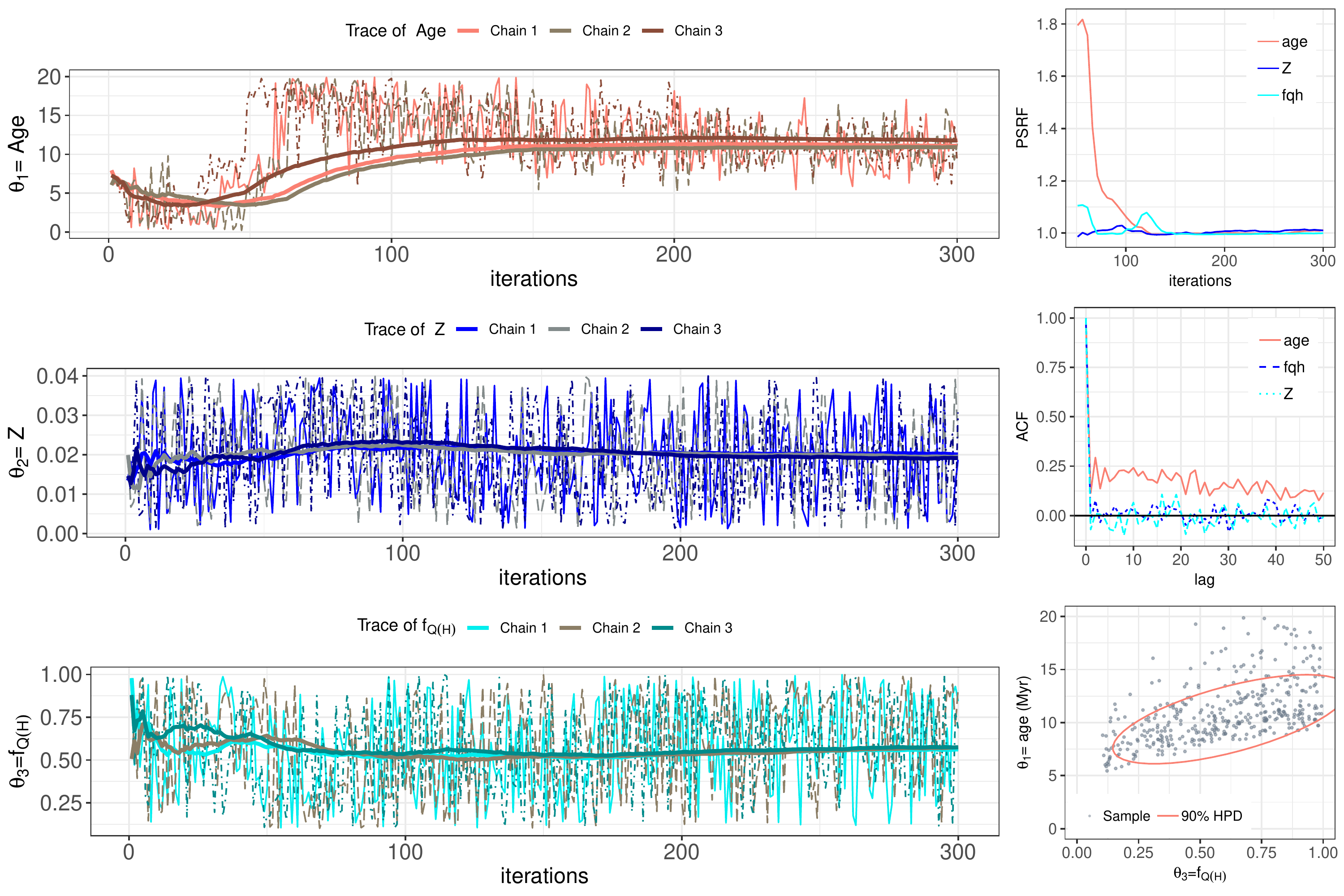}
\caption{
Assessing mixing and convergence. 
\textit{On the left:} Trends of the estimated parameters, for three different chains obtained from the NSA algorithm application. After the first 100 iterations, all the chains are mixed and stable. 
\textit{On the right:} On the top, the Potential Scale Reduction Factor (PSRF) for the three parameters, always lower than 1.1, indicates a fast convergence of the algorithm.  
At the middle a plot with the level of autocorrelation in each property, up to a maximum number of 50 lags.  
At the bottom, an example of the convergence towards the target distribution (red line indicates the $90\%$ Highest Posterior Density interval).
 After a burn-in period of 100 iterations, the chain rapidly converges towards a specific region of the space of parameters in regard of the age, according to the given \Ha\ to FUV flux ratio. Whereas it remains spread on the $f_{Q(H)}$, as it is expected and already shown in the bi-dimensional posteriors distributions  at Figure \ref{fig:Fig6b}.}
\label{fig:FigC1}
\end{figure}

In the Bayesian framework, inference proceeds by
estimating the posterior distribution  $p (\bm{\theta} | \hat{r})$ and then
marginalizing to obtain the age distribution of the region under study
according to Eq. \eqref{eq:marginalizationPost}
\begin{equation*}
p(\mbox{age} | \hat{r}) = p(\theta_1 | \hat{r}) = 
\iint p(\bm{\theta} | \hat{r}) \, d \theta_2 d \theta_3
\end{equation*}
where  the posterior distribution $p(\bm{\theta}|\hat{r})$ was characterized
by an independent and identical distributed sample obtained by using the
Nested  Sampling Algorithm~\citep[NSA;][]{skilling2006}. 

The main objective of
NSA is the estimation of the \emph{evidence}, $p(\hat{r})$, but as a by-product
we can obtain an independent sample of the posterior distribution. 
The NSA is based on the relationship between the likelihood 
$L(\bm{\theta})=p(\hat{r}|\bm{\theta})$ and the \emph{prior volume} 
$X(\lambda)$ defined as the bulk of prior distribution contained within an
iso-contour of the likelihood:
\begin{equation}\label{eq:priorvolume}
X(\lambda) = \int_{p(\hat{r}|\bm{\theta}) > \lambda} p(\bm{\theta}) \, 
d\bm{\theta}.
\end{equation}

As $\lambda$ increases $X(\lambda)$ decreases from $X(0)=1$ to $X(\infty)=0$.
See \citet{skilling2006}, their Figs. 4 and 5 as a perfect example of the nested sampling strategy and how it works. 
The likelihood and the evidence are related by 
\begin{equation}\label{eq:evidence}
p(\hat{r}) = \int p(\hat{r} | \bm{\theta}) dX	
\end{equation}

The key of the NSA is to evaluate numerically integral \eqref{eq:evidence} using 
directly the prior volume $X$, instead of rastering over $\bm{\theta}$, which
becomes impractical for high dimensions ( we only deal with 3 parameters).
For any size of $\bm\theta$, the evaluation of the integral in \eqref{eq:evidence} becomes a one-dimensional integral over the unit range
\begin{equation}\label{eq:1D_Z}
p(\hat{r}) = \int_{0}^1 L(X) \, dX
\end{equation}
being $L(X)$ the inverse function of $X(\lambda)$, and 
$dX=p(\bm{\theta})d\bm{\theta}$. 
The numerical  evaluation of \eqref{eq:1D_Z}
implies dividing the prior $X$ into tiny elements and sorting them by
likelihood volume
\begin{equation*}
0< X_m < \ldots < X_2 < X_1 < 1
\end{equation*}

Thus the integral \eqref{eq:1D_Z} can be estimated as a weighted sum of the
corresponding likelihoods:  
\begin{equation}\label{eq:Zaprox}
p(\hat{r}) \simeq \sum_{i=1}^m L(X_i) w_i
\end{equation}
with $w_i=(X_{i-1}-X_{i+1})/2$.

The resulting sequence of parameters $\bm\theta^1, \bm\theta^2, \ldots,\bm\theta^N$,
for a run with $N$ objects, already allows posterior samples to be extracted
from the evidence calculation as
\begin{equation}
	p(\bm\theta^i | \hat{r}) = \frac{L_i e^{-i/N}}{p(\hat{r})} 
\end{equation}
Thorough explanation and details of the NSA can be found at ~\cite{skilling2006}. 
Algorithm~\ref{Algo:NSA} sketches the NSA applied to each pixel of the observed flux ratio image 
$\hat{r} = \hat{F}_{H{\alpha}} / \hat{F}_{FUV}$. 
The full codes implemented in R can be found in \url{https://github.com/carmensg/Age-maps}. 
{Figure \ref{fig:FigC1} shows several tests for assessing mixing and convergence of the chains.}

\begin{algorithm}
 \caption{Nested Sampling Algorithm}
 \KwData{the observed flux ratio, $\hat{r} = \hat{F}_{H{\alpha}} / \hat{F}_{FUV},$} of a certain pixel of the ratio image. \\
 \KwResult{evidence $p(\hat{r})$ calculation, and the nested sequence of params. $\bm \theta_1,\ldots,\bm \theta_N$}
 initialization\;
 $x_{0} \leftarrow 1$ \\
 $z_{0} \leftarrow 0$ \Comment{ $z$ is the evidence, i.e. $p(\hat{r})$} \\ 
 $j \leftarrow 0$  \\
 \Repeat{$L_{max}x_j < fz_j$ \& $j<M$}{  
 	Draw $N$ objects $\bm \theta_1^{(j)},\ldots,\bm \theta_N^{(j)}$ from prior \eqref{Eq:priors}\\
 	\For{$i=1,\ldots,N$}{
		Use $\bm \theta_i^{(j)}$ to generate the model params.
		$$\{ F_{H\alpha}, f_{FUV}, \sigma_{H \alpha},\sigma_{FUV}, \rho \}_i$$
		with SB99$+$ANN, Eqs. \eqref{eq:derivedfluxes} to \eqref{eq:sigma} \\
		Compute the likelihood $L_i^{(j)}\leftarrow$ Eq.~\eqref{eq:fddR} \\
	}
	$ j \leftarrow j+1$ \\
	Record $L_{j} = \min_{1\leq i \leq N} \{ L_i^{(j-1)}\}$ \\
	$x_j \leftarrow e^{-j/N}$ \Comment{\it crude estimation} \\
	$w_j \leftarrow \frac{1}{2}\left(e^{-(j-1)/N}-e^{-(j+1)/N} \right)$ \\
	$z_j \leftarrow z_{j-1} + L_j w_j$ \Comment{approx. by Eq.~\eqref{eq:Zaprox}}
 }\Comment{terminate the main loop when the largest current likelihood would not increase the current evidence by more than small fraction $f$} \\
 \label{Algo:NSA}
\end{algorithm}

\section{Image Segmentation}\label{Sec:App1}

\begin{algorithm}
 \KwData{$y$ observed flux ratio image, $G$ categories}
 \KwResult{sampling the posterior distr. of $x$ over the full conditional distr. of the model parameters $\{ \bm \mu, \sigma,\beta \}$ }
 initialization\;
 $\bm x, \bm \mu^{(0)}, \sigma^{2(0)} \leftarrow 0$, $\beta^{(0)} \leftarrow 1$ \\ 
  $\{xcum_{ig}\}_{|\mathcal{I}|\times G} \leftarrow 0$ 
  for estimation of the $\widehat{x}_{i}^{MPM} $ \eqref{Eq:MPM}\\ 
 \For{$t=1,\ldots,niter$}{
 	$S(x) \leftarrow 0$ \\
 	\For{$i=1,\ldots,\left|\mathcal{I}\right|$}{
		\For{$g=1,\ldots,G$}{
  			$n_{i,g}= \sum_{j \sim i}{\mathbb{I}_{x_{i}=g}}$ 
			store num. of neighbours of the class $g$\\
		}
		Draw $x_i | \bm y, \bm \mu^{(t-1)},\sigma^{2(t-1)},\beta^{(t-1)}$ 
		from post. d. \eqref{Eq:postpott}\\
		$xcum_{i,x_i} \leftarrow xcum_{i,x_i}+1$\\
		$S(x) \leftarrow S(x) + n_{i,g}$
	}
	\For{$g=1,\ldots,G$}{
  		Draw $\mu_{g}^{(t)}$ from truncated normal \eqref{Eq:postmu}
		$\mathcal{N}\left( \frac{\sum_{i\in\mathcal{I}}{\mathbb{I}_{x_{i}=g}y_{i}}}{\sum_{i\in\mathcal{I}}{\mathbb{I}_{x_{i}=g}}}
		,\frac{\sigma^{2(t-1)}}{\sum_{i\in\mathcal{I}}{\mathbb{I}_{x_{i}=g}}}\right)
		\mathbb{I}_{\left[\mu_{g-1}^{(t)},\mu_{g+1}^{(t-1)}\right]}$\\
	}
	Draw $\sigma^{2(t)}$ from $Inv-\Gamma \left(\nicefrac{\left|\mathcal{I}\right|^{2}}{2},\nicefrac{\sum\limits_{i\in\mathcal{I}}{(y_{i}-\mu_{x_{i}})^{2}}}{2}\right)$ \eqref{Eq:postsigma} \\
	Draw $\beta^{(t)}$ from post. distr. \eqref{Eq:postbeta} with MCMC sampler: \\
	\quad Draw $\tilde{\beta}$ from the uniform proposal: \\
	\qquad $\mathcal{U} \left( \beta^{(t-1)} -h,\beta^{(t-1)}+h \right)$ \\
	\quad Compute the acceptance ratio: \\
	$\rho = \left\{ 
	\qquad \displaystyle \frac{Z(\beta^{(t-1)})}{Z(\tilde{\beta})}
	exp \left( (\tilde{\beta} - \beta^{(t-1)} )S(x) \right) 
	\right\} 
	\wedge 1$ \\
	\quad Set $\beta^{(t)} = \tilde{\beta}$ with probability $\rho$
 }
 \caption{Hybrid Gibbs Algorithm for the Image segmentation}
 \label{Algo:Gibbs}
\end{algorithm}

We describe the fully Bayesian approach for the model presented at Section \ref{sec:ImgSeg}. 
Where the aim is to cluster pixels into homogeneous classes, without prior definition of those classes, based only on spatial coherence.
Suppose that the ``true'' image $\bm x$ can only take a discrete set of values 
$\mu_1,\ldots,\mu_G$, and the assumed prior on  $\bm x$ is a Potts model with $G$ categories, Eq. \eqref{Eq:Potts}. The variables are considered as random bidimensional arrays  $\bm x = \{ x_i, i\in \mathcal{I}\}$,
whose elements are indexed by the lattice $\mathcal{I}$ (the location of the pixels), and related through a neighborhood relation. 

The observed image $\bm y $ is a degraded version of $\bm x$ by additive Gaussian noise,
where $y_i$ are conditional independent given $\bm x$. So the distribution of $\bm y $ is given by Eq. \eqref{Eq:noisyimage}.

Using Bayes theorem we know how to construct the posterior
distribution of $\bm x|\bm y$. The aim is to draw inferences about $\bm x$ based on the posterior $p(\bm x | \bm y)$.

We treat the parameters $\beta, \sigma,\bm \mu$ as variables, this is as hyper parameters. 
As they are nuisance parameters, and there is no additional prior information, the chosen priors are the following uniforms 
\begin{eqnarray*}
\beta &\sim& \mathcal{U}(0,2) \\
\mu_g &\sim& \mathcal{U}(y_{min},y_{max}) \\
p(\sigma^2) &\propto& \sigma^{-2}
\end{eqnarray*}
as described in Section \ref{sec:ImgSeg}.
 
Assuming independent priors for these parameters, the joint posterior distribution is therefore
\begin{eqnarray}
p(\bm x,\beta,\sigma^{2},\bm\mu | \bm y) &\propto& 
p(\bm x,\beta,\sigma^{2},\bm\mu)  \  p(\bm y |\bm x,\beta,\sigma^{2},\bm\mu) 
\nonumber\\[1em]
&\propto& 
p(\beta) \ p(\sigma^{2}) \ p(\bm\mu) \  p(\bm x|\beta) \ p(\bm y |\bm x,\beta,\sigma^{2},\bm\mu)
\nonumber\\[1em]
&\propto& 
\displaystyle\frac{(\sigma^{2})^{-(|\mathcal{I}|/2+1)}}{Z(\beta)}
\ \ \times \ \ 
exp\left\{ \beta\sum\limits_{j\sim i}{\mathbb{I}_{x_{j}=x_{i}}} -
\displaystyle\frac{1}{2\sigma^{2}}\sum\limits_{i \in \mathcal{I}}(y_{i}-\mu_{x_{i}})^{2}
\right\}
\end{eqnarray}

For sampling purposes (implemented in the hybrid Gibbs  Algorithm~\ref{Algo:Gibbs}), 
it is necessary the full conditionals distributions of the parameters 
\citep[see ][for more information]{10.1007/978-1-4614-8687-9}.
The conditional distribution for $\bm x$ is given in Eq. \ref{Eq:postpott},
%
$
P(x_{i}=g|\bm y,\beta,\sigma^{2},\bm \mu) \propto 
exp \left\{ 
\beta \sum_{j\sim i} \mathbb{I}_{x_j = g}
- \frac{1}{2\sigma^2}(y_i - {\mu_g})^2
\right\}
$.

Given $\bm x$ the pixels associated to each category $g$ can be separated, 
so the parameters $\mu_g$ can be simulated indepentdently as 
\begin{eqnarray}
P(\mu_{g}|\bm y,\bm x,\sigma^{2}) &\propto& 
exp \left\{
-\displaystyle\frac{1}{2\sigma^{2}}
\sum\limits_{i : x_i=g}(y_{i}-\mu_{g})^{2}
\right\}
\nonumber\\[1em]
&\propto&  exp \left\{ 
-\displaystyle\frac{n_g}{2\sigma^{2}}
\left(\mu_{g}-\frac{s_g}{n_g}\right)^{2}
\right\} 
\label{Eq:postmu}
\\ \nonumber
\end{eqnarray}
a truncated normal on $\left[y_{min},y_{max}\right]$, 
or $\left[\mu_{g-1},\mu_{g+1}\right]$ if we establish the ordering $y_{min}\leq \mu_1 \leq \ldots \leq \mu_G \leq y_{max}$ (setting $\mu_0=y_{min}$ and $\mu_{G+1}=y_{max}$). 
Where $n_{g}=\sum_{i\in\mathcal{I}}{\mathbb{I}_{x_{i}=g}}$
and $s_{g}=\sum_{i\in\mathcal{I}}{\mathbb{I}_{x_{i}=g}y_{i}}$. 

The conditional distribution of $\sigma^2$ is 
\begin{eqnarray}
P(\sigma^{2}|\bm y,\bm x) 
&\propto&
(\sigma^{2})^{-(|\mathcal{I}|/2+1)}
exp \left\{
\displaystyle\frac{-1}{2\sigma^{2}}\sum\limits_{i \in \mathcal{I}}(y_{i}-\mu_{x_{i}})^{2}
\right\}
\label{Eq:postsigma}
\end{eqnarray}
an inverse gamma distribution with parameters 
$\nicefrac{\left|\mathcal{I}\right|^{2}}{2}$ and $\nicefrac{\sum\limits_{i\in\mathcal{I}}{(y_{i}-\mu_{x_{i}})^{2}}}{2}$. 

And finally, the conditional distribution of $\beta$ is 
\begin{eqnarray}
p(\beta|\bm x)
&\propto&
\displaystyle\frac{1}{Z(\beta)}exp\left(\beta\sum\limits_{j\sim i}{\mathbb{I}_{x_{j}=x_{i}}}\right)
\mathbb{I}_{\left[0,\beta_{crit}\right]}
\label{Eq:postbeta}
\end{eqnarray}
since $\beta$ only depends on $\bm x$. Together the difficulty of the calculation of the normalising coonstant $Z(\beta)$, we have this is no for longer a Potts model, or some known distribution function. 
We follow the \textit{path sampling} methodology to estimate $Z(\beta)$, described in  \citet{10.1007/978-1-4614-8687-9}, in order to implement the MCMC sampler at the end of the Algorithm \ref{Algo:Gibbs}. 

To give an estimation of the corresponding image $\bm x$ of the flux ratio, given the observed one $\bm y$, classified by $G$ homogeneous regions with a common flux ratio mean $\mu_g$ we use the associated  marginal posterior mode (MPM) estimator. 
\begin{eqnarray}
\widehat{x}_{i}^{MPM} &=&
arg max_{1\leq g\leq G}\mathbb{P}^{\pi}(x_{i}=g|y), i\in\mathcal{I} 
\nonumber \\
&\simeq&  \max_{1\leq g\leq G} \sum_{n=1}^{N} {\mathbb{I}_{x_{i}^{(n)}=g}}
\label{Eq:MPM}
\end{eqnarray}
The later equation gives an approximation of the MPM based on a simulation $\{ \bm x^{(n)} \}_{n=1,\ldots,N}$ from the posterior distribution of $\bm x$, \citet{10.1007/978-1-4614-8687-9}. 

All the codes implemented in R can be found in \url{https://github.com/carmensg/Age-maps}.

\section{Age maps for the galaxies from Paper I}\label{Sec:App3}

\begin{figure}
\begin{center}
\includegraphics[width=0.85\textwidth]{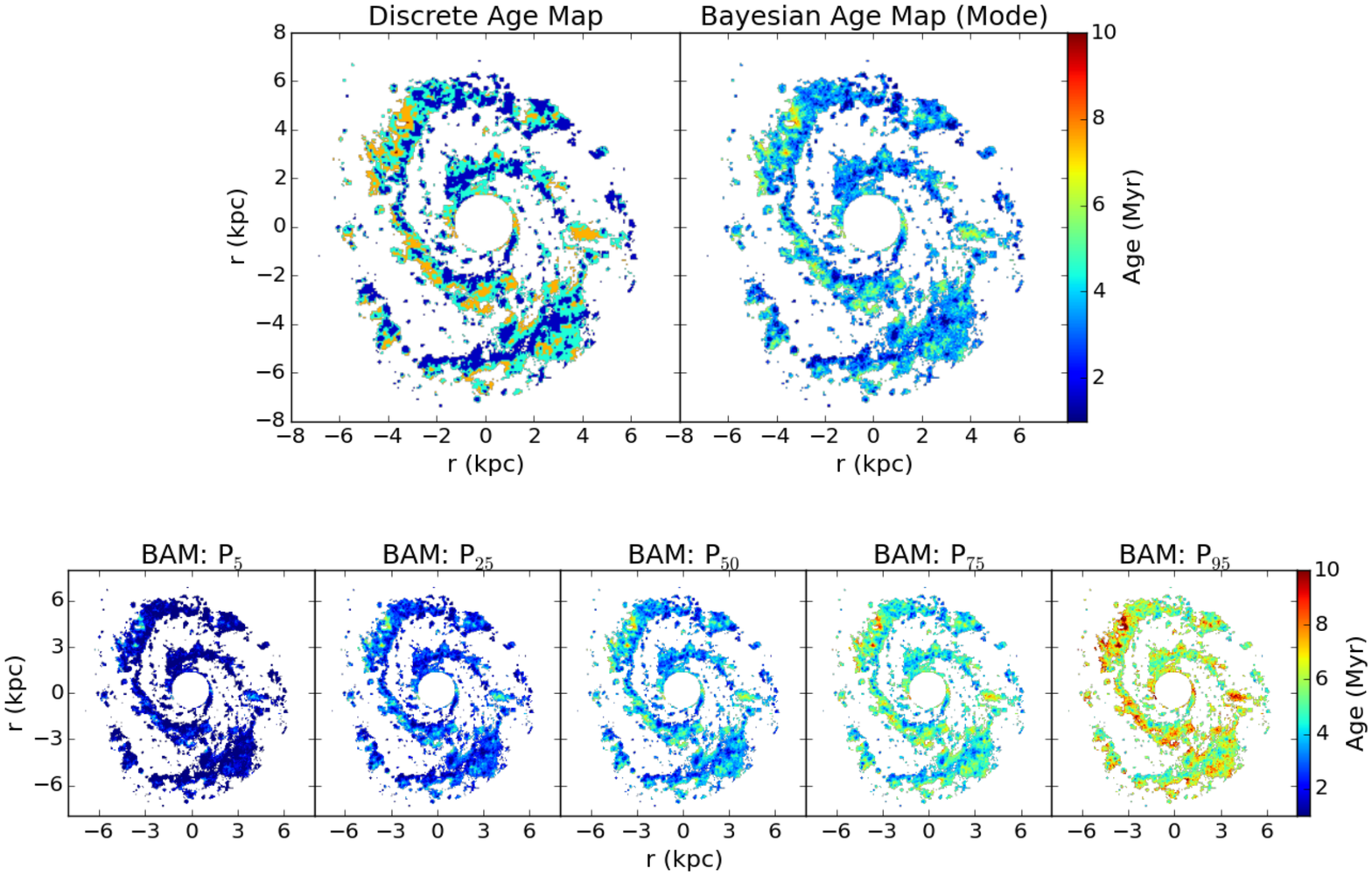}
\end{center}
\caption{M51 age maps. 
\textit{Top left}: The discrete age map obtained in Paper I.
\textit{Top right}: Bayesian age map, calculated by method of Sec. \ref{sec:HBM}.
\textit{Bottom}: Different Bayesian age maps (BAM), calculated in Sec. \ref{sec:HBM}, but taken this time 5\%, 25\%, 50\%, 75\% and 95\% percentiles from the age posterior probability distribution. 
Axis are the distance to the galactic center (kpc units). }
\label{fig:FigD1}
 \end{figure}

\begin{figure}
\centering
\includegraphics[width=0.8\textwidth]{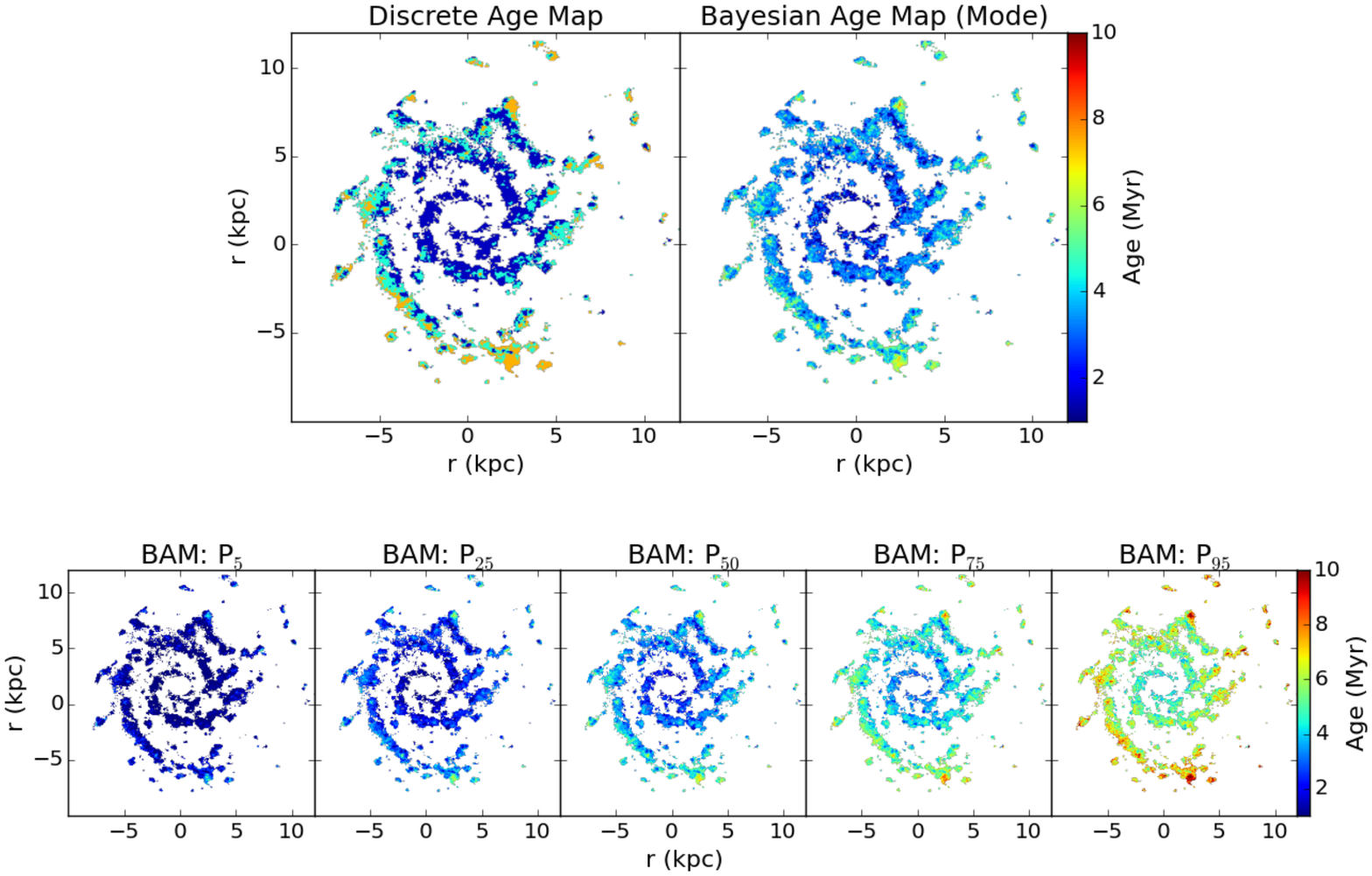}	
\caption{Same as Fig.\ref{fig:FigD1}, but for M74}
\label{fig:FigD2}
 \end{figure}

\begin{figure}
\centering
\includegraphics[width=0.85\textwidth]{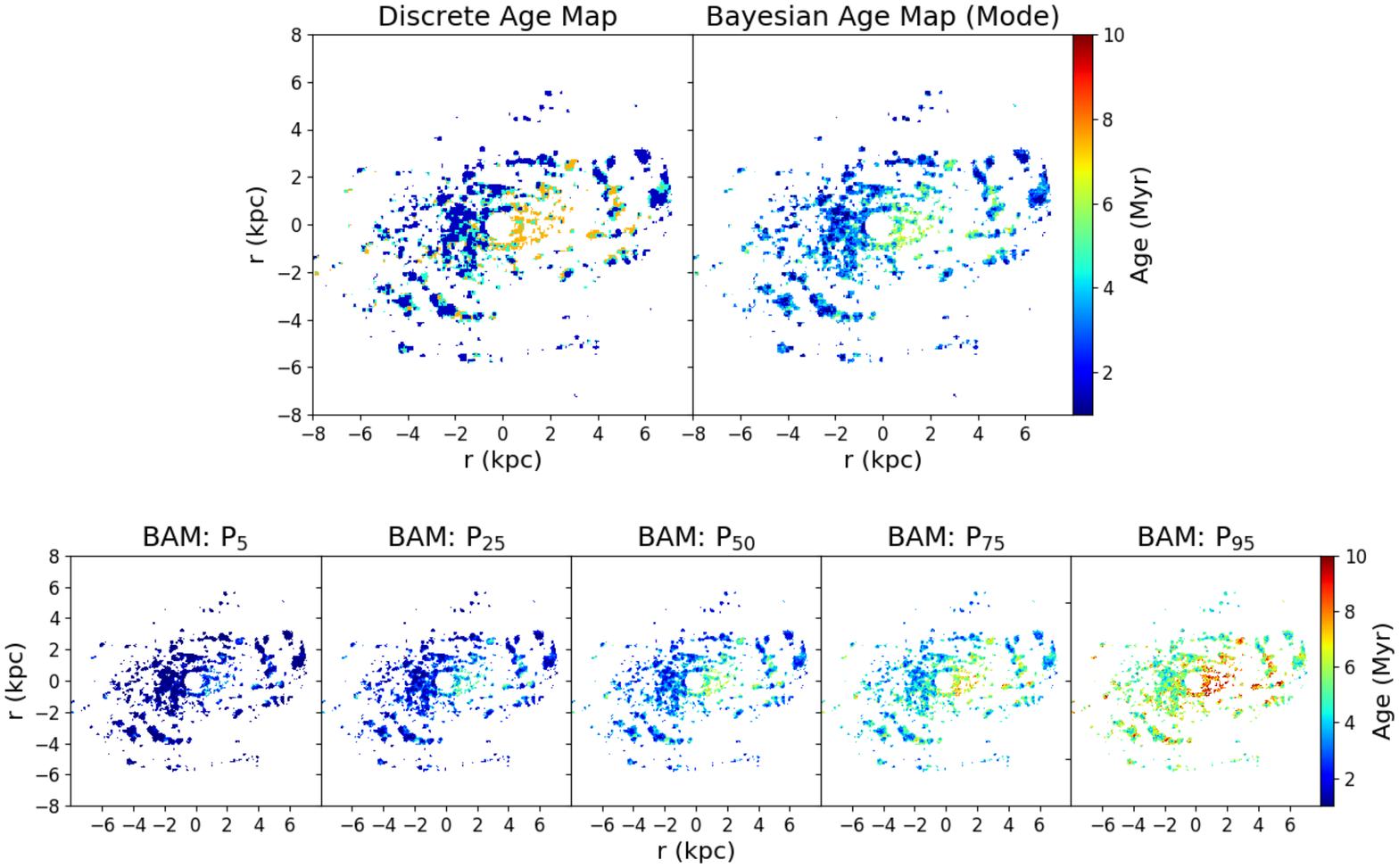}	
\caption{Same as Fig.\ref{fig:FigD1}, but for M63}
\label{fig:FigD3}
 \end{figure}

\begin{figure}
\centering
\includegraphics[width=0.85\textwidth]{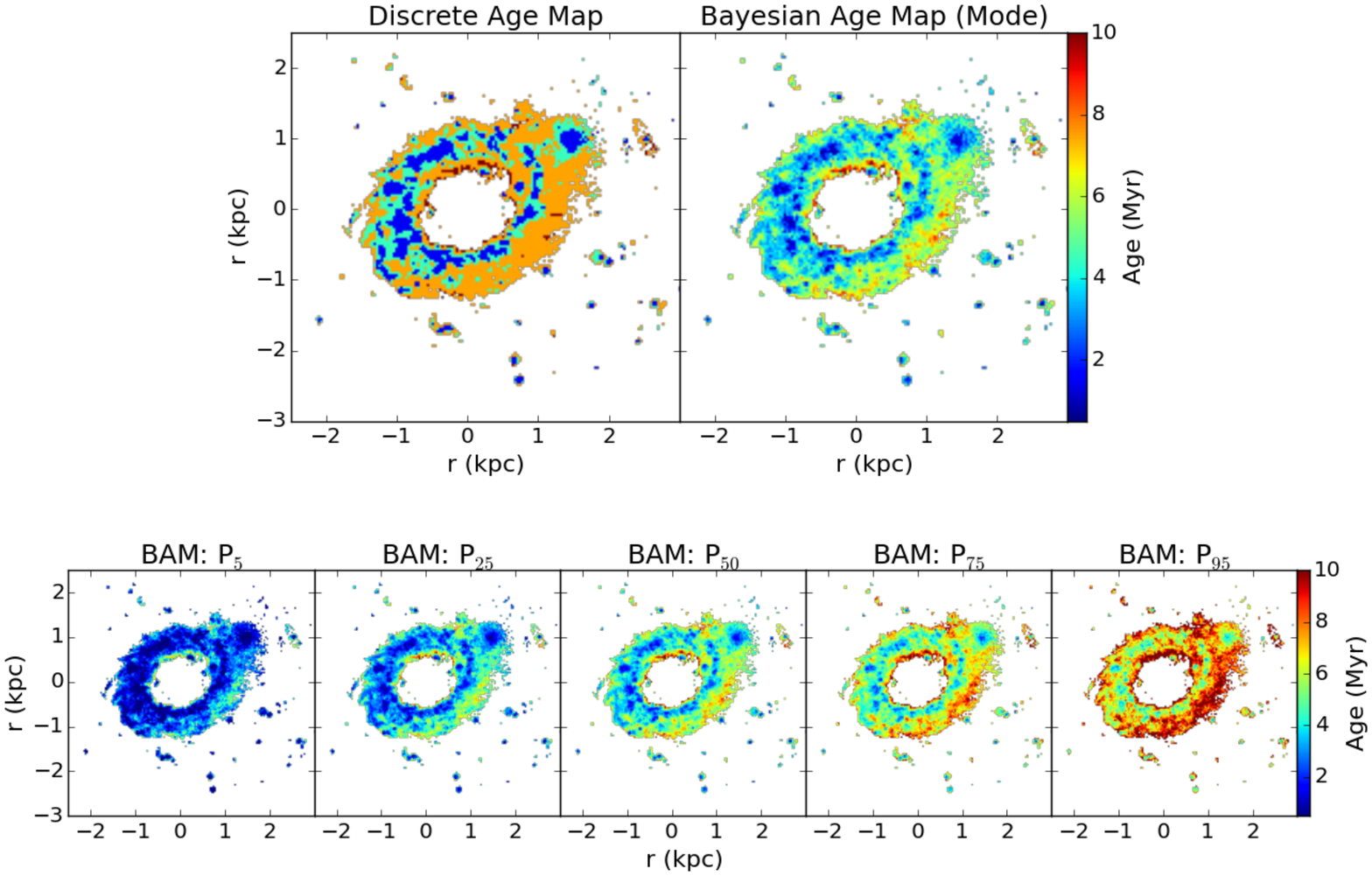}	
\caption{Same as Fig.\ref{fig:FigD1}, but for M94}
\label{fig:FigD4}
 \end{figure}

\begin{figure}
\centering
\includegraphics[width=0.85\textwidth]{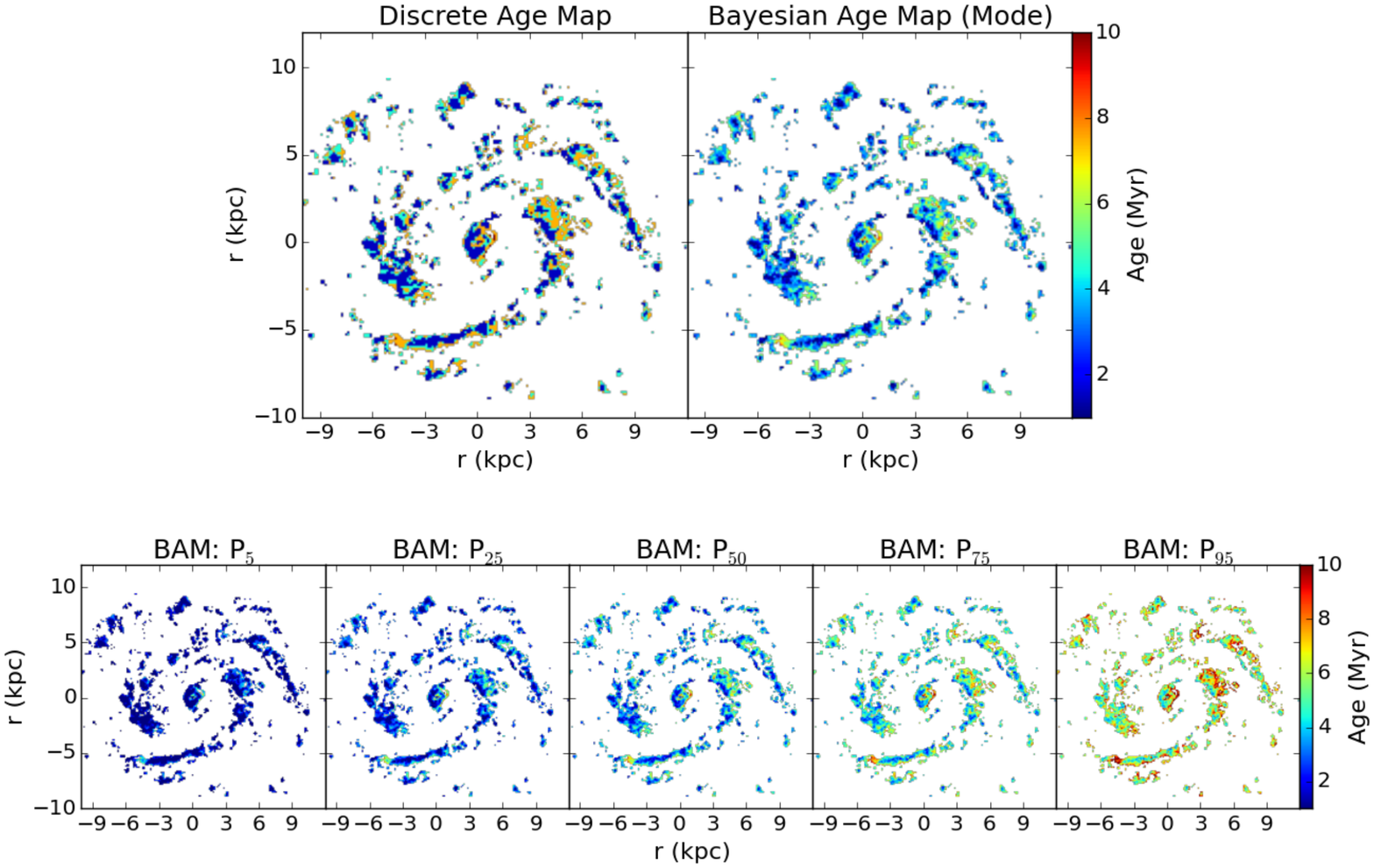}	
\caption{Same as Fig.\ref{fig:FigD1}, but for M100}
\label{fig:FigD5}
\end{figure}

\begin{figure}
\centering
\includegraphics[width=0.85\textwidth]{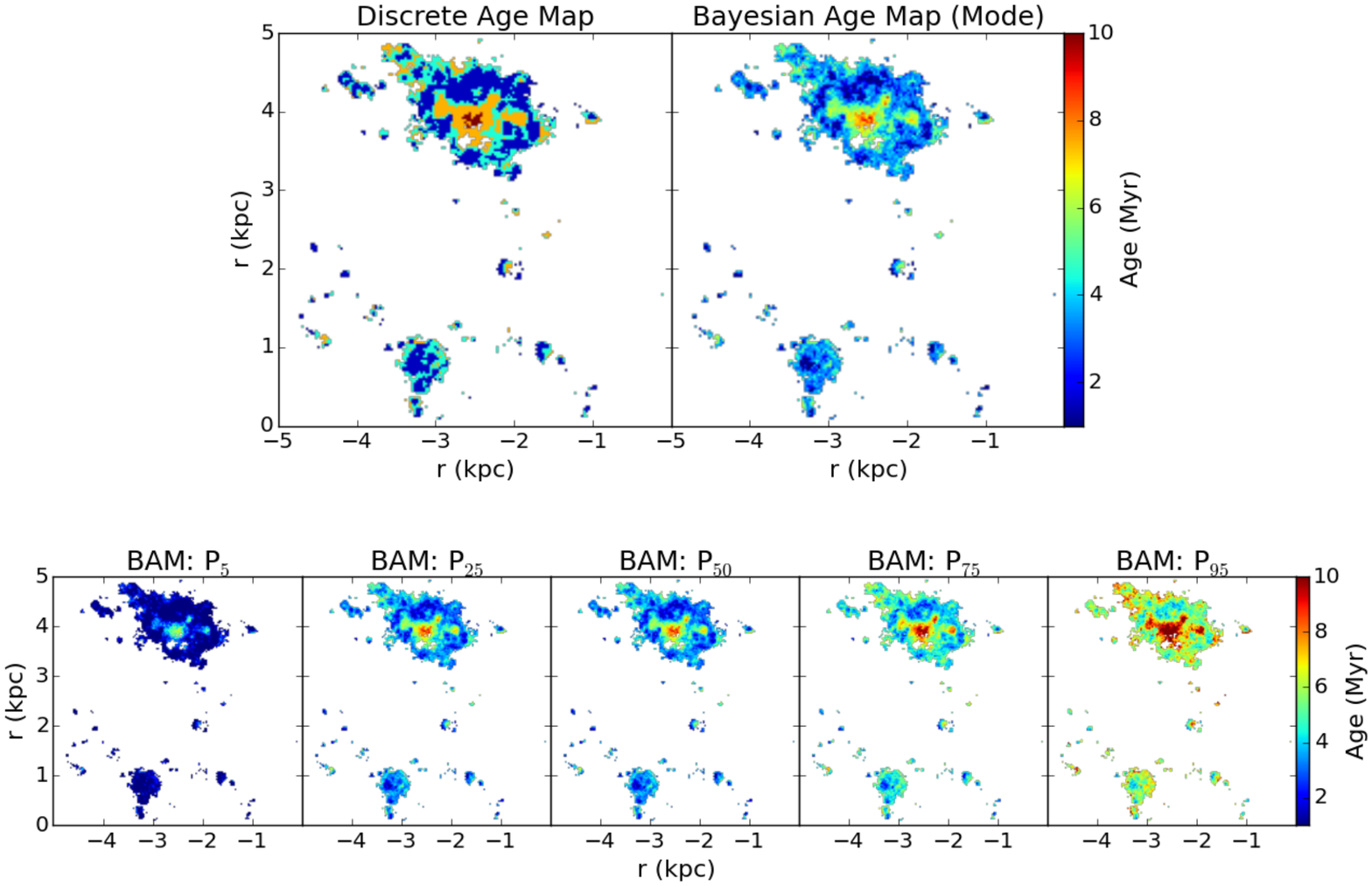}	
\caption{Same as Fig.\ref{fig:FigD1}, but for IC2574}
\label{fig:FigD6}
\end{figure}

Next, the age maps for the galaxies sample in  Paper I are also included. 
In Figs. \ref{fig:FigD1} to \ref{fig:FigD6}, top plots show the comparison  between the age maps obtained in Paper I with those applying the Bayesian approach, from Section \ref{sec:HBM}.

At bottom plots it is checked the uncertainties in age estimation when it is used another estimator than the Mode. For example the percentiles of the resulting sample from the age posterior probability function.  The extreme cases determine a 90\% posterior credible interval for the age estimation.  

\bsp	
\label{lastpage}
\end{document}